\documentclass[referee]{aa} 
\usepackage{graphicx}
\usepackage{txfonts}
\def\me{M$_{\it Earth}$\  }
\def\dt{$\Delta$T$_{s}$\ }
\def\R'HK{R'$_{HK}$\ }

\begin{document}
   \title{Using the Sun to estimate Earth-like planets detection capabilities.
 \thanks{}  }

   \subtitle{I. Impact of cold spots. }

   \author{
     A.-M. Lagrange \inst{1}
     \and
 M. Desort \inst{1}
     \and
   N. Meunier \inst{1}
   }

   \offprints{
     A.M. Lagrange,\\
     \email{Lagrange@obs.ujf-grenoble.fr}
   }

   \institute{
     Laboratoire d'Astrophysique de l'Observatoire de Grenoble,
     Universit\'e Joseph Fourier, BP 53, 38041 Grenoble, France
   }

   \date{August, 08, 2009; accepted October, 28, 2009}

   
   \abstract
   {}
   {It is known that stellar spots may in some cases produce radial velocity (RV) signatures similar to those of exoplanets. so far, the Sun is the star on which we have the biggest set of data on spots, active regions, and activity in general. To further investigate the impact of spots, we aim at studying the detectability of Earth mass planets in the habitable zone (HZ) of solar type stars, if covered by spots similar to the Sunspots.
   }
   {We have used the Sunspots properties recorded over one solar cycle between 1993 and 
2003 to build the RV curve that a solar type star seen edge-on would show, if covered 
by such spots. We also derive interesting parameters such as bisector velocity span (BVS) and photometric curves, commonly used in the analysis of RV data. We compare the 
obtained data with archival solar data available for the same epoch (irradiance, Ca index). We also simulate the RV of such  a spotted 
star surrounded by an Earth mass planet located in the HZ.
   }
   {The RV of the spotted  star appears to be very variable, in a complex way, depending on the activity level, with amplitudes from a few tens cm/s  up  to  5 m/s   
(assuming  \dt = T$_{\odot}$ - T$_{spot}$ =  550K ). A correlation between the BVS and the RV is observed even when several spots are present with a slope nevertheless so small that only very high precisions (better than 5 cm/s) can enable its detection. 
Photometric variations up to 0.5 $\%$  are predicted, depending on the level of activity, in agreement with actually measured Sun photometric 
variations. 

Under present assumptions, the detection of a 1\me planet located between 0.8 and 1.2 AU requires an intensive monitoring (weekly or better), during 
several years. The temporal sampling is more crucial than the precision of the data (assuming precisions in the range [1-10] cm/s). 
Cooler spots may become a problem for such detections.  Also, we anticipate that plages, not considered in this paper, could further complicate or even compromise the detections.

}
   {} 

   \maketitle
%

\section{Introduction}
Most of the exoplanets  so far detected around Main Sequence stars (http://exoplanet.eu) have been identified thanks to the  RV  technique. Their periods range 
between a few days and a few  (up to 5) years; even though most of them are giant planets, a population of super Earths (less than 10 M$_{\it Earth}$ , down to 2 \me , \cite{mayor09}) planets have been detected thanks to very precise RV measurements. In the following 
decade, many efforts will be devoted to the search for even lighter planets, using improved instruments and sometimes larger 
telescopes such as the ELTs (\cite{dodorico07}). 

However, it has also been suspected for a long time that stellar activity and pulsations could 
also produce RV variations which could in some cases, mimic 
those of RV planets. There might then be risks of mis-interpretations of RV variations, 
especially when these variations have periods less or 
equal to the star rotational period. This actually happened recently in the case of 
TW Hydrae (\cite{huelamo08}) or LkCa19 (\cite{huerta08}). 
In order to investigate the impact of stellar spots, \cite{saar97} made first  
estimations of the impact of a stellar spot and of convective inhomogeneities on the RV curve of FeI lines. In the case of an equatorial spot of a star seen edge-on, 
they showed that RV amplitudes up to 50 m/s could be produced by spots or convective inhomogeneities, depending on the spot size, 
 the projected rotational velocity of the star and star age. The impact of spots was also investigated by 
\cite{hatzes02} using the CaI 643.9 nm line, which lead to the conclusions that RV variations in the range of a few m/s could be observed due to spots. Note that \cite{saar97} also pointed out that convective inhomogeneities can lead to  even 
larger RV variations, especially for G2V type stars.

More recently, we investigated in detail the impact of stellar spots on the RV and photometric  curves, and on 
other diagnostics commonly used to disantangle stellar activity and planets in RV surveys (\cite{desort07}). We computed the 
visible spectra of stars 
with various spectral types from F to K type, various projected rotational velocities and various orientations covered with spots 
with different sizes and latitudes. From the obtained, simulated spectra, we derived the RV, photometric and astrometric variations, 
using the same tools as those used in the exoplanets searches, and assuming that the spectrograph used was HARPS (\cite{mayor03}). 
 We showed that the impact on the RV studies is indeed far from negligeable and that in the case of stars with low v sin(i), RV curve, 
bisector/span variations and photometric variations may be not enough to clearly rule out spots as explanations of the observed variations. 
 More quantitatively, spots with typical sizes of 1\% can mimic both radial  velocity curves and bisector behaviour of short periods 
giant planets around G-K stars with $v\sin{i}$ lower than the spectrograph resolution. For stars with intermediate $v\sin{i}$,
 smaller spots may produce similar features. Such spots may therefore complicate the search for low mass planets on orbits with 
periods of the order of the star rational period. Additional observables (e.g., photometry, spectroscopic diagnostics) are in such cases mandatory 
to confirm the presence of short period planets, but unfortunately, photometric variations may in some cases be too 
small to clearly rule out spots as explanations of the observed variations. This is particularly important when searching for super-Earth planets.

Another criterium often used to distinguish spots or more generally stellar activity from planets is the so called Ca activity indicator, either 
expressed in terms of S index or in terms of R'$_{HK}$. Stars with log(R'$_{HK}$) smaller than  about -5.0 (which corresponds to the value measured 
for the Sun at its minimum of activity, see below) are usually considered to be unactive at levels of RV variations of 2-10 m/s. Many of the recent detections 
of very light planets rely on this assumption.

The \cite{desort07} study, as well as the ones quoted above assumed one spot at the surface of the star. The 
example of our Sun shows that the spot pattern is much more complex, with, often, several 
spots, with 
different sizes and timelifes present at a given time on the visible hemisphere. 
The number of spots and their size (which determine the filling factor) also strongly 
vary along the solar cycle. An interesting piece of information would be the knowledge of 
the 
so-called integrated Sun RV variations, moreover analyzed in the same way as in the RV 
searches for exoplanets. However,  RVs of the integrated solar disc  are intrinsically very difficult to obtain, and to our knowledge, the rare available 
measurements were performed in individual lines rather than using a large number of lines 
(which averages out individual line contributions) and the resulting RV curves differ from one author to the other, with in particular amplitudes 
ranging between less than 4 m/s (\cite{mcmillan93}) to 16 m/s (\cite{jimenez86}) or more than 25 m/s (\cite{deming94}). Hence, 
quantitative precise results on the integrated Sun RV variations are still lacking, in contrast with the precise 
and well documented integrated Sun brightness variations recorded over several cycles 
(see e.g. \cite{frohlich98}; \cite{lockwood07} and \cite{livingston07}). 
 
In order to  test the impact of  
spots on stars RV and photometric variations in a more realistic way, we decided to 
use actual  Sunspot data to compute the observables that would be derived when observing a solar type star covered 
with spots identical to the ones observed on the Sun. To do so, we have used the reported Sunspot patterns observed during a full solar cycle to 
 synthetise the spectra of this ``integrated'' Sun and measure 
a number of observables (RV, BVS, periodograms), and to estimate 
the associated photometric variations. Our approach is described in Section 2. The 
results are provided in Section 3 and compared to available data on the Sun variability in Section 4. We then simulate the RV 
curve of such a spotted star surrounded by a 1 \me planet located in the habitable zone (Section 5), and derive some conclusions on the 
detectability of such planets under the present assumptions and in the framework of forthcoming RV instruments (Section 6). 


  \section{Description of the simulations}
  \subsection{Input solar data}


To construct our  spot patterns, we use the  Debrecen Heliographic Data (\cite{gyori03}), a catalogue 
of positions and areas of Sunspots. These have been extracted from white light full-disk images from several observatories,
including the Debrecen observatory. For this work, we have used data covering more than a 
whole solar cycle, between Jan, 1st, 1993 and Dec, 31, 2003: They represent 3700 days of observations.  
The temporal sampling is about 1 day. All spots with 
area larger than 10$^{-6}$ the solar hemisphere (1 micro hemisphere) were taken into account. 
Figure~\ref{example_spots} shows an 
example of an observed map and the simulated spotted surface for JD 2451686. 
Over the whole period, more than 160000 spots were then considered. 
Their position (latitude, longitude) as well as their surface assumed to be circular were recorded and used as 
inputs of our simulation tool. The projected filling factor  f$_{p}$  (i.e. the projection 
of the spot surface over the Sun hemisphere) over the whole period is provided in Figure~\ref{fp_all}.

  
\subsection{Simulated spectra}

For the 1993-2003 period, we then used all spots larger than one micro hemisphere and 
produced 3700 integrated spectra, as described in \cite{desort07}.  
We assume that the star is seen edge-on (indeed, it is not possible to add an inclination 
as we wish to rely only on observed spots). The visible, spotted 3D hemisphere is divided 
into cells 
with sizes adapted both to the spectral resolution of an HARPS like spectrograph, and to 
the minimum size of the spots considered. We then compute the resulting spectrum, assuming that each 
unspotted cell emits a solar-like spectrum and that each spotted cell emits like a black body with a temperature T$_s$.  The spots are then assumed to have a uniform temperature.  Most of the time, we will assume T$_{\Theta}$ - T$_s$ = $\Delta _T$ = 550 K.
 This temperature is compatible with the bolometric spot contrast of 0.32 usually used in irradiance reconstructions (see \cite{chapman94}), who observed contrasts ranging from 0.21 to 0.38), which corresponds to a $\Delta _T$  of about 500K. \cite{krivova03} used a $\Delta$T of 1070 K for spot umbra and 370 K for the penumbra, which is also compatible with our value assuming a realistic umbra to penumbra ratio. 
The ratio between the total area of Sunspots to the umbral area is variable in the literature and is typically  between 4 and 6 (Solanki 2003). Assuming a typical temperature contrast of 1500 K for the umbra (range 1000-1900 K in Solanki (2003)) and of 300 K for the penumbra (250-400K in Solanki (2003)) gives an average temperature contrast between 600 K and 504 K respectively, so our value is consistent with these results. 
We point out however that 1) the precise temperature contrast that reproduces the  correct photometric contribution of the spots to the total solar irradiance depends on the actual spot data set used, and 2) the spot temperature may differ from one spot to the other (\cite{chapman94}. We will also consider different spot temperatures, 
representative of sellar spots temperatures (\cite{berd05}). 
Summing up the contribution of all the cells provides the spectrum of an integrated ``spotted Sun'' (we will refer 
in the following to a spotted Sun to indicate that we took into account only the spots at the surface of the Sun). 

Our first purpose is to identify the specific noise (jitter) induced by the spots; we therefore do not add any instrumental or photon noise to the simulated data in a first step (Sections 3 and 4). Noise will be considered when simulating a planet signal around the spotted star (Section 5).

\subsection{Computation of the various observables}

We then used the SAFIR software (see \cite{galland05}) to compute the RV, the CCF, the corresponding bisector 
span and bisector curvature (see again \cite{galland05}). We  therefore use the same 
procedure as the one we use to search for exoplanets around stars with RV technics. We also compute the 
associated photometric variations at 550 nm.

As the whole process is quite time consuming and as we deal with a huge 
number of spectra (10 to 100 times larger than the number of data usually used to 
identify and publish exoplanet detections), we used only one order per spectrum 
(order 31) instead of the whole spectrum. It is acceptable to do so as considering one order or the 
whole spectrum  gives similar velocities (within 10 $\%$; \cite{desort07}).
 
\subsection{Simulation of a spotted star with planets}
Finally, we compute the RV curve obtained from a star covered by such spots, and surrounded by a planet in the HZ (see Section 5). To do so, we simply add 
up the spots and the planet contribution. In order to study the planet detectability, we have to take into account a noise contribution; 
we therefore  add random noise to the radial velocities, with levels corresponding to the precision expected for the forthcoming instruments.

  \section{Results: spotted Sun variability over the period 1993-2003}
In this section, we first provide the results of the simulations in terms of RV, BVS, and photometric variations, assuming  that the spot temperature is 550K less than the Sun effective temperature. We then briefly study the impact of the spot temperature on these results. 

In order to compare our results with published data on the Sun variability, we chose  one period of reference for low activity : from July, 1, 1996 (JD 2450266) 
to April, 1, 1997 (JD 2450540), and one period of reference for high activity: from February, 1st, 2000 (JD 2451576) 
to November, 1st, 2000 (JD 2451850).  The low and high activity periods last then approximately 10 solar rotational periods. They respectively correspond to average values of R'$_{HK}$ of approximately -4.85 and -5.0  (assuming an average S index 
of 0.170, an amplitude of variations of 0.017 for the S index, as derived from \cite{lockwood07}, a B-V of 0.66 and using the \cite{noyes84} 
empirical relation, we get R'$_{HK}$ between -4.88 and -4.97; these values should be regarded as indicative because the average S values depend on the dates of measurements and to second order, they also depend on assumptions on B-V; this is why the published values generally vary from one author to the other).

\subsection{RV variations }

The {resulting RVs} are clearly variable over the whole cycle, as can be seen in 
Figure~\ref{rv_all}. The RV curve is moreover much more 
complex than predicted by simple modeling of one spot located on a star viewed edge-on 
(see e.g \cite{saar97} or \cite{desort07}), in which case 
the spot signature is present over half the star rotation period (when the spot is on the visible hemisphere).

The amplitude of the RV strongly depends on the filling factor of the spots, as seen in
 Figure~\ref{rv_fp_all}. During the low activity phasis  previously defined, apart from very few (3) peaks at velocities larger than 40 cm/s (absolute value), the RV signal is quite "flat". The rms of the RV during this low activity period is about 16 cm/s, however it is dominated by the few peaks present during the period.
During the defined high activity phasis, the RV signal is much more variable, and on much higher amplitudes: the rms is more than 3 times higher 
(60 cm/s), and peaks as high as 2 m/s (absolute value) are computed (note however that the highest RV peak during the whole cycle is almost at 5m/s). 
Table~\ref{summary} provides the RV rms for the whole cycle as well as for these reference periods. 

Figure~\ref{rv_all} also shows the Lomb-Scargle periodograms of the RV when considering respectively 
the whole cycle, and the low activity and the high activity periods. When considering the whole cycle, we observe a 
peak at the Sun rotation period and one at half a rotation period  with a FAP $\leq$ 1$\%$. The period at half a rotation period can be 
interpreted as the result of ``active longitudes'' (spots are not distributed randomly in longitude, but it seems that two persistent active longitudes separated by 180$^{\circ}$ have been observed to persist over more than one century  (\cite{BU03})). It may also be due to the lifetimes of the large spots, which are often less than 2 solar periods (lifetimes of about 1.5 rotation period for instance could produce an additional peak at half a period). Those peaks are not detected in the yet noisier periodograms of the low and high activity periods.  
 We note that the highest peak detected in the periodogram of the high activity period is at 20.8 days, ie  considerably  different from the solar rotation period. This shows that even when considering 10 solar rotation periods, the periodogram does not reveal the solar rotation period. We  verified 
 that considering a longer period (2000 days) enables the recovery of  the two peaks seen in the periodogram of the whole cycle (see also Section 5).

\begin{table*}[htp!]
  \caption{Summary of measured rms values when modeling all Sunspots: RV (col 2); BVS (col 3); fr: fraction of the Sun covered by spots (col 3); Phot: relative photometry (col 4). The subscript ${\it all}$ refers to the whole cycle; the subscript ${\it low}$ (resp. ${\it high}$) refers to the low (resp. high) activity period.}
\label{summary}
\begin{center}
    \begin{tabular}{l l l l}
\hline
      & RV (m/s)      &  BSV (m/s)      & Phot \\
\hline
rms$_{all }$& 4.35 10$^{-1}$ &        2.0 10$^{-2}$  &    6.1 10$^{-4}$ \\
rms$_{low} $ & 1.60 10$^{-1}$    &   7.9 10$^{-3}$   &      1.9 10$^{-4}$ \\
rms$_{high}$ &        5.88 10$^{-1}$     &   2.6 10$^{-2}$  &    6.9 10$^{-4}$\\
 \hline
    \end{tabular}
  \end{center}
\end{table*}

\subsection{Bisector variations }

Spots induce variations in the lineshapes. One way of quantifying the lineshape variations 
is to measure for each spectrum, the bisector velocity span (hereafter BVS) of a line, or 
a set of lines or the CCF. 
While individual lines are often considered in solar studies or in stellar studies, 
exoplanet searches generally consider the bisector of the whole CCF, in order to get the 
best possible precision on its shape. In such studies, a 
correlation between the RV and the BVS variations is a good indicator that the RV 
variations are due to spots rather than planets. However the absence of correlation (or a 
flat bisector) does not necessarily mean that the RV variations are not due to spots, as 
 shown in \cite{desort07}. The level of correlation indeed depends on the star projected 
rotational velocity and the spectrograph resolution. 

We plot in Figure~\ref{bsv_all} the variations of the BVS of our simulated spectra over 
the whole period. As expected, the BVSs vary with time, over the whole period.  
The RV and BVS variations appear moreover well correlated (see Figure~\ref{bsv_rv}), as 
in the case of a single spot. This is true for low as well as high activity periods. The BVS/RV slope is found to 
be -0.037, irrespective of the activity level. The amplitude of the BVS variations is much smaller (25 times, at 
the level of a few cm/s) than 
the amplitude of the RV variations. The BVS variations will  then be much more difficult to detect 
than the RV variations. Of course, it will be even more difficult to detect the BVS variations in noisy data; we will see below that a precision of 1 or 5 cm/s 
 may enable the detection of the RV-BVS correlation, whereas a precision of 10 cm/s will not.
 The periodogram of the BVSs, when considering the whole cycle, reveals a peak at $\simeq$ 14 days and one at $\simeq$ 7 days, i.e. at periods half of those found when considering the  RV variations. A peak in the BVS periodogram at half the period of the peak observed in the RV periodogram can indeed be expected in the case of equatorial spots at the surface of a star seen edge-on (\cite{desort09}); briefly, this is due to the fact that during each quarter of a period, while the RVs keep the same sign, the BVS sign changes (see for instance Figure 4, bottom left in \cite{desort07}); the number of extrema in the BVS variation curve is then twice the number of extrema in the RV variation curve. In the present case, the spots are equatorial or close to equatorial, and the Sun is seen edge-on, so the same reasonning applies.
Finally, the periodograms of the BVS variations obtained when considering the low and high activity periods are, as in the case of the RVs, very noisy.

\subsection{Photometric variations }


The computed photometry of our spotted Sun also shows temporal variations 
(see Figure~\ref{phot_all}), with amplitudes as high as 0.5 $\%$ in one extreme case, but 
generally in the range 0-0.1 $\%$. The amplitude of the variations strongly vary according to the cycle phase 
 (see Table\ref{summary}). Photometric variations at 10$^{-3}$ level such as the ones predicted during 
the high activity period are detectable from 
the ground, and actually achieved when searching for transits of short period planets. 
They are easily detected with space born telescopes (see e.g \cite{mosser09} in the case of Corot detection of spots around one F8 type star, with a photometric variation level of about 0.2 $\%$). The photometric variations observed during the 
low activity period are on the contrary not detectable from the ground: indeed, the strongest 
peak, responsible for a RV peak at 90 cm/s on JD2450416 produces a photometric minimum of 
0.1 $\%$ (on JD2450413), and the rms of the photometric variations during the low activity period is 0.02 $\%$. 



Figure~\ref{phot_all} shows the periodograms of the photometry of our 
spotted Sun when considering respectively the whole period, the low activity and the high 
activity periods. The periodogram corresponding to the whole cycle reveals the peak at the Sun rotation period, but also a peak at about 110 days. This peak is not seen in the periodogram corresponding to the low and high activity periods. We note a peak at  about twice the rotation period in the case of the high activity period.


\subsection{Impact of the spot temperature}

\cite{desort07} showed in the case of a single spot that its temperature has an impact on 
the resulting RV curve, and also on the BVS and photometric curves. 
 Our aim is here to briefly investigate in a simple way the impact of different temperatures  
on the observables when considering all Sunspots. This is motivated by the fact that in the case of the Sun, the temperature may vary from one spot to another 
(e.g. \cite{chapman94}), with a significant dependence on the spot size (e.g. \cite{mathew07}, 
\cite{wesol08}), with a dispersion of the order of a few hundreds of K. Also, in the case of 
other stars, different spot temperatures are inferred, with a possible trend between the 
spot temperature and the stars effective temperature (see \cite{berd05} for a review).

 
\subsubsection{Spots temperature 1200K less than the Sun effective temperature}
We ran then another complete simulation, as described in the previous section, assuming that the spot 
temperature is 1200K (instead of 550K) below that of the Sun. { The RV variations due to spots of temperature 1200K lower than the Sun effective 
temperature are larger by a factor of 1.75 than those obtained assuming \dt = 550K. They will then induce a stellar jitter 1.75 times larger, which is not negligeable.

Under the same assumption, we find a BVS/RV slope of -0.037, similar to the one found assuming \dt = 550K. We conclude that the BVS/RV 
slope is not sensitive to the temperature within the considered range of temperatures.
 Also, the ratio BVS(1200)/BVS(550) is found to be 1.75, ie.. similar to the one obtained for the RVs. 
 Finally, the amplitude of the photometric variations for the cooler spot is found to be larger by a 
factor of 1.8 than the warmer spot. We note that the amplitude of the photometric variations 
during the low activity period is still below the precision achievable in ground based observations.

\subsubsection{Various spots temperatures}

To investigate other temperatures, we limited our simulations to the case of one single spot, with various temperatures lower by 200, 400, 600, 1000 and 1200K from the Sun effective 
temperature. We first check that the ratio V$_{Max}$~(1200)/V$_{Max}$~(550) where V$_{Max}$~(550) (resp. V$_{Max}$~1200)) is the maximum RV amplitude assuming \dt = 550K (resp. 1200K), is consistent with the ratio obtained when 
considering the whole set of spots. The maximum RV obtained in each case is plotted as a function of the spot 
temperature in Figure~\ref{Ts_vmax_bsvmax_abs}.  
Figure~\ref{Ts_vmax_bsvmax_abs} also gives the maximum BVS and the maximum flux absorption as a function of the spot temperature. We see 
that the spot temperature impacts significantly these observables, with 
larger amplitude signals in the case of cooler spots. These figures allow to estimate the RV, BVS and photometric variations for any spot temperature. 

\section{Comparison with published data on the integrated Sun over the period 1993-2003}
\subsection{Radial velocity data}
The amplitude of our computed RV variations has an rms of 60 cm/s and may be as high as 
5-6 m/s during the high activity period. As already mentionned a few peaks 
at 1m/s may be found during low activity periods, but the rms of the variations is much smaller 
during the low activity period.

 \cite{jimenez86} published actual RV measurements on the integrated Sun between 
1976 and 1985 (cycle 21), collected with a resonant scattering spectrometer. The strongest 
variations are obtained in 1982, with amplitude of variations of about 16 m/s at maximum activity. 

 \cite{mcmillan93} published RV measurements in the period 1987-1992, measured on spectra of 
the Sunlit surface of the Moon, in the blue part of the spectrum, between 425 and 475 nm. They did 
not see variations larger than 4 m/s over this 5 year period. However, they noted that the lines
 present in the considered spectral range could be less sensitive to activity 
than other lines, such as those considered by \cite{jimenez86}. During the same period, between 1984 and 1992, 
\cite{deming94} also recorded RV variations of the integrated Sun, using the CO lines at 2.3 microns. 
Their measurements revealed variations, with peak to peak values of about 25 m/s. 

Our simulated RV amplitudes are smaller than those measured by \cite{jimenez86} or \cite{deming94}.
 This might be due to the fact that the Sun was less active in cycle 23 than in the previous 
cycles. It might also be due to the fact that we take into account only spots and not plages and convection 
in the present simulations (but see below). It might  also be due to the fact that our predicted 
RVs are ``measured'' on the CCF and not on single lines. However, the strong discrepancies between 
the published measurements  prevent us  from deriving firm conclusions, and from validating or not 
our simulations. 
 
\subsection{Photometric data}

In this section we compare our simulated photometric variations 
due to spots to the measured photometric data such as the Total 
Solar Irradiance (hereafter TSI), the measured Spectral Solar Irrandiance (hereafter SSI). The aim is 
to test our spot model, 
even though we do not expect this model to fully reproduce the Sun
photometric variations as it is well known that plages strongly contribute to the Sun's brightness 
variations (see below). 


\subsubsection{Total Solar Irradiance (TSI):}

We first compare our computed photometric variations 
due to spots to the observed Total Solar Irradiance (TSI) 
variations during the period 1993-2003, compiled by Claus Fr\"ohlich and Judith Lean, and available 
at http://www.ngdc.noaa.gov/stp/SOLAR/. Figure~\ref{jd_tsi} provides the measurements recorded over 
the whole cycle, as well as zooms on the variations during our two reference activity periods. It 
also shows the corresponding periodograms.

Over its cycle, the Sun's observed brightness (see e.g. \cite{frohlich98} and \cite{lanza07}) increases 
with the activity due to the plages. In addition, sharp brightness decreases are sometimes
 observed, due to spots. A relative difference of about 1 W/m$^2$ is observed in the TSI 
between the solar minimum average and the maximum average. From \cite{frohlich98} reconstruction of 
the plage and spot contributions, it seems that plages dominate the long time scale variations, 
producing photometric amplitudes of about 2 W/m$^2$ during this cycle 
and that the spots contribution to the 
long term variations is twice smaller (relative intensity of about 0.9993). In addition, the 
spot contribution to short time scales variations may be significant, with transient absorption 
peaks up to 3-4 W/m$^2$. Such values for the peaks due to spots are in excellent agreement with 
our simulated photometric variations as can be seen when comparing Figure~\ref{phot_all} 
to Figure~\ref{jd_tsi}, especially during the low and high activity periods. For instance, the 
isolated 
simulated narrow peak (during low activity), due to a spot, is simultaneously observed in the TSI 
curve, and the simulated intensity on JD 2450413 with respective intensities of $\simeq$ 0.07 $\%$ 
and $\simeq$ 0.11 $\%$.  The discrepancy between the simulated photometric variations (0.11\% ) and the TSI curve variations (0.07\% ) is rather small, given the fact that the TSI variations are not only due to spots, but also to plages; we therefore do not try to attribute it to physical effects (such as a lower spot temperature in our simulated spots). Also,   
the temporal variations of the TSI and of the simulated photometry are  remarkably similar 
during the high activity period, even when several spots are present on the visible hemisphere 
at a given time. 
 We conclude then that the present simulations of the Sunspots are correct and that 
the assumption of a single temperature for the Sunspots is acceptable. Therefore, we can apply them 
safely to stars  for which activity is dominated by spots.

The comparison between simulated and observed data also confirms that the Sun short 
time scale photometric variations during high activity are quite 
sensitive to the spots and that the variations of the Sun brightness during the low 
activity period are not dominated  by the spots. Consequently, we can stipulate 
that, as far as the Sun is concerned, our simulations provide a good enough description of 
its RV variations during the high activity period but may not provide a good estimation of the 
Sun RV variations during the low activity period.

%

The good agreement between the TSI variations and the simulated photometric variations during the high activity period can also be seen in Figure~\ref{tsi_rv_phot}, where we have plotted the TSI and simulated spots flux absorptions when considering the days (3568 days over the cycle) for which we have both spots identifications (hence photometric data from our spot simulations) and actual TSI measurements. Whereas we do not see any general correlation between both quantities if we consider the whole cycle (or the low activity period, except in the rare cases when spots are present), we see a rather good correlation when we focus on the high activity period.  

The periodogram of the TSI (Figure~\ref{jd_tsi}) over the whole cycle is quite noisy; it reveals one peak at the Sun rotation period, also seen in the simulated data (Figure~\ref{phot_all}), with a false alarm probability better than  but close to  1 $\%$.
It reveals also other peaks, with however lower confidence levels, in particular at about 20 days; the discussion of these peaks is out of the scope of the present paper, and we refer to 
 \cite{hempelmann03} for their discussion. 
The periodogram corresponding to the low activity period appears to be quite different. In particular, the  very clear (FAP much better than 1$\%$ ) peak observed at 26 days 
in the TSI periodograms is not seen in the periodogram of the corresponding simulated data; which is quite coherent with what is directly seen on the photometric curves (strongly modulated signal). 
On the contrary, the periodogram corresponding to the high activity period appears to be more similar  to the simulated one; in particular, they both reveal a peak at about twice  the solar rotation period. The similarity between both periodograms is again coherent with the fact that the short  timescale variations are more sensitive to the spots during the high activity period.


\subsubsection{Spectral Solar Irradiance (SSI):} 

Similar conclusions are reached when we compare the simulated data (Figure~\ref{phot_all})
with the SSI data at 402, 500 and 862 nm during the reference activity periods (Figure 6 of \cite{lanza04}, 
see also \cite{fligge98}, \cite{fligge00}). 
We note moreover that the simulated data are closer to the 402 nm SSI data than to the longer wavelength data. This shows that the 
short wavelengths may be more sensitive to the spots than the longer wavelengths.
The SSI indeed exhibits a larger amplitude of the spot signal in the blue part of the spectrum compared to the red. 
We note that this is also true for the plage signal, therefore it is clear that 
the shorter wavelengths are a better indicator of activity than the longer wavelengths.  

\subsection{Ca index}

While the TSI variations are due to both plages and spots (although dominated by plages on long 
time scales), Ca variations are due to the chromospheric emission of plages. 
\cite{lockwood07} reported on 
the Ca variations (S index and Ca core emission) over 3 cycles, 
from 1975 to 2004. As expected, the Ca variations over the 3 cycles roughly follow 
the solar cycles, as seen from the comparison of the TSI temporal variations and the Ca 
temporal variations (see also \cite{livingston07}). We note that the amplitudes of both the 
TSI and the Ca variations significantly vary depending on the cycles; in particular, the last one (cycle 23) corresponding to the data used in the present paper show the smallest amplitude among the 3 
cycles, indicating an activity level lower than during the previous cycles. 
Importantly in the present context, we note the presence of high frequency variations in the 
measured Ca indexes all over cycle 23 (see Figure~\ref{jd_caindex}), at maximum activity 
(with peak to peak variations larger than 0.01), but also at minimum activity (peak to peak variations: 
0.005). 


The periodogram of the Ca index over the whole cycle does not reveal any significant peak apart 
from one at 7 days (and 3.5 days) which we checked are also present in the 
periodogram of the temporal sampling of the data. Some peaks may be present at about 26 days, 
but the associated false alarm probability is  in the range 1-10$\%$ . The periodogram corresponding to 
the low activity period reveals a peak at the Sun rotation period, with however a 10$\%$ false 
alarm probability. Finally, the periodogram of the high activity period does not bring valuable 
information. This shows that the Ca index periodogram 
hardly provides valuable quantitative information on the periodicities involved, even though we are 
dealing with a relatively large number of data.


As the Ca index variations are mostly controled by the plages, we do not expect to find a clear correlation 
between the Ca index and the presence of spots. This is confirmed in Figure~\ref{rv_caindex} where we show the 
(RV - Ca index)  diagram corresponding to the days (1134 in total) in period 1993-2003 for which are available at the same time 
the spots simulations and the Sacramento Peak Solar Observatory Ca index measurements from 
the http://www.ngdc.noaa.gov/stp/SOLAR/ database. However, we see that the higher RV amplitudes are found during periods of high Ca indexes. This is consistent with the fact that the largest spots are present during high activity  periods. 




\subsection{Notes on the plages}

Plages are large structures slightly hotter than the Sun surface. The long timescale variations of the Sun photometry are dominated by plages, even though spots do have clear 
signatures in short timescale variations.  

Spots have been inferred since several decades at the surface of stars, 
either dwarfs or young stars, due to rotation-modulated flux decreases in the 
photometric curve. Also, Doppler imaging has enabled maps of such 
spots to be produced in the case of high  projected equatorial velocities stars (early type stars or active stars; 
see e.g. \cite{donati97}; \cite{skelly08}). Plages have been recently inferred 
from long term (years or decades) photometric surveys (\cite{lockwood07}) and from  
line profile studies of young spotted stars (\cite{skelly08}; \cite{skelly09}). According to 
\cite{lockwood07}, the long term photometric variability of 
stars with $<$ log(\R'HK) $>$ lower than $\simeq$ -4.75, is plage-dominated, with an increase of brightness at maxmimum activity, like in the 
case of the Sun, where as stars with $<$ log(\R'HK) $>$ larger than $\simeq$ -4.75 are 
spot dominated. Note however that 1) this limit value of \R'HK is 
approximative, and 2) \cite{hall09} have recently questioned the correlation between activity and brigntness in the case of very low activity stars, as well as the pertinence of \R'HK (or S index) for stars with low activity.

 
In any case, if present on the surface of stars in addition to spots, plages  
could a priori induce RV variations as spots do. As our paper is devoted to the impact of spots on the RV variations, a detailed study including the plages is not presented here. However, it is interesting to make a rough 
estimation of their contribution. 

To do so, we computed the contribution of a plage to the RV variations. Following \cite{lanza07}, we 
assume that the temperature difference between the plage and the Sun surface varies as 1-$\mu$, where $\mu$ = cos($\theta$) and $\theta$ is the angle between the normal to the surface and the line of sight; using the plage 
contrasts observed by \cite{ermolli07}, the plage temperature is then Tp = 5800 + 85 (1-$\mu$) K. 
We furthermore assume a filling factor   fp = 20 $\% $  for the plage, 10 times more than the spot filling factor. 
Figure~\ref{contrib_spot_plage_tot} provides the RV variations induced by such a plage located at the equator, as well as 
those produced by a spot 550 K cooler than the Sun. The total RV variations, due to the spot and the 
plage is also provided. 
In this simple example, the RV signal due to the plage appears to be slightly lower in amplitude and of opposite 
sign wrt the RV signal produced by spots. 
When adding up both signals, assuming then that the spot and plage are seen at the same time and are centered at the same place, 
the effect of the plages is therefore to reduce slightly the amplitude of the RV variations due to spots alone, 
and to modify slightly the position of the RV extrema.

Of course, this is a very simplified case and the real situation is more complex, as the ratio between  
the plage filling factor and the spot filling factor exhibits in practice a large 
dispersion (\cite{chapman01}), and as many plages are seen without beeing associated to spots (this is clearly seen 
for instance in Figure~\ref{jd_tsi} during the low activity period). The actual total RV variations 
when considering spots and plages will therefore be actually more complex. We can anticipate that the low activity period will be particularly affected. If we assume that the relative impact of the spots compared to plages on the RV will be comparable to the one observed on the TSI curve during the low activity period, then the plages would induce a pattern of variations with amplitudes of about 20 cm/s, strongly modulated by the Sun rotation period. As already mentioned, we expect the high activity period to be comparatively less affected by the plages, even if they are of course present.

We will  develop  a complete and consistent set of simulations taking into account the plages in a subsequent paper. The possible impact of convective downflows, not considered here within the plages will be also investigated in this paper.

\section{Earth mass planets in the HZ of a spotted solar-type star}

In this section, we wish to illustrate the impact of spots on the detectability of Earth mass 
planets in the HZ of solar stars observed with the future generations of high precision spectrographs, such as Espresso on the VLT or Codex on the E-ELT (\cite{dodorico07}) .
We then assume that the spotted solar type star is surrounded by a 1 \me planet on a circular orbit, located at a = 1.2, 1 or 0.8 AU.  Such values are  representative of the inner and outer boundaries of the HZ for a $\simeq$ 1 solar mass MS star (see e.g \cite{jones06}). For instance, in the case of the Sun, \cite{kasting93} find HZ boundaries of 0.95 and 1.37 AU, but the limits of the continuously HZ (CHZ), which takes into account the variations of the solar luminosity with its age are more restrictive : 0.95 to 1.15 AU.  
 The RV amplitudes associated to these radii are respectively 8.1, 8.9 and 9.5 $^+_-$ 0.05 cm/s, and the periods associated with  these radii are respectively about 480, 365 and 261 days.  We assume that the precision of the RV data is either 10 cm/s, 
which corresponds to the precision requirement of the Codex instrument on the ELT and 5 cm/s or even 1 cm/s which corresponds to the goal on the Espresso instrument on the VLT. Otherwise mentionned, we assume that the spot 
temperature is 550 K below the Sun effective temperature. 
We assume that the star coordinates are (RA, dec) = (00:00:00, -45:00:00) and we consider that it is observable under acceptable 
conditions whenever its airmass is lower than 1.5, hence, given its declination, during 8 months. 
We then assume that the star is actually observed either every night during these 8 months (best temporal sampling), or every 20 nights, or 
8 nights or 4 nights. We computed all RV curves and associated periodograms corresponding to these cases (orbital radius {\it a}, temporal sampling, precisions), and tried to fit the RV curves in a number of interesting cases. We detail here the results obtained in the case of a planet 
located at {\it a} = 1.2 AU, and then discuss the impact of the planet orbital radius and the impact of the spot 
temperature.
 

\subsection{Case {\it a} = 1.2 AU} 
 Figure~\ref{1p2_4m} provides the resulting RVs for a one-day interval of observation, as described above, and for precisions of 10 cm/s, 5 cm/s and 1 cm/s.  The associated periodograms are also provided. These results are given for the whole cycle, as well as for low and high activity periods. Note that in this section, we extend as much as possible the duration of the low and long activity periods so as to cover several planet orbital periods: the low activity period now extends from JD 2449500 to JD 2450900 and the high activity period  from JD 2451000 and JD 2453000. These periods last then 1400 and 2000 days respectively. 

 We see in Figure~\ref{1p2_4m}  that the planet signal is clearly 
visible in the RV curve at most during about 3-4 years, during the lowest activity period; it is much less visible in the data corresponding to 
the high activity period. The periodogram corresponding to the low activity period 
clearly reveals the planet orbital period in addition to the activity-related peak at about 14 days (which we note, was not detected when considering a 400 day period, but is well detected now on this 1400 day period considered). The  FAP of the planet peak is  smaller than 1 $\% $. We note that this peak is higher than the activity-induced ones, whatever precision is assumed. When considering the whole cycle, 
the planet-induced peak intensity gets fainter than the activity-related peaks, but has still a FAP  smaller than 1 $\% $. When considering the high activity period, the peak associated to the planet has a FAP larger than 10 $\% $. (We note nonetheless that as it is well detached from the other peaks, it can still be identified).
This is coherent with the fact that the low activity period is not strongly affected by the spots, conversely to the high activity period. Importantly, we see that the peak detectability very marginally depends on the precision of the data in the present case. 

In the case where the star is observed every 4 days (see for instance  Figure~\ref{1p2_tempsampling} in the case of a 5 cm/s temporal sampling), the conclusions remain qualitatively similar; however, the planet peak in the periodogram is lower than in the previous case (daily observations), and only in the case of the low activity period is its FAP smaller than 1 $\% $. Note that we detect several additional peaks that are also present on the periodograms of the observing dates, and are signatures of the actual sampling. Again, the precision does not have a significant impact on the results. 

 When the temporal sampling is worse (Figure~\ref{1p2_tempsampling}), the peak corresponding to the planet period  becomes much less clear in the case of a temporal sampling of 8 days, with  a FAP larger than 10 $\% $, and is no longer detectable in the case of a temporal sampling of 20 days during the low activity period.  During the high activity period, it cannot be identified already for  an 8 day sampling. Again, the precision of the data does not significantly impact the conclusions. 

The precision has nevertheless a major impact on the capability to detect the signature of the spots, as seen in Figure~\ref{rvtot_bsv_noise_550} where we plot the (RV;BVS) diagrams assuming precisions of 1cm/s, 5 cm/s and 10 cm/s, as well as an infinite precision in the planet+spots case and in the spots-only case for comparison. When considering the planet+spots case, the RV-BVS correlation is quite clear when assuming no noise, or even a 1 cm/s noise. The correlation is not so clear when assuming a 5 cm/s noise and is not clear at all with a 10 cm/s noise. 
In the spots-only case, the conclusions are similar, as the (RV;BSV) curves are quite similar. This is because the 
planet induces variations with  RV amplitudes smaller than 8.1 cm/s, which vary with a significantly larger amplitude due to the spots (see Section 3), and of course, the planet does not induce BVS variations. The impact of the planet is therefore 
 quite negligeable.  With a much closer or more 
massive planet, the conclusions would naturally be quite different.

 We then tried to fit the RV curves to check whether the planet parameters (orbital radius, mass) could be retrieved. We first assumed that the star is observed with the best temporal sampling. In this case, the fit of the RV data corresponding to the whole cycle provides a minimum $\chi^2$ corresponding to parameters close to the input values, within the uncertainties. This means that the data enable detection of  the planet and to derive a proper estimate of its orbital parameters. This is also true when considering the RV data corresponding to the low activity period. 
During the high activity period however, the minimum $\chi^2$ found corresponds
to small periods, typically in the range [0-30] days (corresponding to the largest peaks in the periodograms). We can still find a reasonably good fit of the RV curve providing a range of periods, including the planet period, roughly estimated from the periodogram, is given. Similar conclusions are reached when considering data obtained with a 4 or 8 days temporal sampling. We show an example in Figure~\ref{fits_1p2_4m_4d_1cms} in the case of a 4 day temporal sampling and a noise of 1 cm/s. We have performed 10$^6$ realizations of a planet RV signal with 3 parameters: the period is taken randomly between 0 and 1000 days, the phase between 0 and 2$\pi$, and the amplitude between 0 and the maximum of the observed signal. For each realization, the $\chi^2$ is computed. The upper panels of the figure  show the smallest values of the $\chi^2$. In this example, we obtain at a 1$\sigma$ level: a = 1.175$^{+0.022}_{-0.020}$\ AU and M = 1.21$^{+0.16}_{-0.15}$ \ \me, and at a 2$\sigma$ level: a = 1.175$^{+0.034}_{-0.032}$\ AU and M = 1.21$^{+0.25}_{-0.25}$\ \me. If the temporal sampling is worse (every 20 days), no good fit of the RV curve can be obtained, even when considering the whole cycle or the low activity period.

We conclude that under the simple assumption of a planet on the circular orbit, the planet signal can be detected, provided the star is observed with a very good temporal sampling (better than every 8 days), and over 4 periods at least, during the low activity period. 
During high activity period (log(\R'HK) $\simeq$ -4.85), only a very high temporal sampling may allow to detect the planet signal, but deriving its orbital parameter becomes very difficult. The temporal sampling is crucial to detect the planet signal,  whereas  the precision is not so important. The interest of the RV precision remains to identify stellar activity.

\subsection{Impact of the orbital radius}
Similar conclusions are reached when considering a planet orbiting at 1 or 0.8 AU. Of course, the smaller the orbital radius, the larger the planet signal, and the easier it will be to detect the planet, for a given temporal sampling. We nevertheless note that the peak at the planet orbital period appears at a somewhat smaller intensity in the case of {\it a} = 1 AU; this is due to the fact that the planet orbit cannot be completely sampled because the planet period exactly matches 
the observability period of the star. This is illustrated in Figure~\ref{4m_4d_5cms} where we show the periodograms obtained in the case of a temporal sampling of 4 days and a precision of 5 cm/s. 

 From the former two sections, we conclude that under the present assumptions (\dt =  550K), a 1\me planet on a circular orbit in the HZ can be detected providing the  temporal  sampling is good enough (better than every 8 days), and long enough (during at least 4 periods). The temporal sampling is therefore crucial to reveal the planet period and to allow a proper fit. This is very demanding in terms of telescope time but seems to be mandatory.
 On the contrary, the precision (ranging between 1 cm/s and 10 cm/s) does not significantly impact the detectability of the planet.

\subsection{Impact of the spot temperature}
We then consider a spot temperature 1200K below the Sun temperature, instead of 550K. The cooler spots induce RV amplitude higher (factor 1.75) than those induced by the former spots. The effect is that for a given sampling, the peak due to the 
planet in the periodograms is lower than for the \dt = 550K. Again the impact is stronger
 when considering the high activity period than the low activity period and the precision does not 
significantly impact on the peak detectability. We show in 
Figure~\ref{4m_4d_impacttemp} the comparison between both cases, assuming the star is
 observed every 4 days, and the precision is 5 cm/s. 

The RV fitting now provides good results only when considering the low activity period and if the temporal sampling is the best one (every day over the observability period) or every 4 days.

\section{Concluding remarks}
\subsection{Summary}
We computed the RV variations due to the set of spots observed on the Sun surface between 1993 and 2003. These variations are 
representative of the RV signatures that would be measured on a G2V star with a spot pattern and a rotation period 
identical to that of the Sun, and seen edge-on. The main results are the following:
\begin{itemize}
 \item[-] Due to the large number of spots, RV variations are always present, with amplitudes up 
to a few m/s if we assume the the spots are about 550 K cooler than the Sun photosphere. 
\item[-] The RV amplitudes are very variable, between 0.2 and 5 m/s, 
depending on the activity level. Such amplitudes are larger than the ones estimated from single 
spot consideration, and are  much larger than the  $\simeq$ 9 cm/s RV induced by an Earth 
mass planet orbiting at 1 AU from the star. During the low activity period, the RV curve is 
significantly more quiet and with much smaller amplitudes than during the high activity period.
\item[-] BVS variations do occur, but the amplitude of the (RV,BVS) slope is quite small, 
so that such variations would require very high precisions (better than 5 cm/s) to be detectable during 
the low activity period.
\item[-] RV and BVS periodograms require a lot of data to provide significant results. They reveal 
peaks at periods sometimes very different from the solar rotation period. This should be taken into account before attributing peaks at periods different from the star rotation period to planets rather than to spots. 
\item[-] Assuming spot 1200K cooler than the Sun increase the amplitude of the RV or BVS curves
  by a factor 1.75.
\item[-] The simulated photometric variations are in very good agreement with the spots signatures effectively observed on the Sun during the same period. The amplitude of the variations is relatively low, and in particular, would not be detectable from the ground during the low activity period. The comparison with the observed TSI moreover shows that during the low activity period, even though the highest variations are due to spots, the plages produce, a modulation of the TSI on timescales of the Sun rotation velocity. These plages could affect the RV as well.  On the other hand, during the high activity period, the TSI variations seem to be quite well matched by the spots induced variations. 
 
\end{itemize}

Simulations of a spotted solar-type star surrounded by a 1 \me planet located in the HZ show that 
the planet period can be  detected (with $\leq$  1 $\% $ false alarm probability) in the periodograms of the RV, provided the star is observed 
enough (more often than every 8 days) during a long period of time (at least 4 periods). The signal is seen much more clearly during the low activity period (much fewer spots), and RV fitting enables satisfactory recovery of  the orbital parameters. During the high activity period, the planet signature can be identified,  providing  excellent temporal coverage is obtained. The  orbital parameters will be derived with more difficulty. This means that detecting Earth mass planets may in some cases (e.g. around stars with activity levels comparable to that of the active Sun, or stars with short observability periods) 
be hardly achievable and in any case demand large allocations of telescope time. The  precision of the data in a range 1-10 cm/s  is not found to impact significantly these results, but on the other hand, helps identifying spots signatures.

The case developped in this paper is however very favorable, for several reasons: 
\begin{itemize}
\item[-] the planet is alone and is on a circular orbit
\item[-] the star is seen edge-on, hence the long lived spots are hidden half of the time. In case where the 
star is seen inclined, the spots may be observed all the time and their signature, even though smaller 
in amplitude, is more comparable in shape to the one expected from planets (see Desort et al, 2008).
\item[-]  the level of activity of the Sun was rather low  the cycle considered
\item[-]  the spot temperature was assumed to be constant, only 550 K cooler than the Sun temperature, a value which seems to fit quite well the Sun data, but may be in the lowest range of possible values if we consider stars instead of the Sun (\cite{berd05})
\item[-]  no plages or convective flows were considered; where as we have shown that the impact of the plages on the RV of the Sun is not negligeable, expecially during the low activity period. Hence, if solar type stars with similar activity are also covered by plages similar to the Sun in addition to spots, their RV jitter will be higher at least during the low activity period, than in the case described in this paper, which would have a strong and negative impact on Earth mass planet detectability.
\end{itemize}

\begin{acknowledgements}

  We acknowledge support from the French CNRS. We are grateful to 
  Programme National de Plan\'etologie ({\small PNP, INSU}), as well
  as to french Agence Nationale pour la Recherche, ANR. We thank J\'er\^ome Bouvier for fruitful 
discussions, and Severine Pouchot for her participation. The spot data have been
  provided by the Debrecen Observatory. We acknowledge the Sacramento Peak
Observatory of the U.S. Air Force Phillips Laboratory for providing the Ca index. The irradiance data set (version $\#$25) have been provided by PMOD/WRC, Davos, Switzerland and we acknowledge unpublished data from the VIRGO experiment on the cooperative ESA/NASA mission SOHO. We also made use of the INSU/CNRS database BASS2000.

\end{acknowledgements}

\vfill\eject
\begin{figure}[ht!]
  \centering
  \includegraphics[width=.8\hsize]{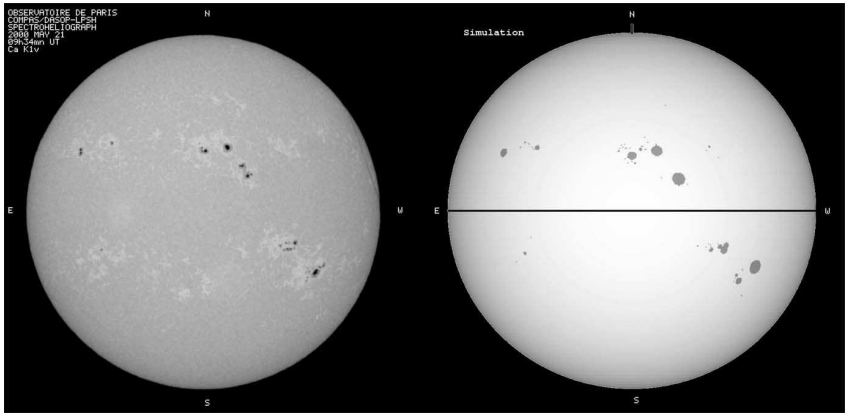}
  \caption{ Left: Sun surface observed with the Paris Observatory Spectroheliograph on JD2451686. Right: simulated surface with spots derived from Debrecen Heliographic Data on the same day (see text).} 
  \label{example_spots}
\end{figure}


\begin{figure}[ht!]
  \centering
  \includegraphics[angle=0,width=\hsize]{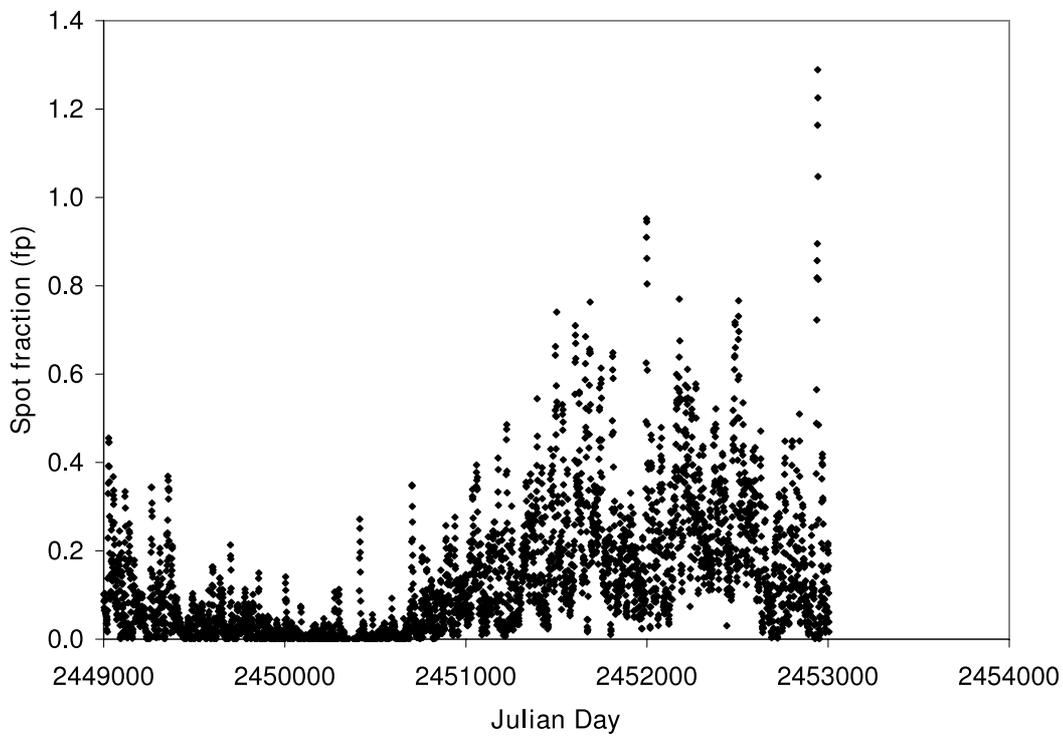}
  \caption{ Temporal variations of the projected filling factor over the whole period (see text).} 
  \label{fp_all}
\end{figure}

\begin{figure}[ht!]
  \centering
 \begin{tabular}{cc}
\includegraphics[angle=90,width=.3\hsize]{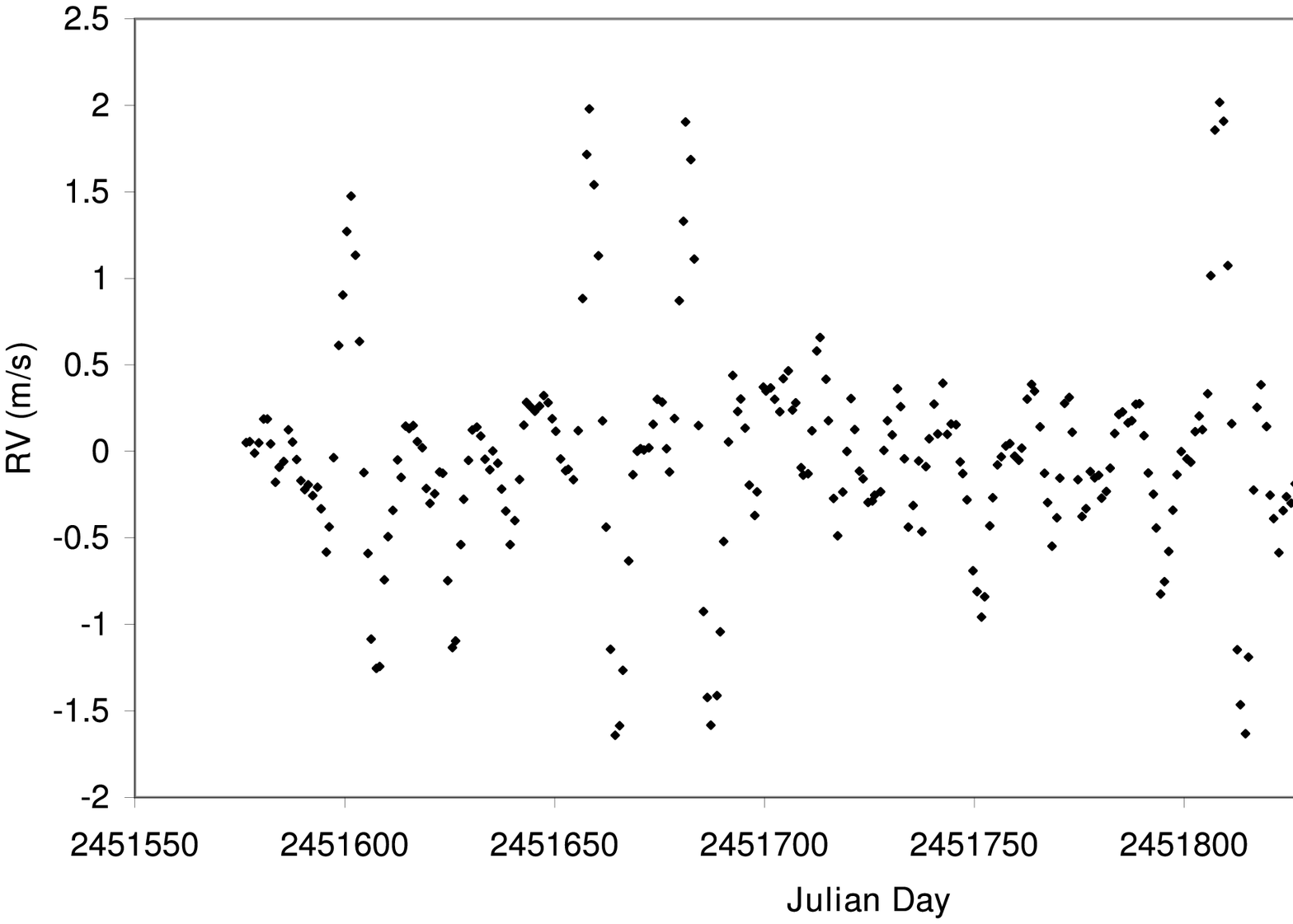} & \includegraphics[angle=90,width=0.3\hsize]{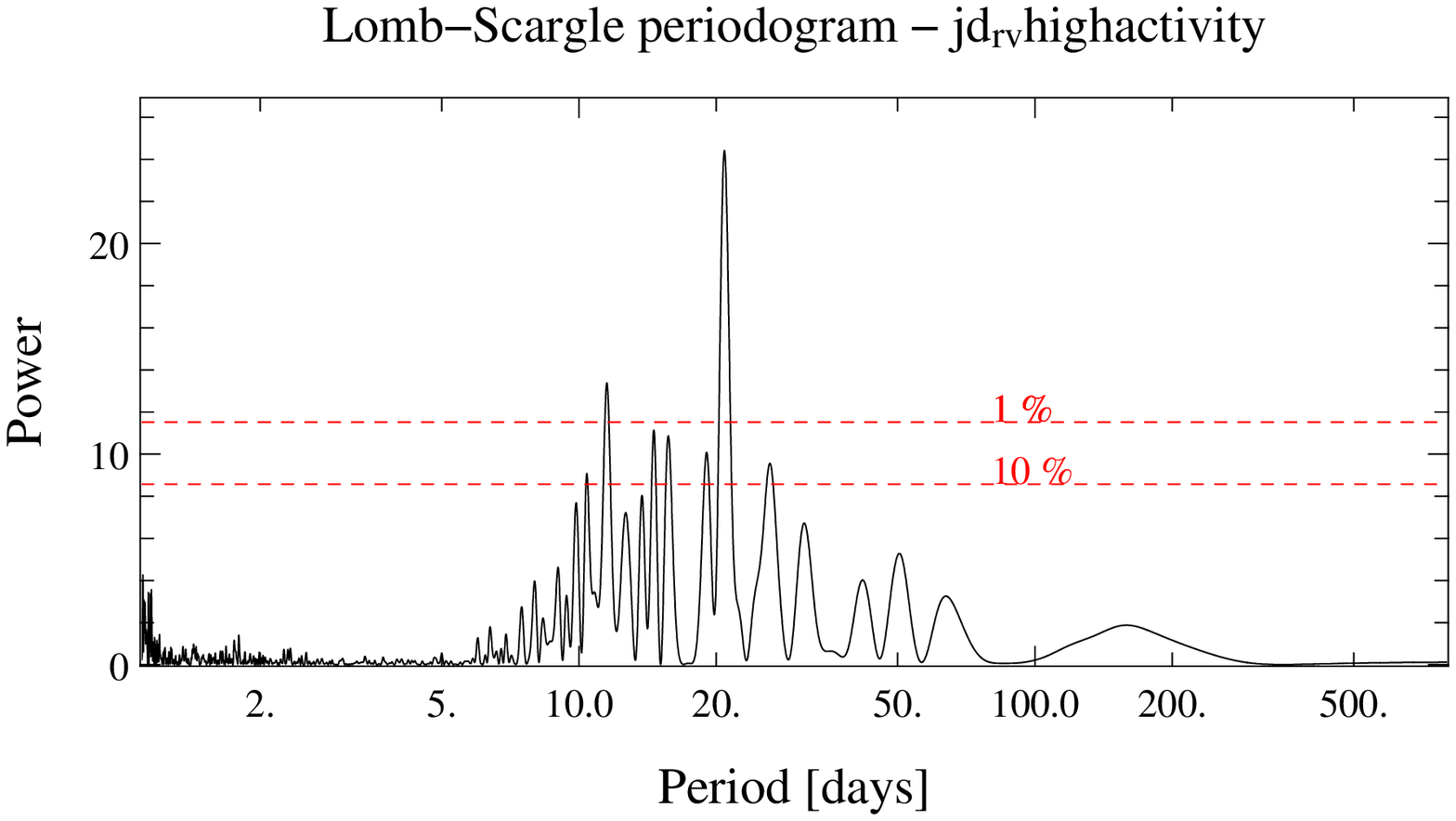}\\
 \includegraphics[angle=90,width=.3\hsize]{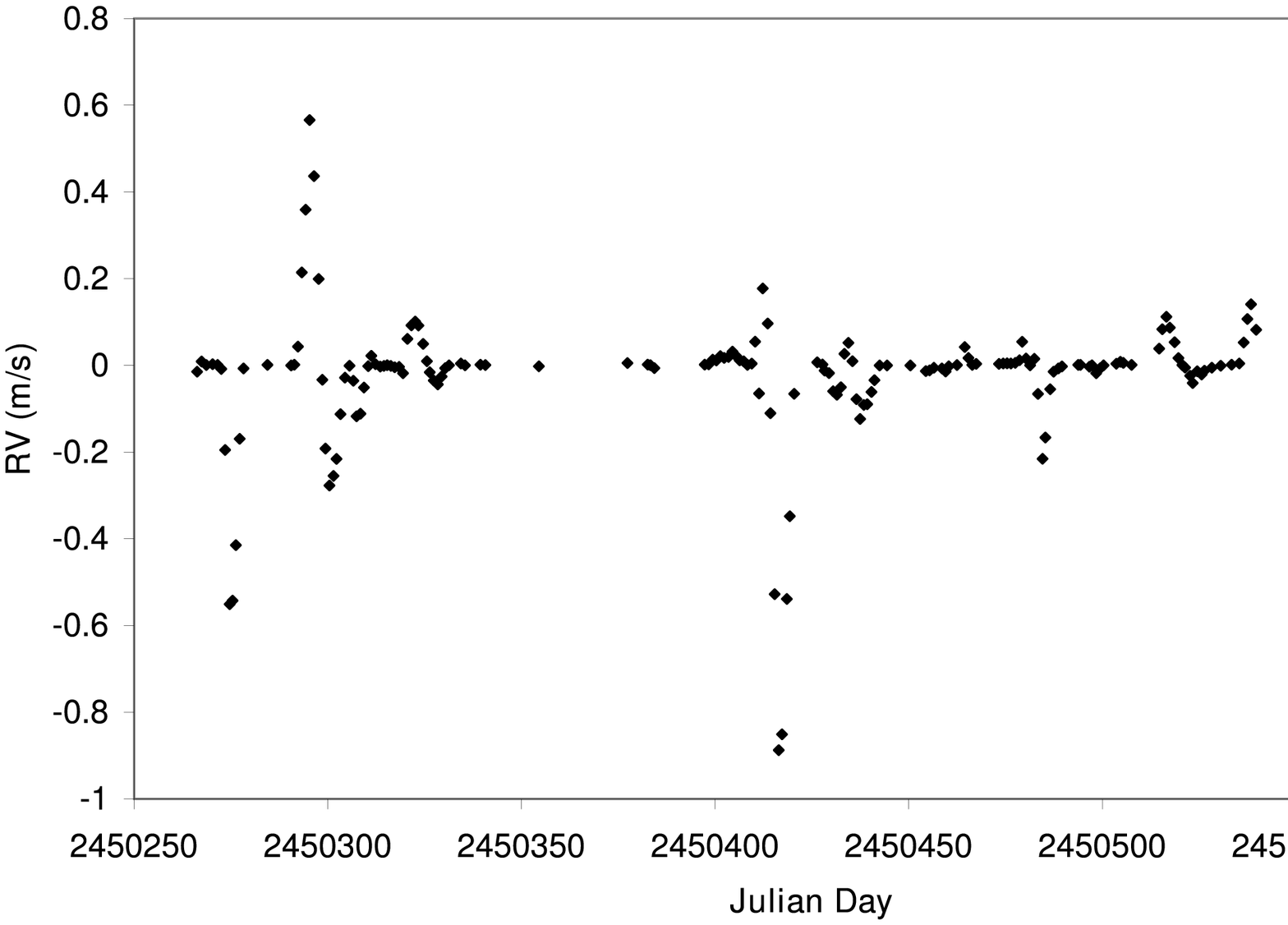} & 
\includegraphics[angle=90,width=0.3\hsize]{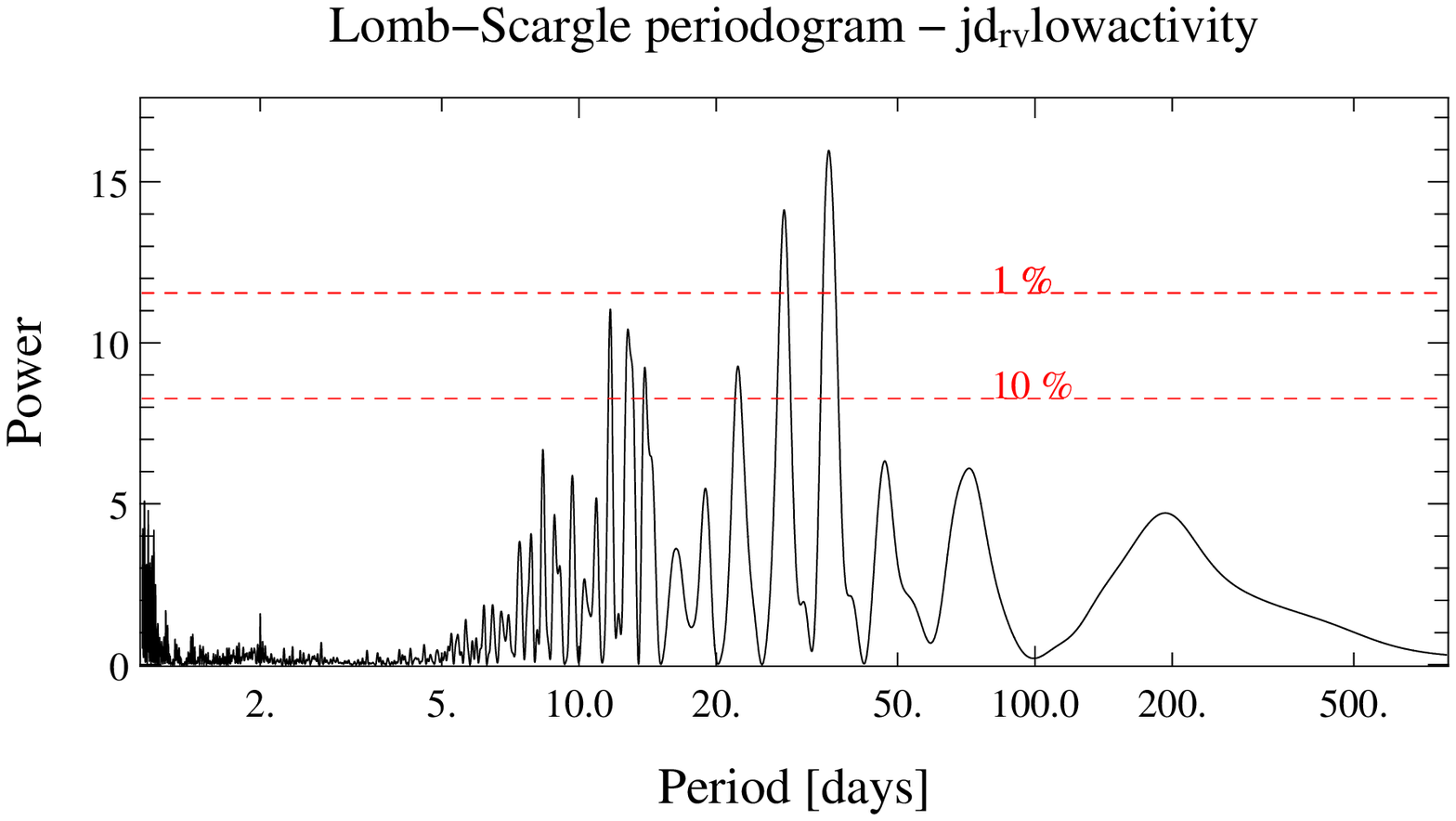}\\
 \includegraphics[angle=90,width=0.3\hsize]{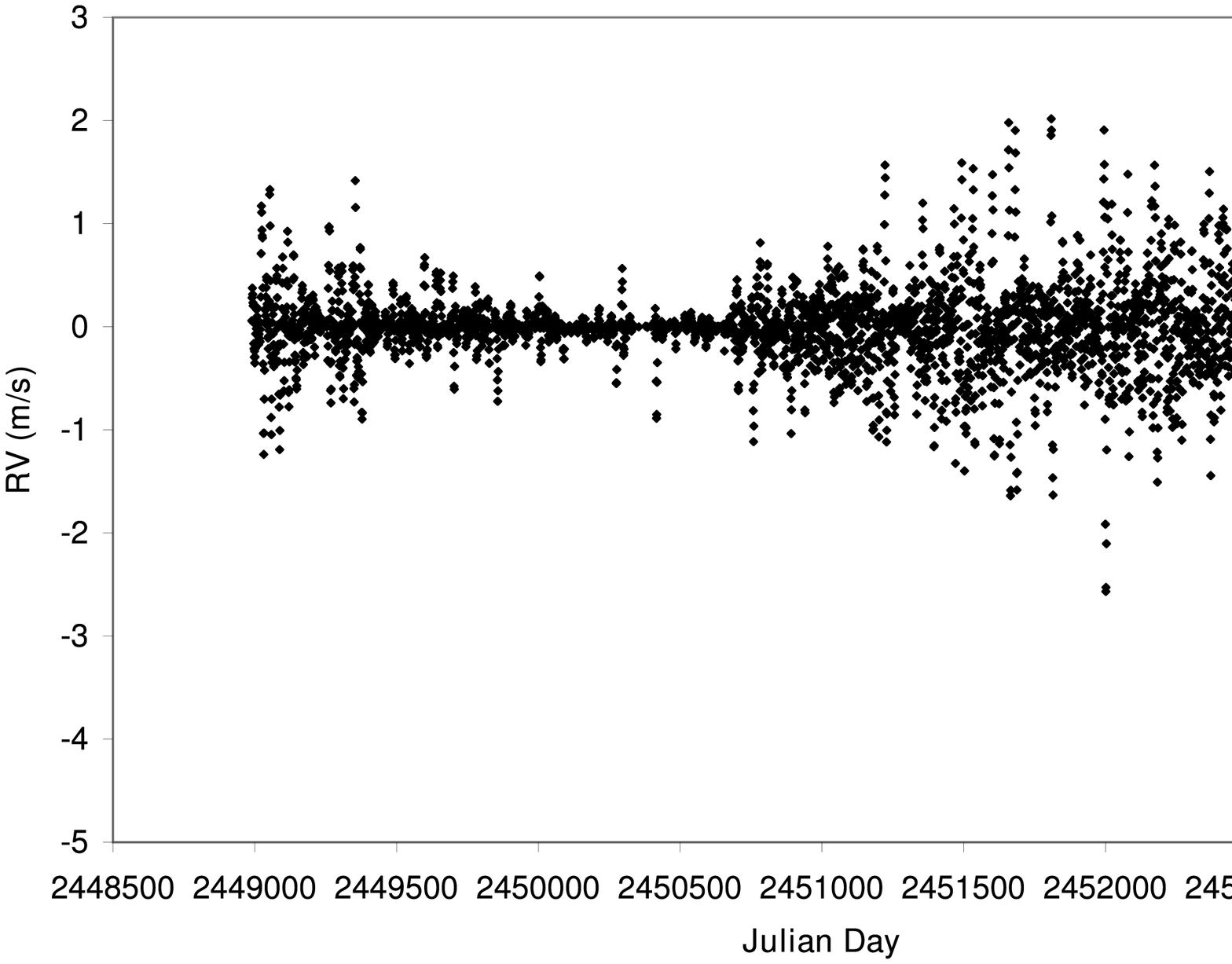} 
& \includegraphics[angle=90,width=0.3\hsize]{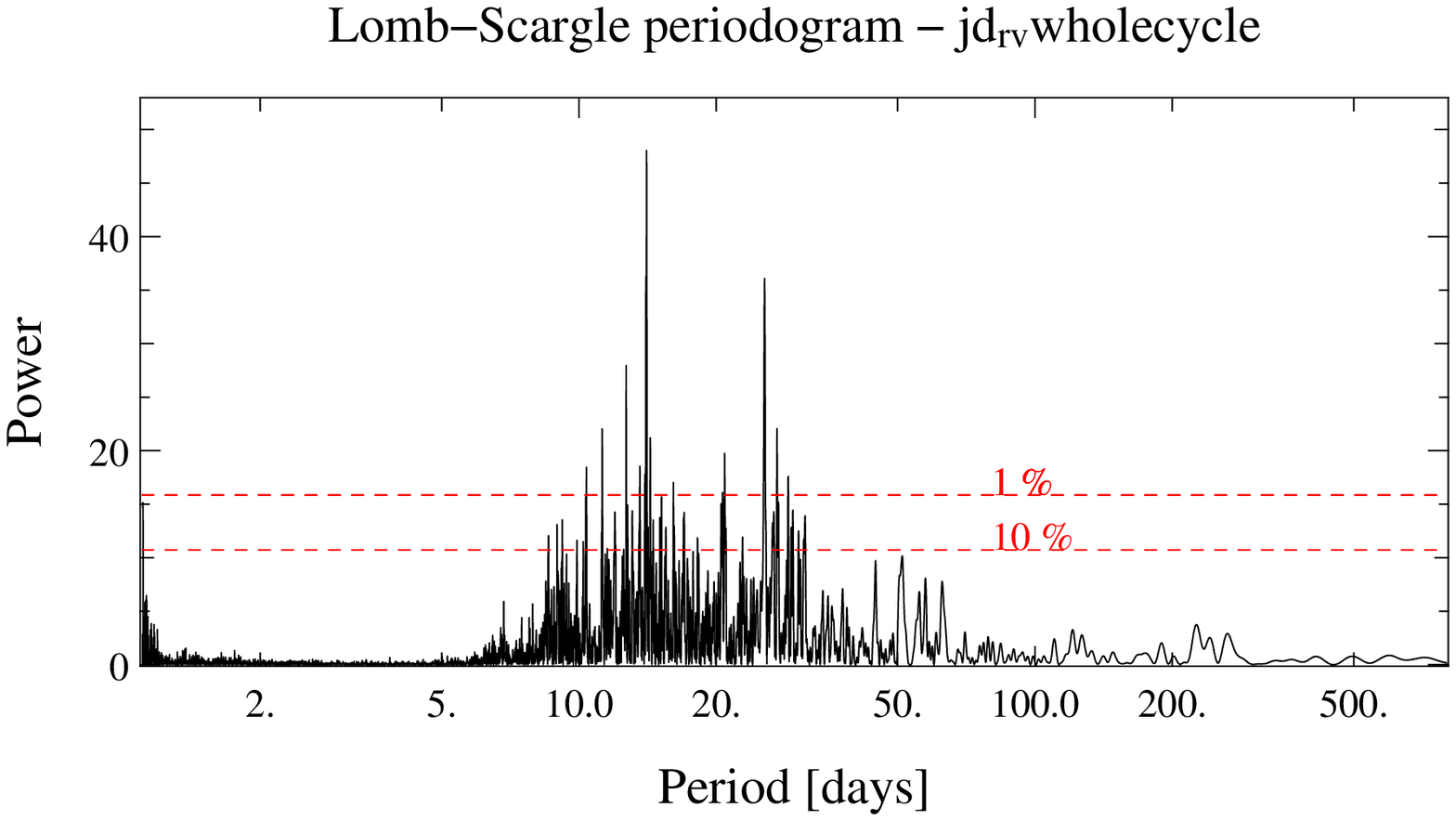} \\
 \end{tabular} 
 \caption{Temporal variations of the RVs, and corresponding periodograms. Left) the whole period (JD 2448500 to JD 2453500) is considered; Middle) the low activity period (JD 2450250 to JD 2450600) is considered and Right) the high activity period selected (JD 2451550 to JD 2451900) is considered. }
  \label{rv_all}
\end{figure}


\begin{figure}[ht!]
  \centering
  \includegraphics[angle=0,width=\hsize]{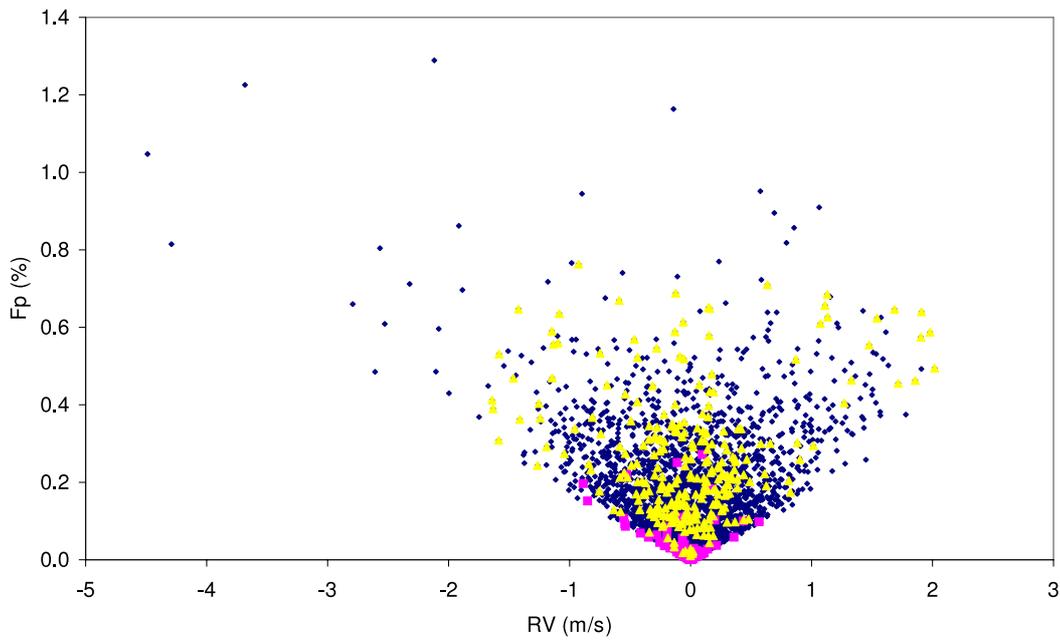}
  \caption{RV and projected filling factor (fp) variations over the cycle. The values corresponding to the high and low activity periods selected for reference (see text) are indicated resp. by triangles and squares.  We see that the spots projected filling factors over the whole cycle vary mainly between 0 and 0.6 $\% $. The corresponding RV amplitudes vary linearly with fp as in the case of a single spot (see \cite{desort08}. This explains the V-shape contour of the cloud of points.} 
  \label{rv_fp_all}
\end{figure}

\begin{figure}[ht!]
  \centering
 \begin{tabular}{cc}
\includegraphics[angle=90,width=.3\hsize]{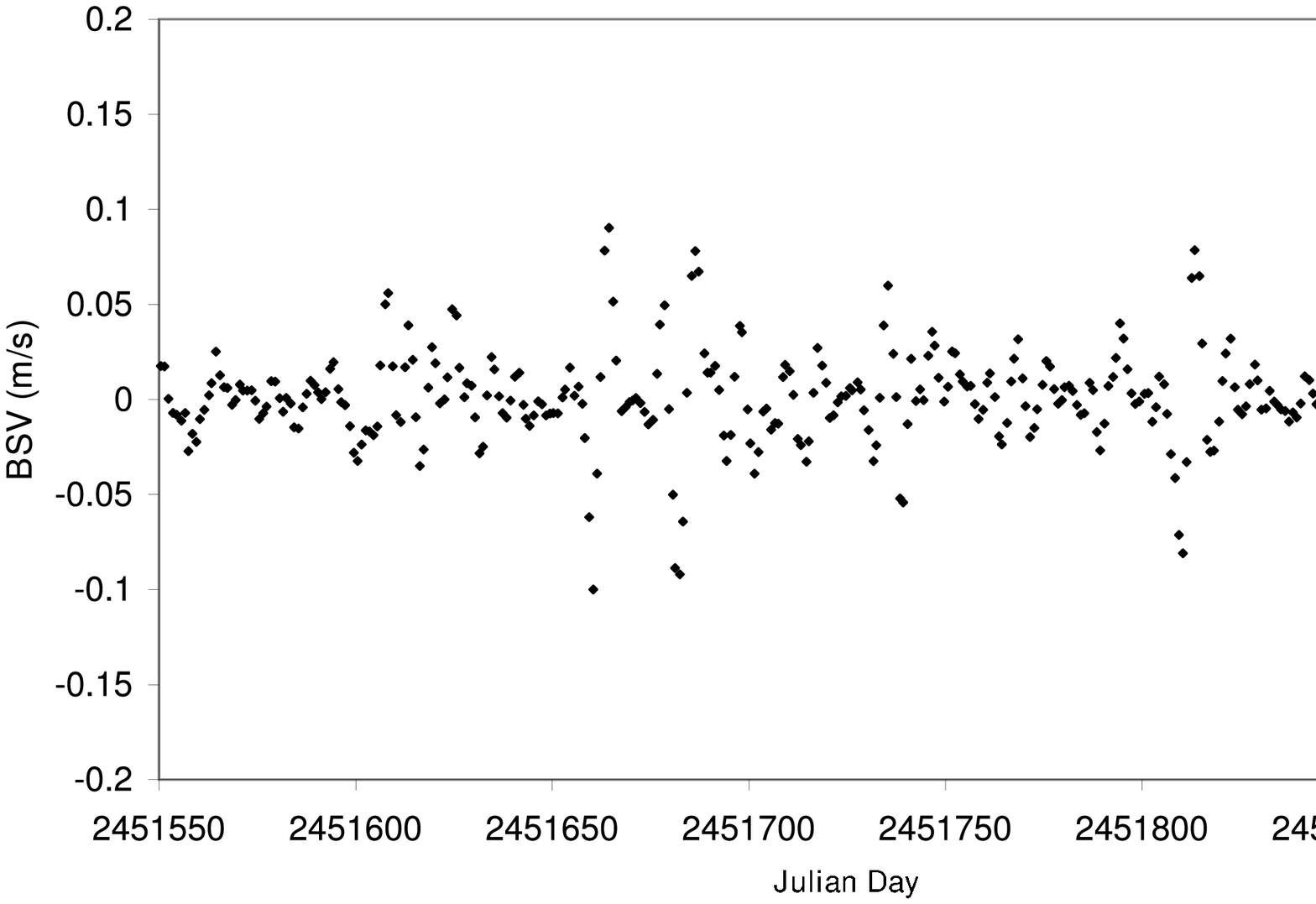} & \includegraphics[angle=90,width=0.3\hsize]{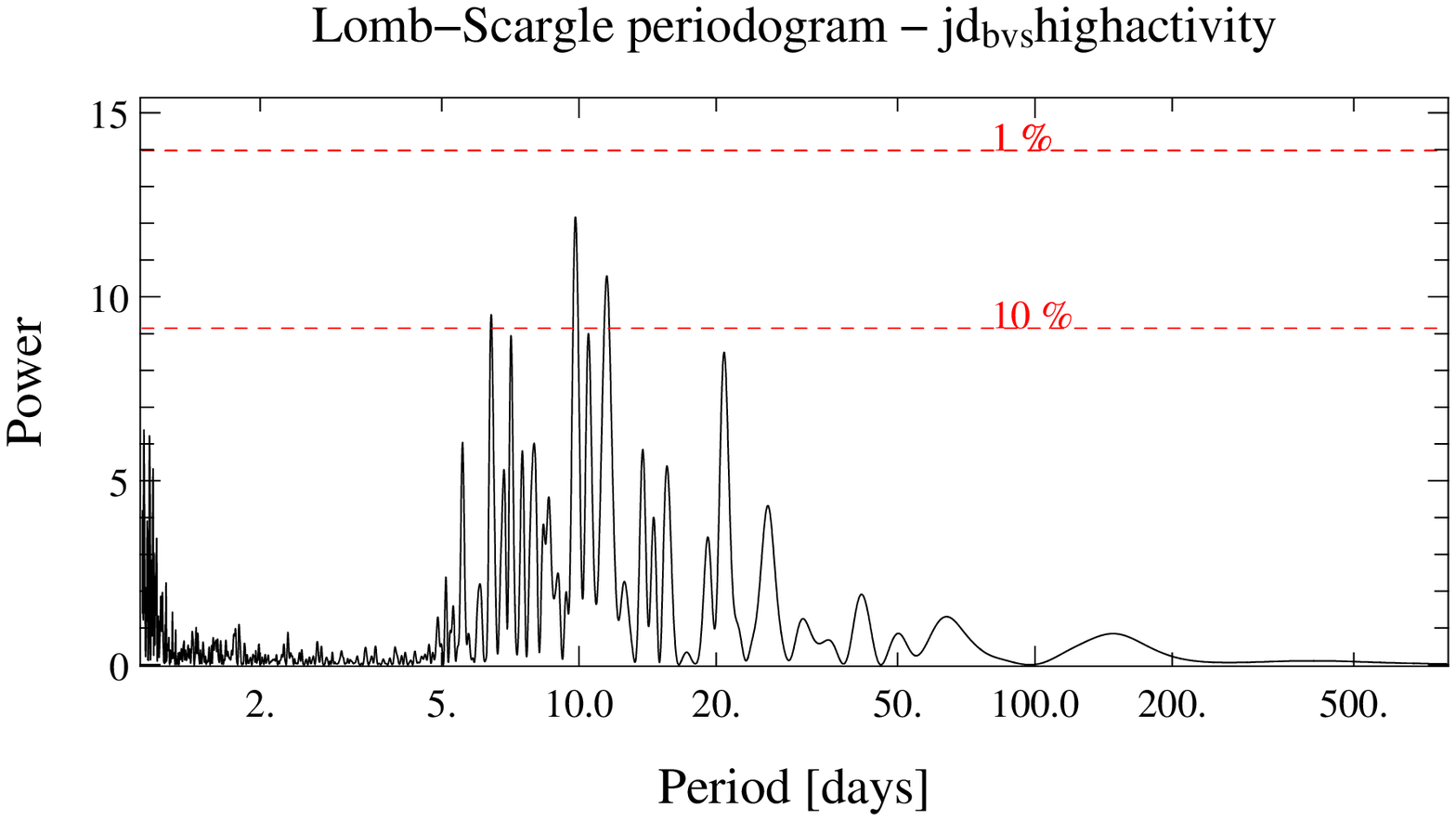}\\
 \includegraphics[angle=90,width=.3\hsize]{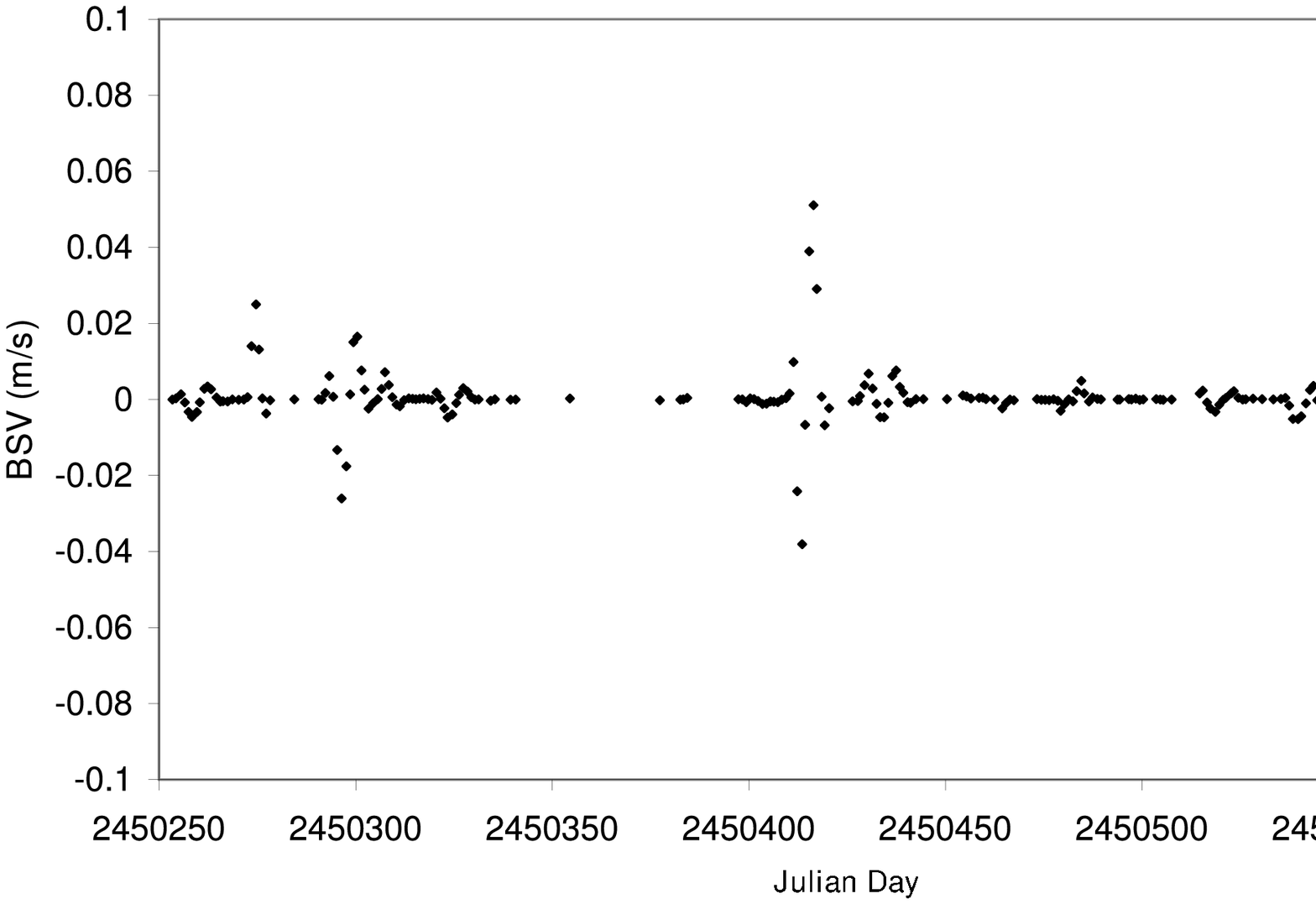} & \includegraphics[angle=90,width=0.3\hsize]{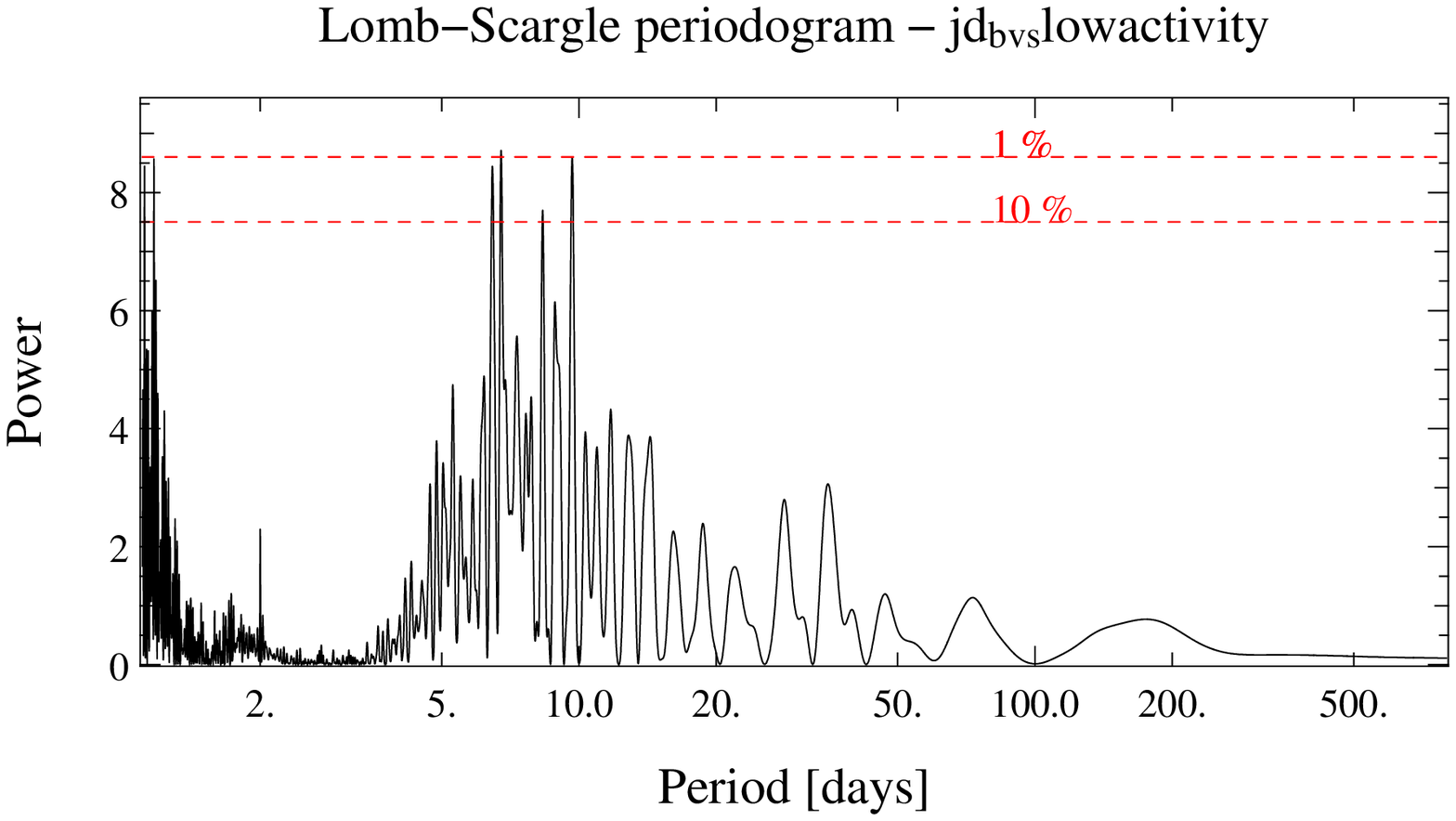}\\
 \includegraphics[angle=90,width=0.3\hsize]{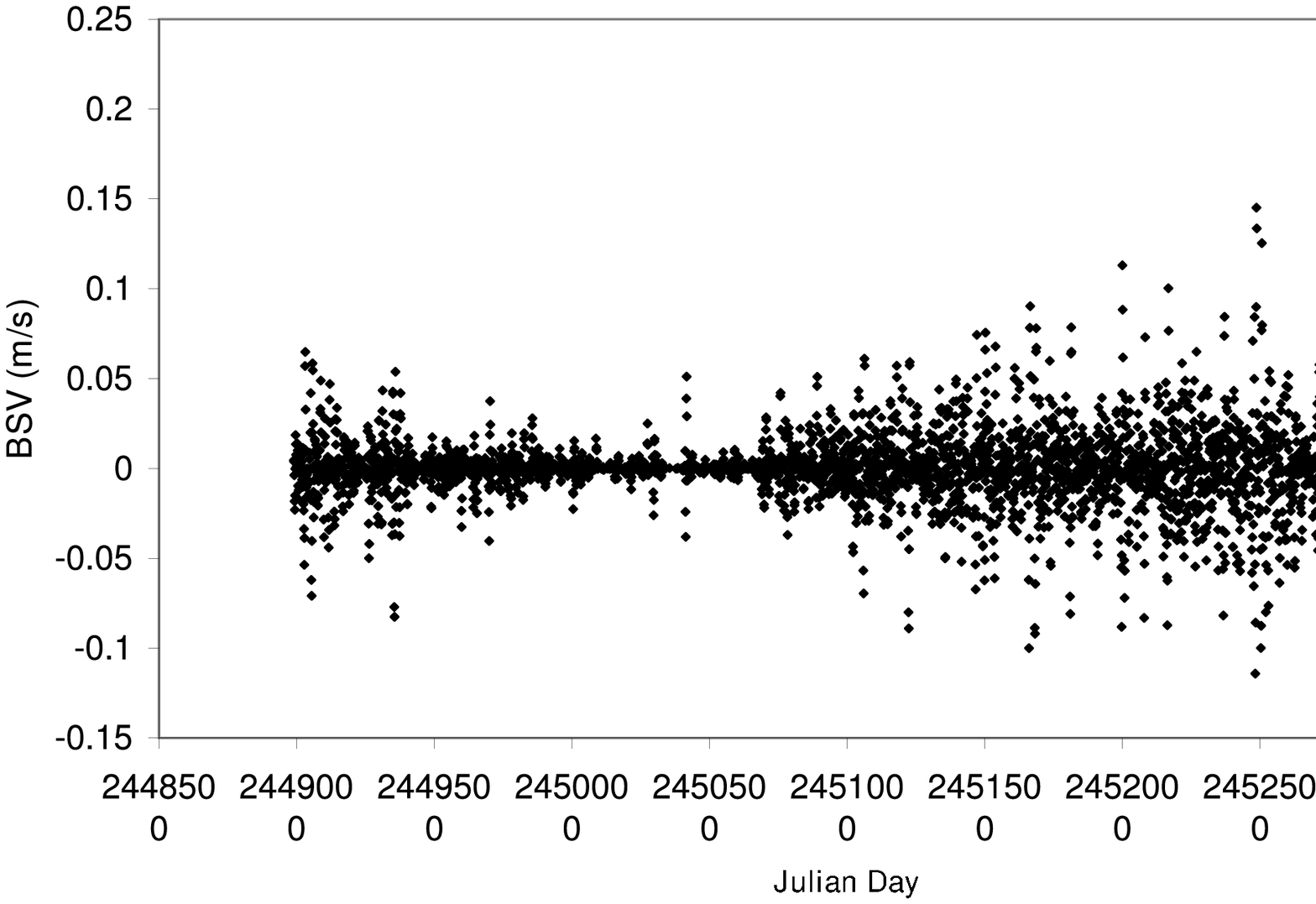} & \includegraphics[angle=90,width=0.3\hsize]{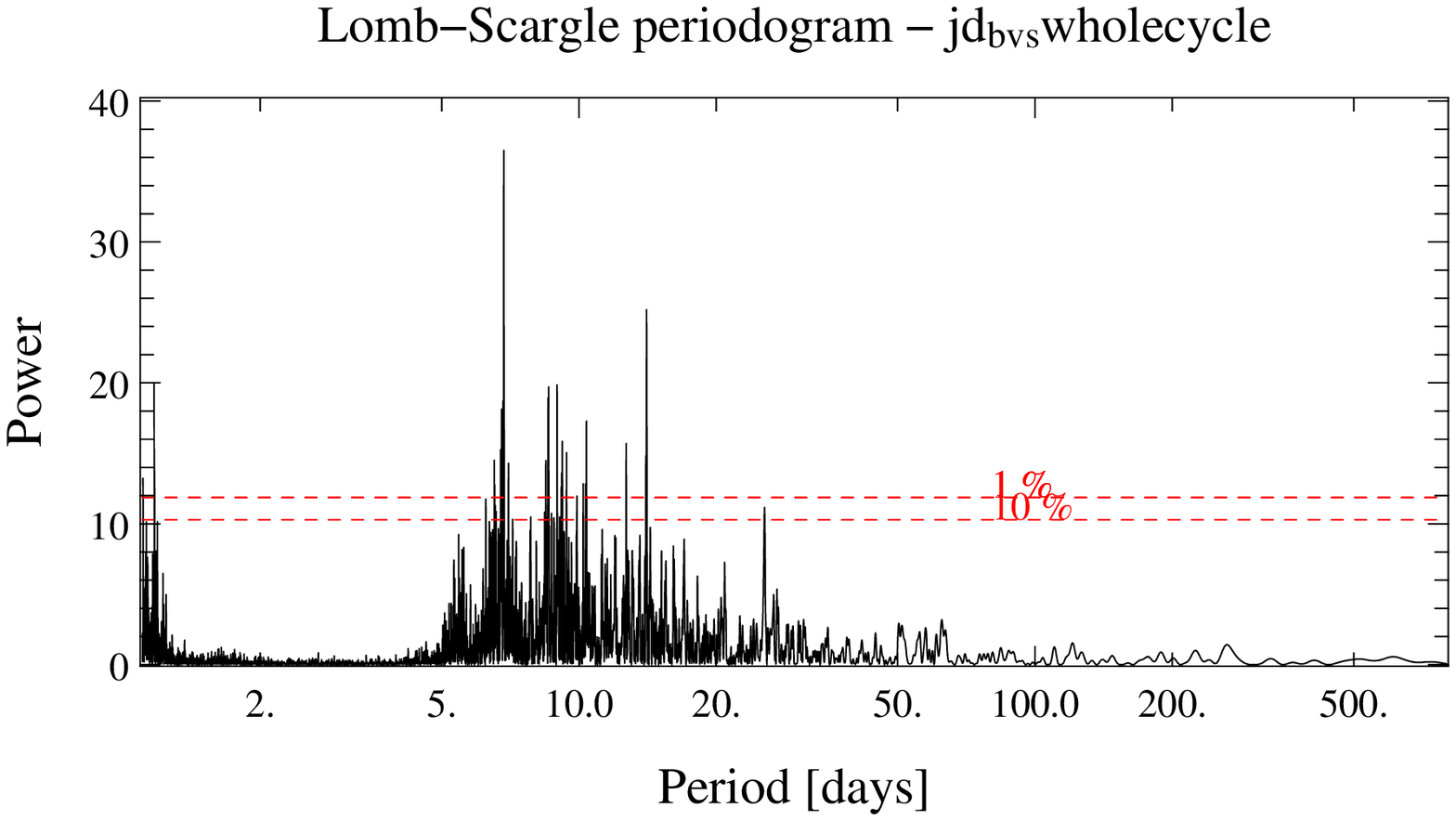}\\
 \end{tabular} 
 \caption{Temporal variations of the BVS and corresponding periodograms, when considering the whole period (Left), the low activity period selected (Middle) and the high activity period selected (Right).} 
  \label{bsv_all}
\end{figure}

\begin{figure}[ht!]
  \centering
  \includegraphics[angle=0,width=\hsize]{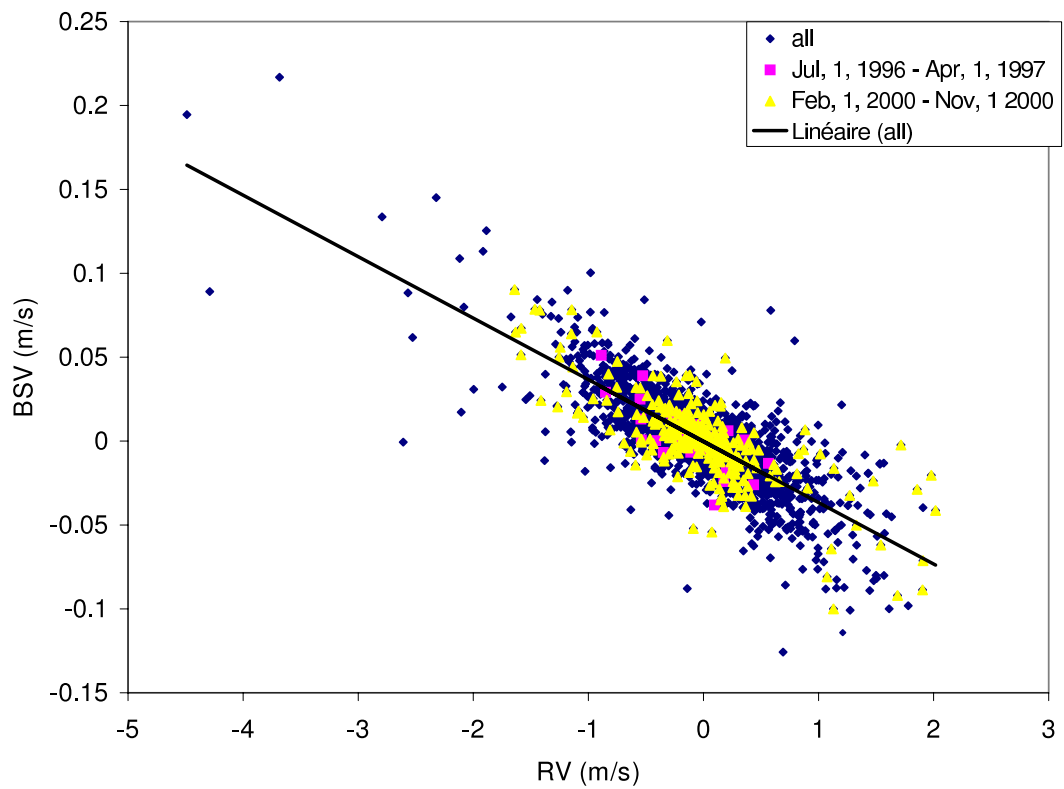}
  \caption{Correlation between the RV and the BVS over the whole period (see text). The values corresponding to the high and low activity periods selected for reference (see text) are indicated resp. by triangles and squares. } 
  \label{bsv_rv}
\end{figure}

\begin{figure}[ht!]
  \centering
 \begin{tabular}{cc}
\includegraphics[angle=90,width=.3\hsize]{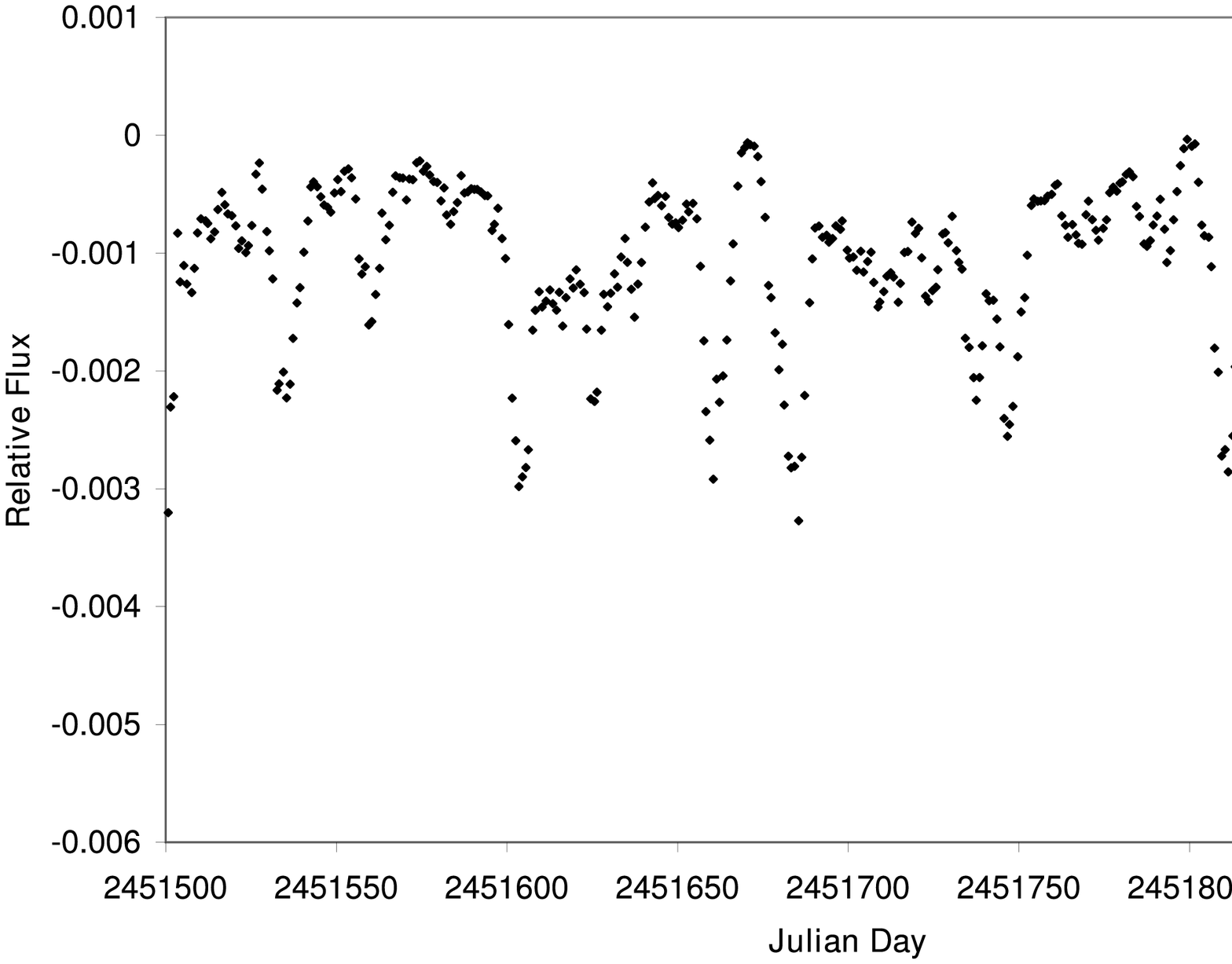} & \includegraphics[angle=90,width=0.3\hsize]{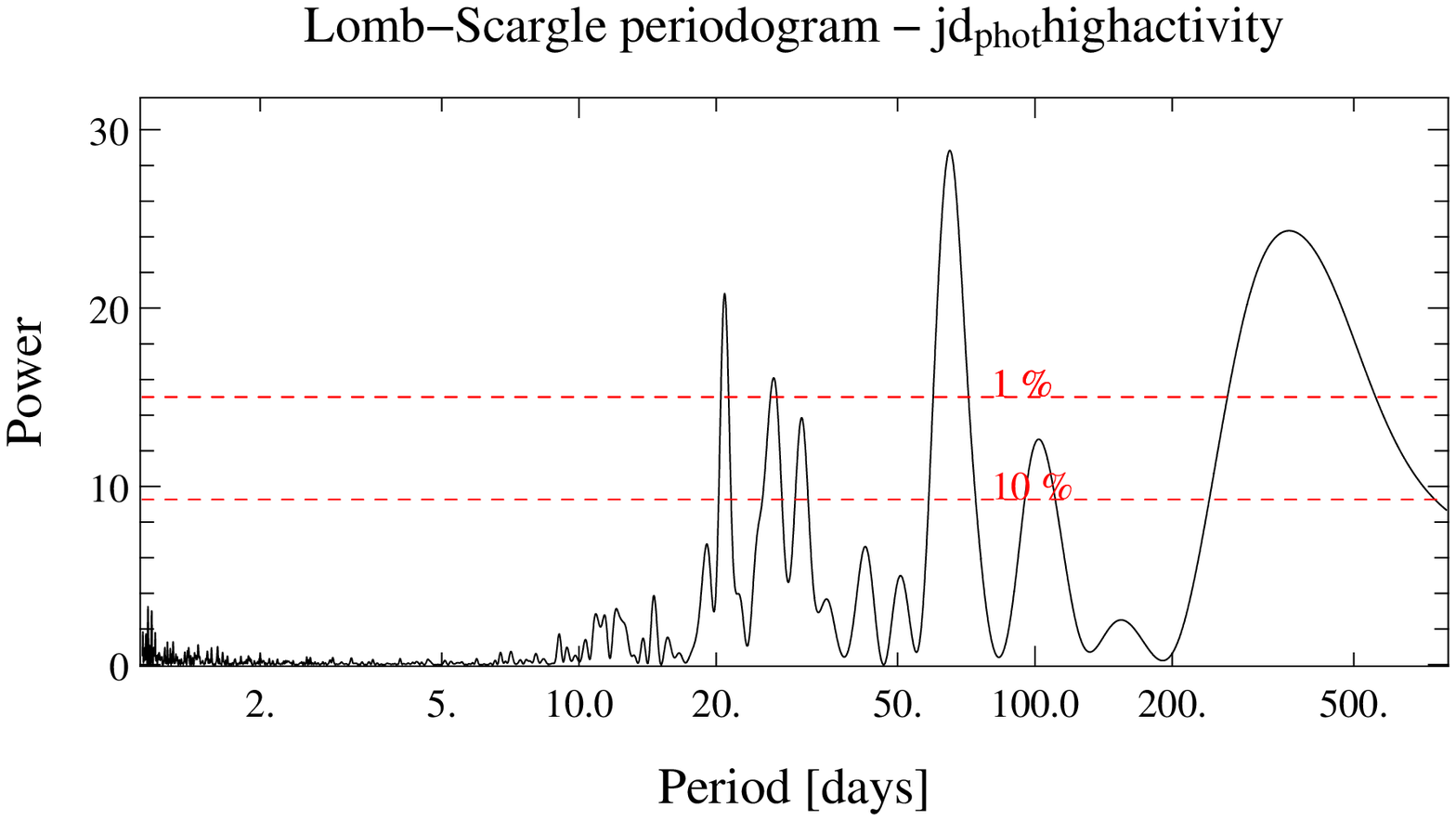}\\
 \includegraphics[angle=90,width=.3\hsize]{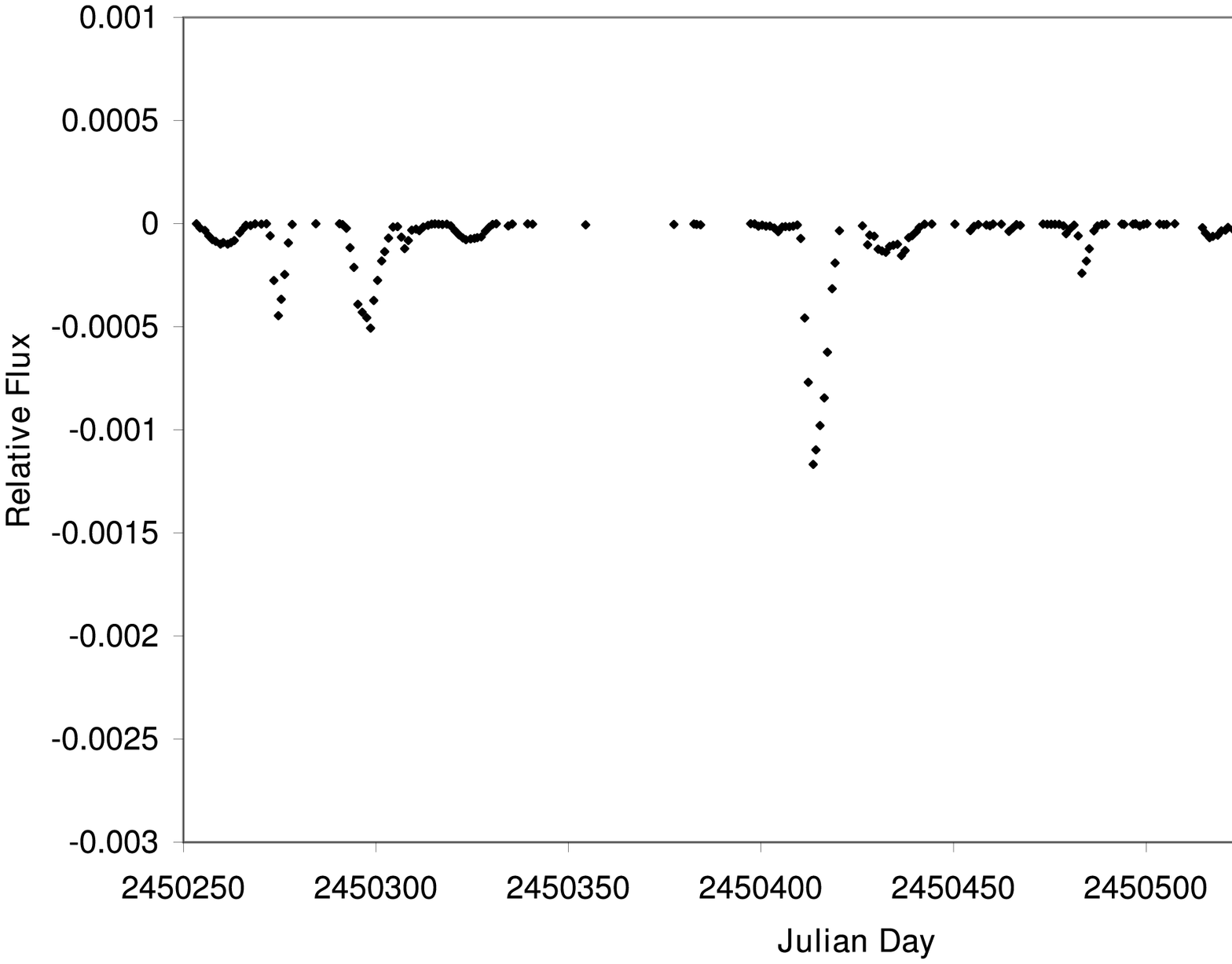} & \includegraphics[angle=90,width=0.3\hsize]{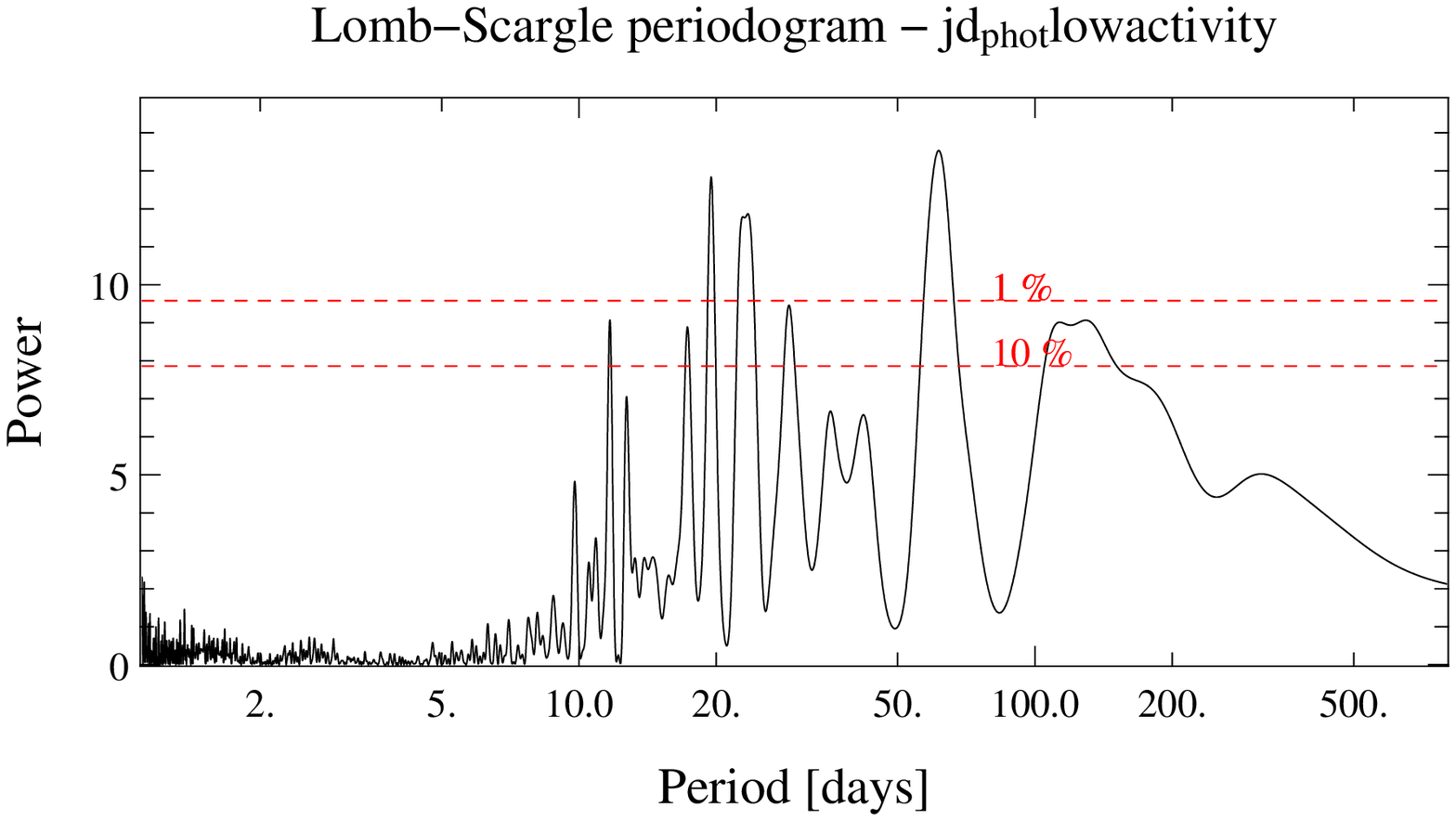}  \\
 \includegraphics[angle=90,width=0.3\hsize]{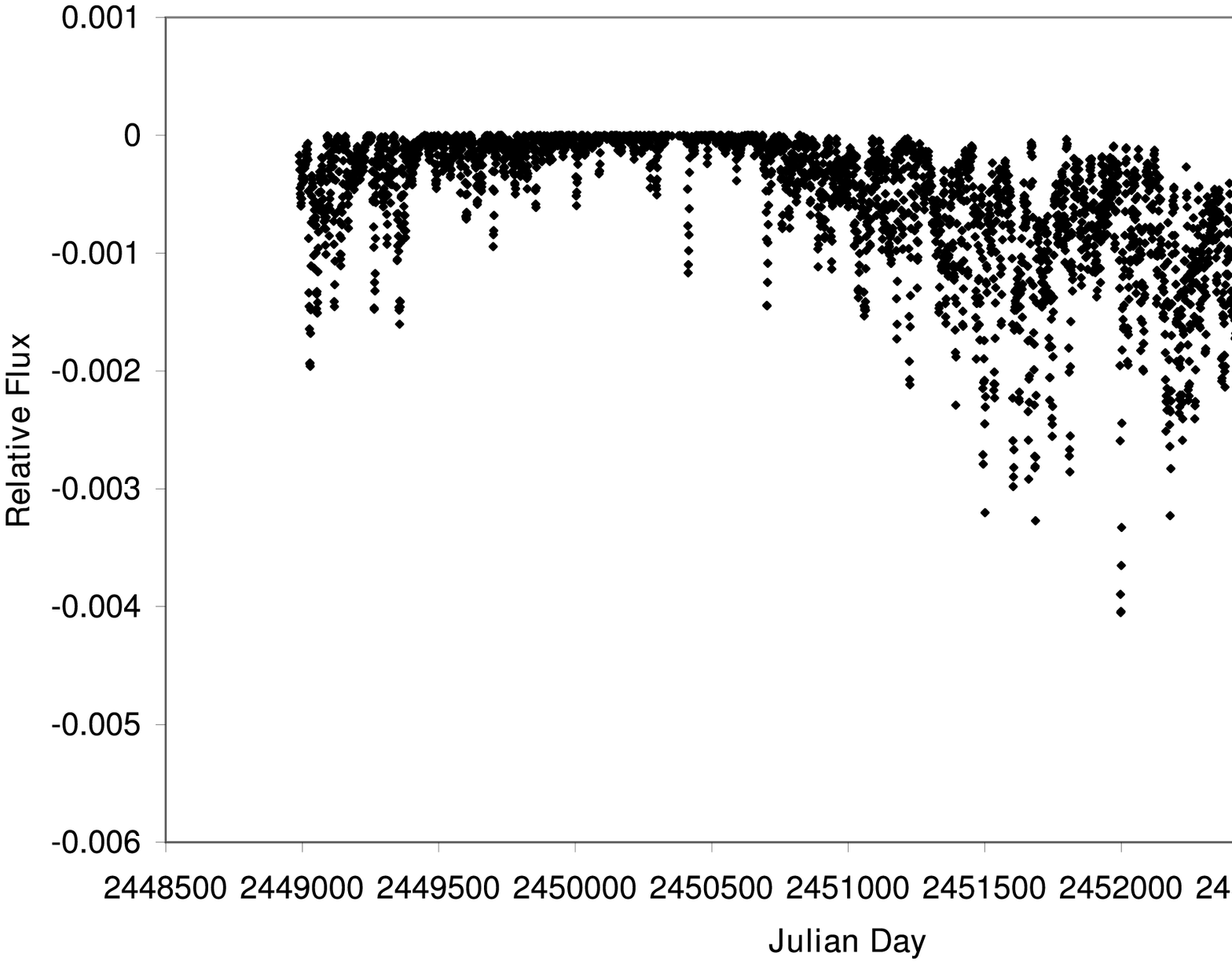} & \includegraphics[angle=90,width=0.3\hsize]{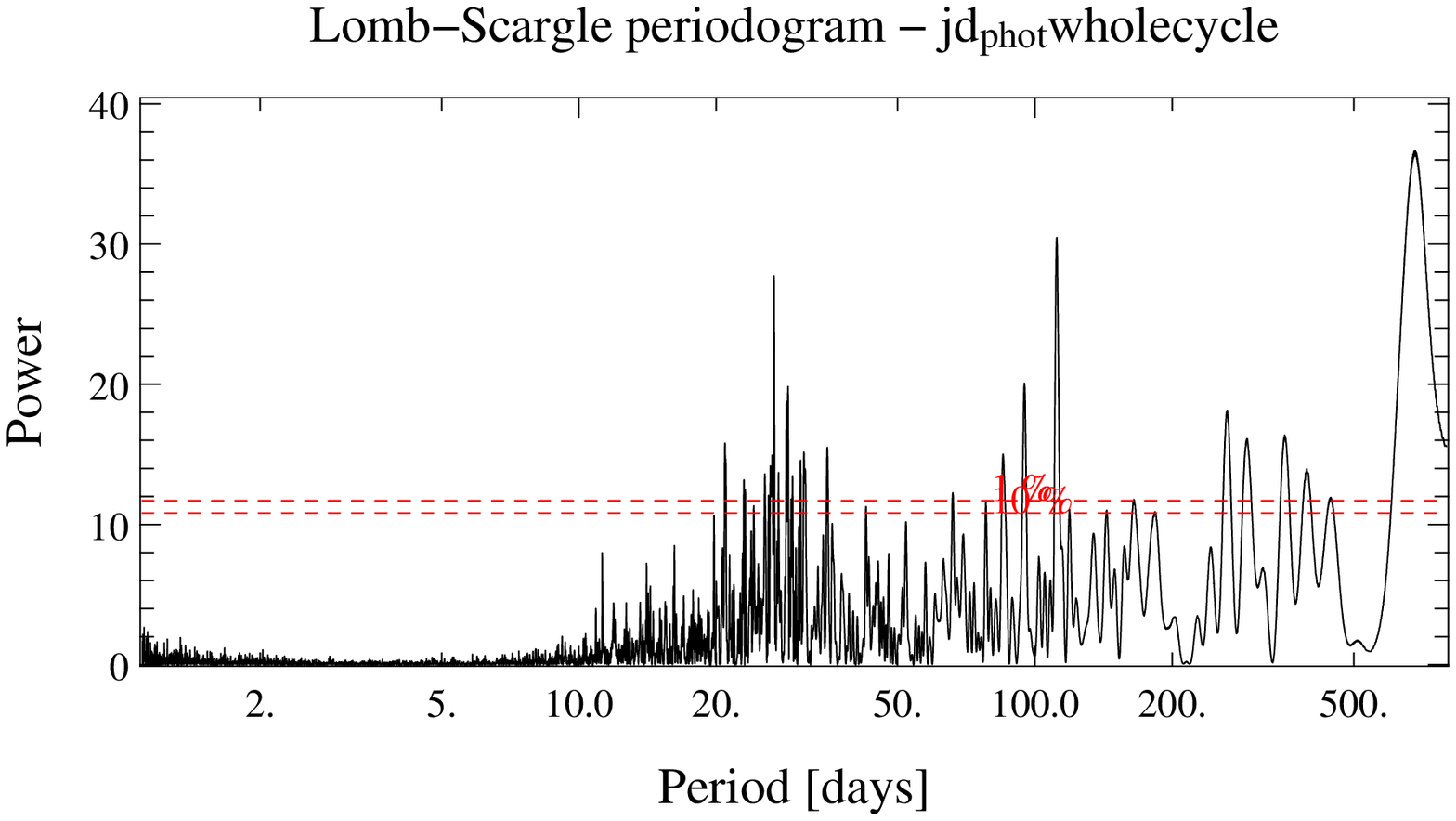} \\
 \end{tabular} 
 \caption{Temporal variations of the spotted Sun relative photometry (ie compared to a Sun without any spot) and corresponding periodograms, when considering the whole period (Left), the low activity period selected (Middle) and the high activity period selected (Right).} 
  \label{phot_all}
\end{figure}


\begin{figure}[ht!]
  \centering
  \includegraphics[angle=0,width=.8\hsize]{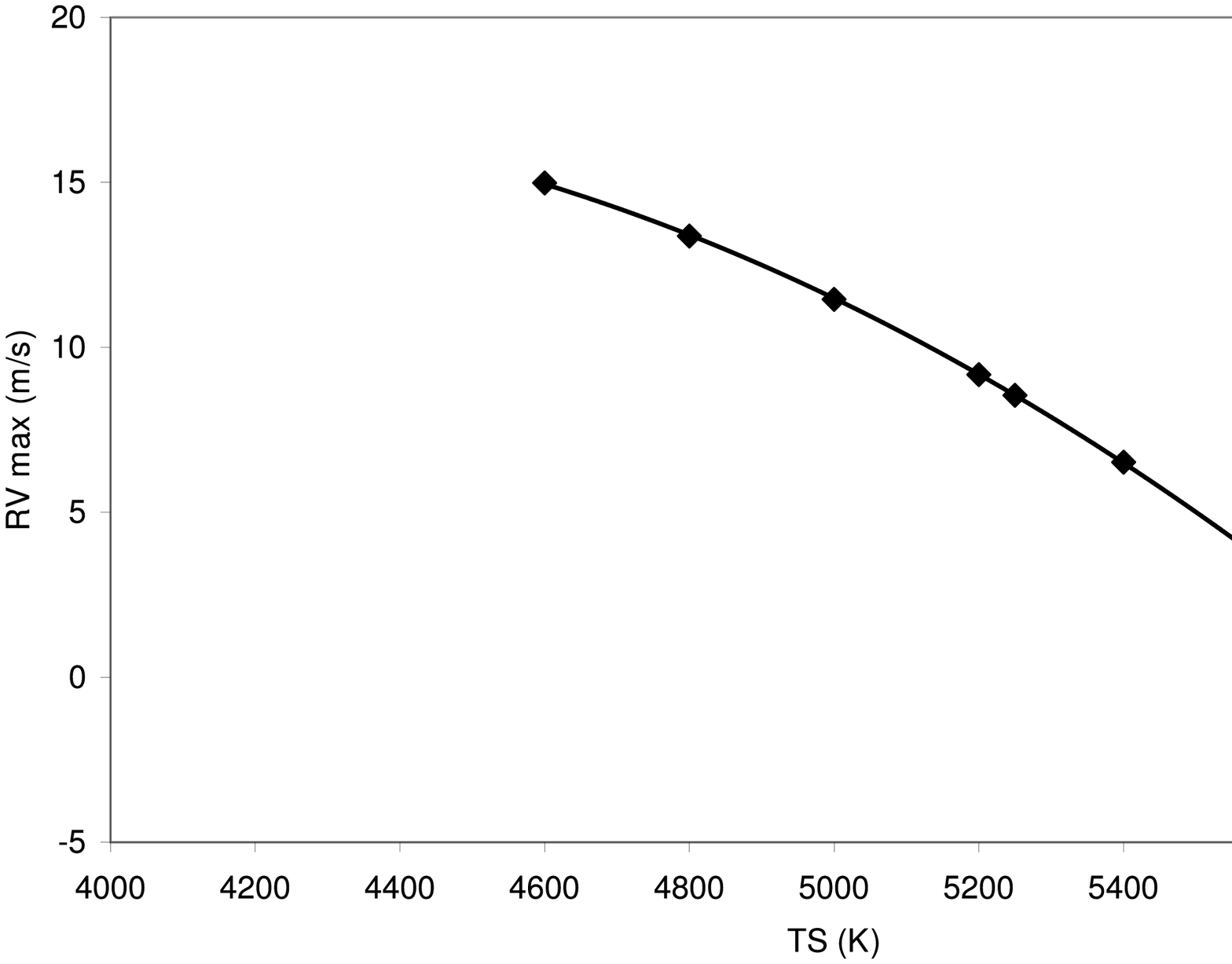}\\
 \includegraphics[angle=0,width=.8\hsize]{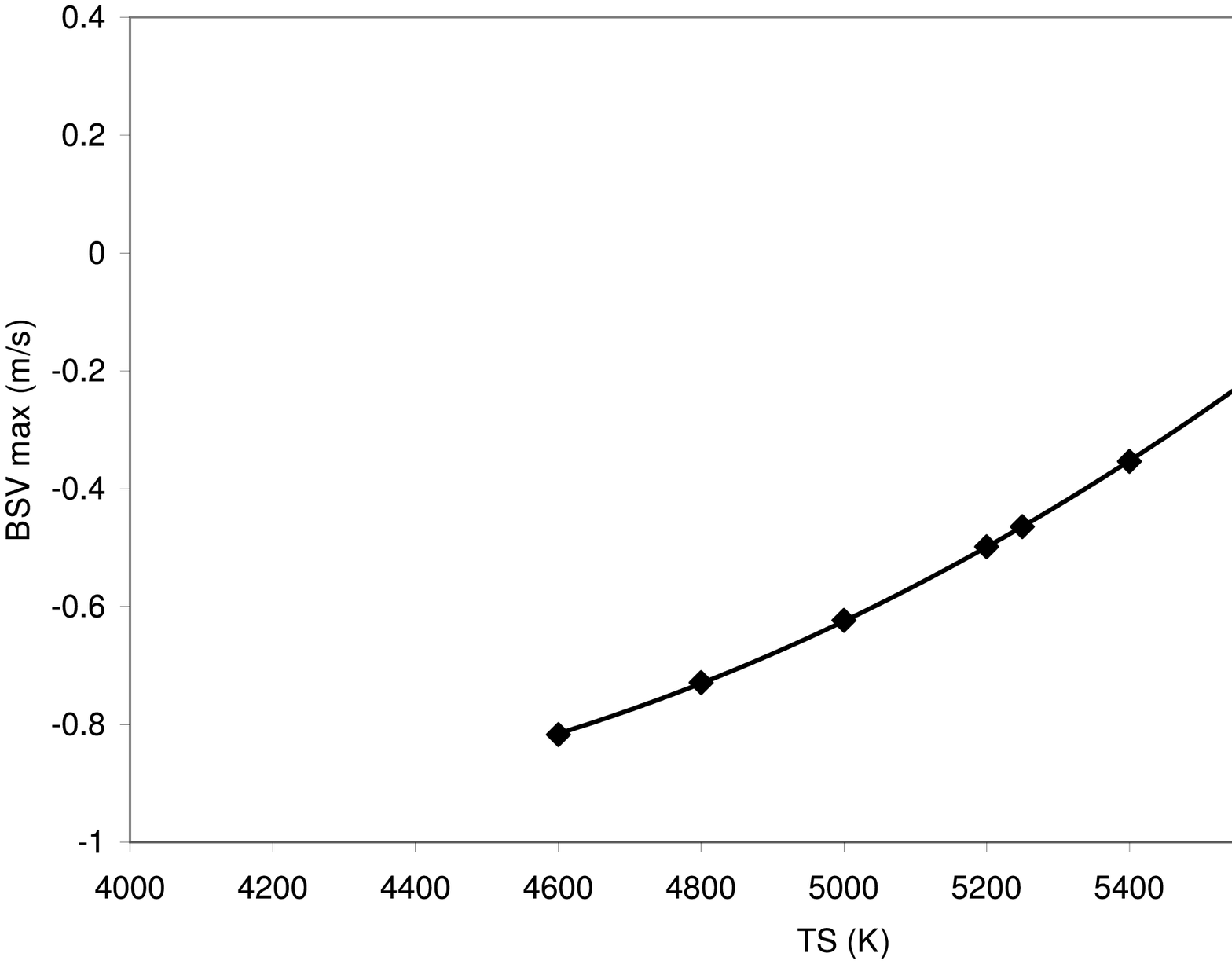}\\
   \includegraphics[angle=0,width=.8\hsize]{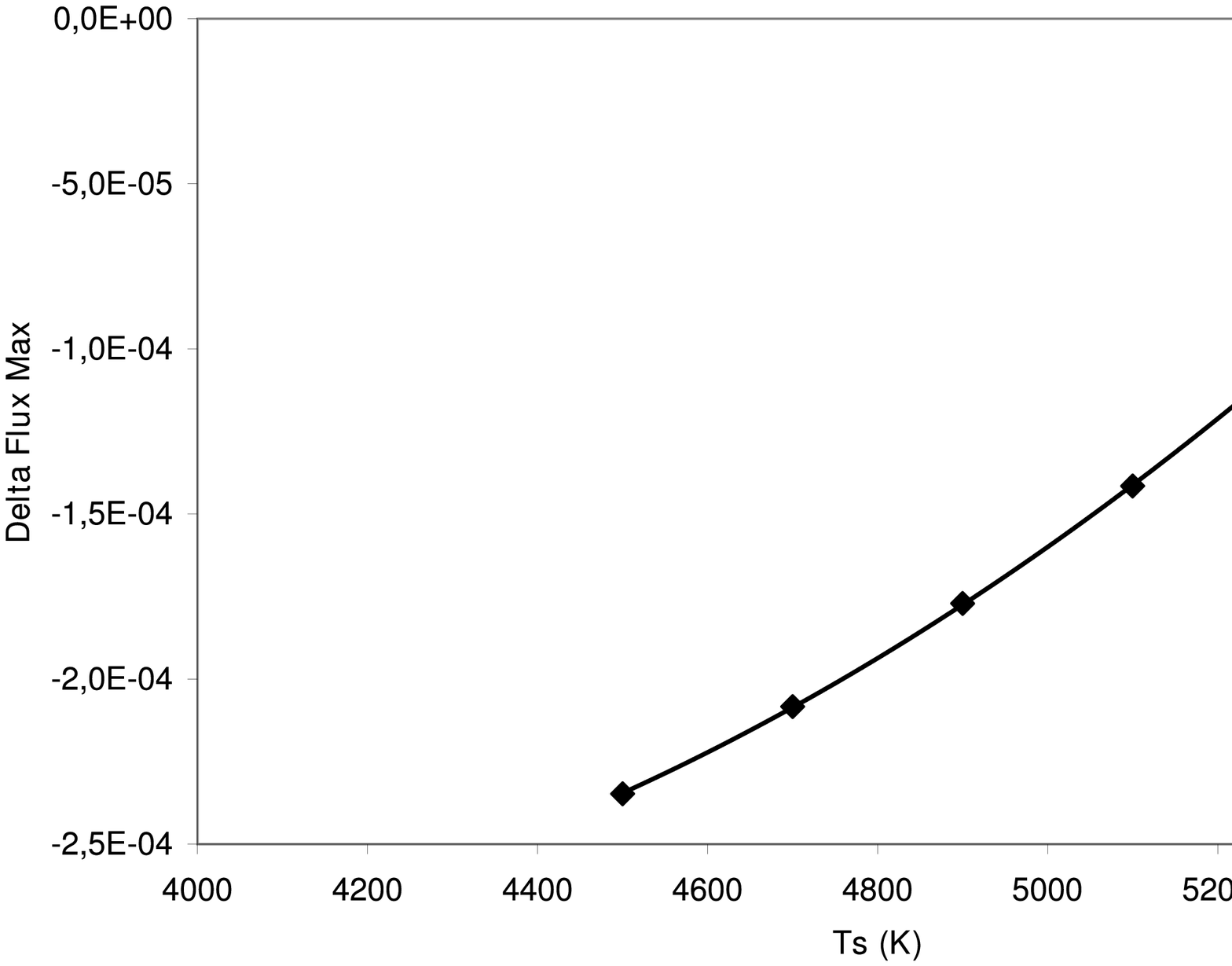}
 \caption{Maximum RV (Top), BVS (Middle) and Spot Absorption (Bottom) assuming different spot temperatures. } 
  \label{Ts_vmax_bsvmax_abs}
\end{figure}

\begin{figure}[ht!]
  \centering
  \begin{tabular}{cc}
 \includegraphics[angle=90,width=.3\hsize]{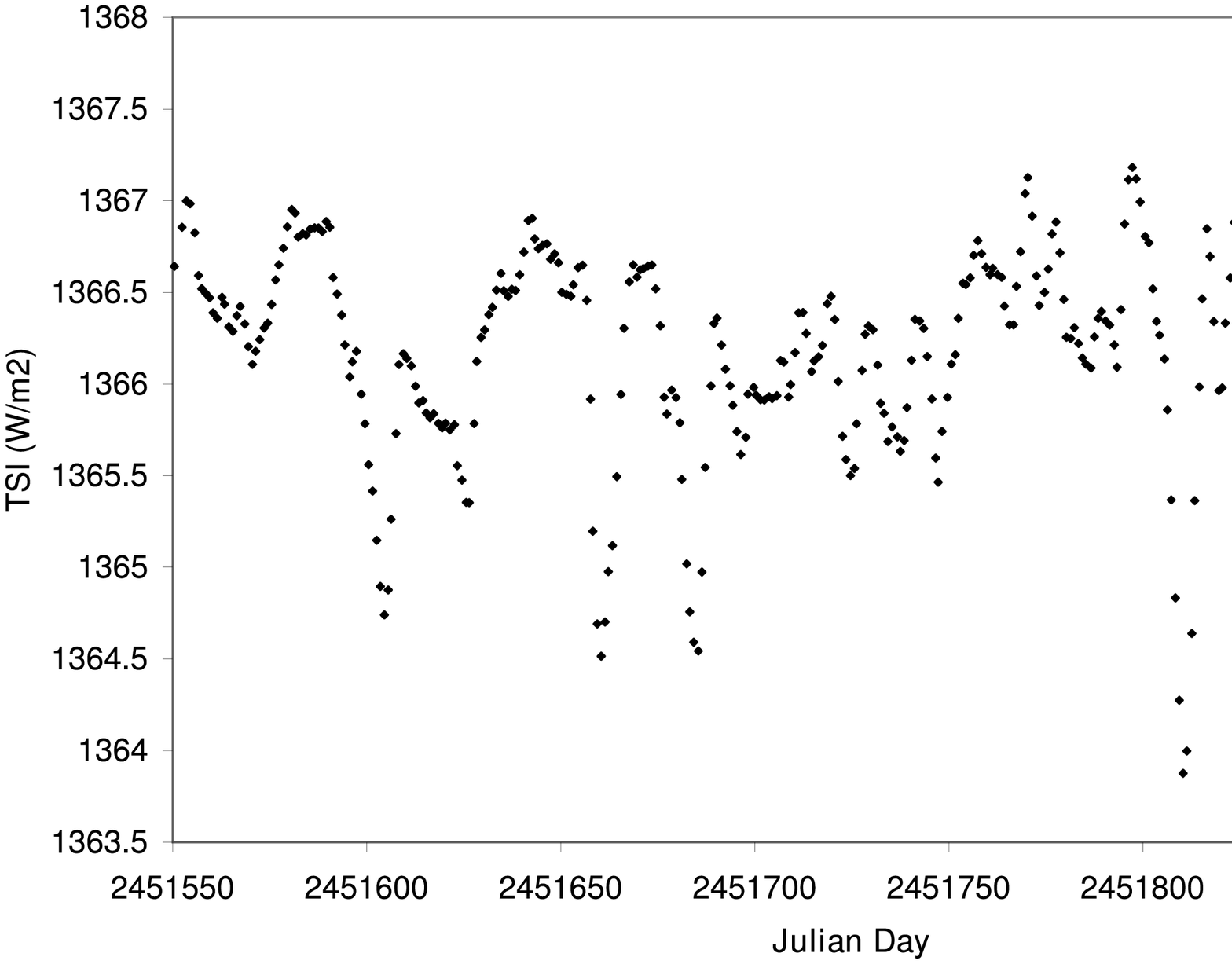}& \includegraphics[angle=90,width=0.3\hsize]{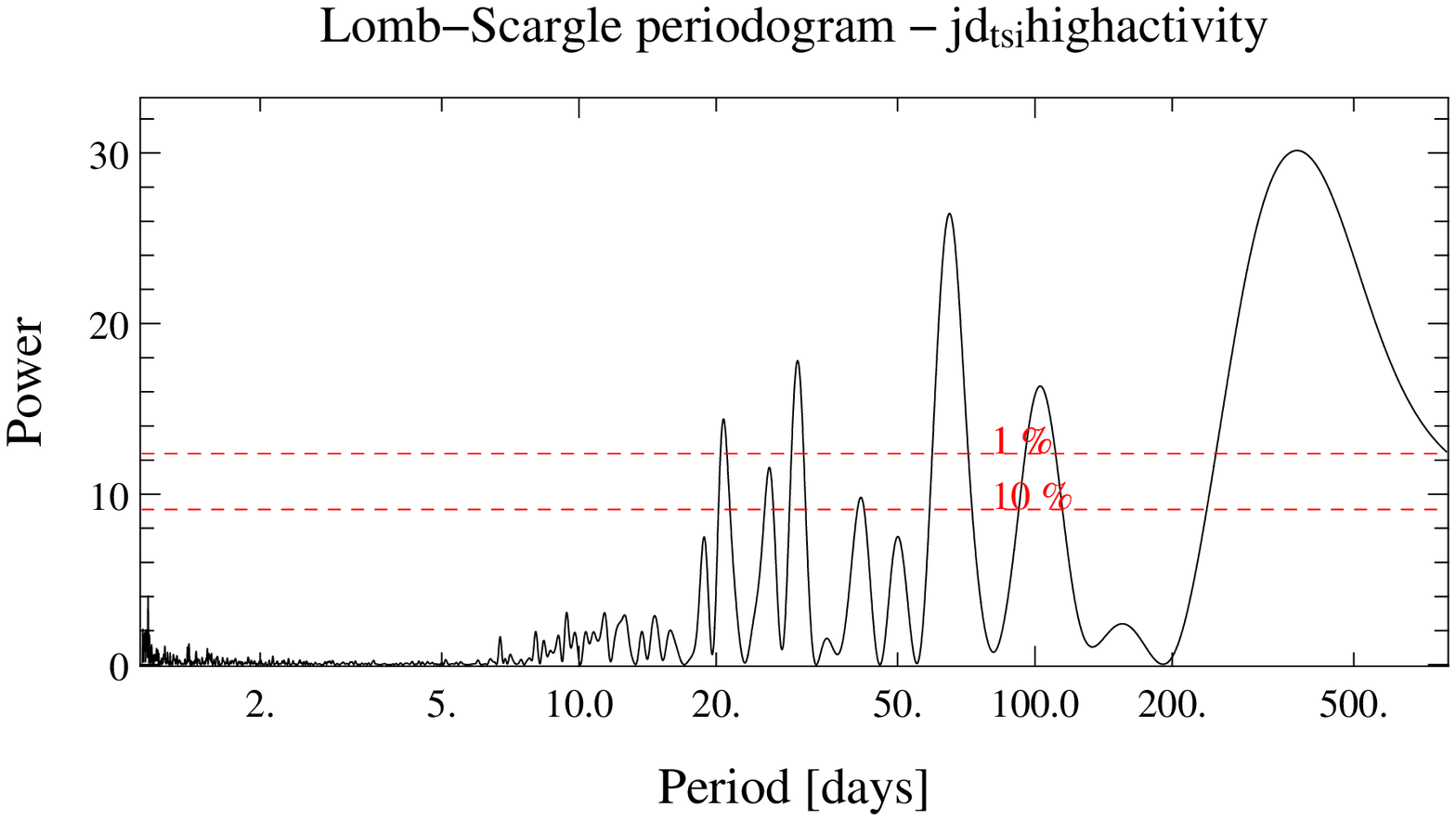}\\
\includegraphics[angle=90,width=.3\hsize]{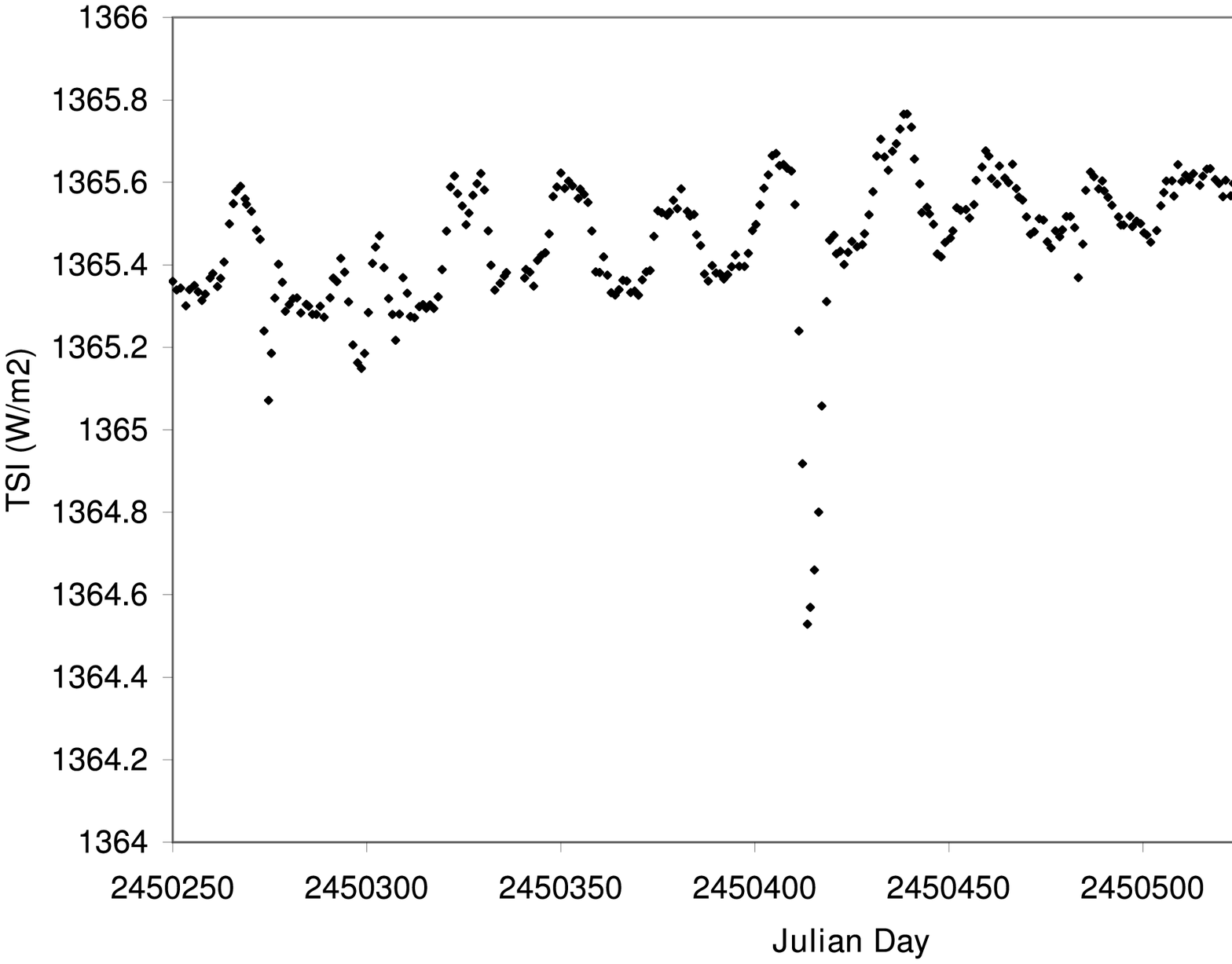}& \includegraphics[angle=90,width=0.3\hsize]{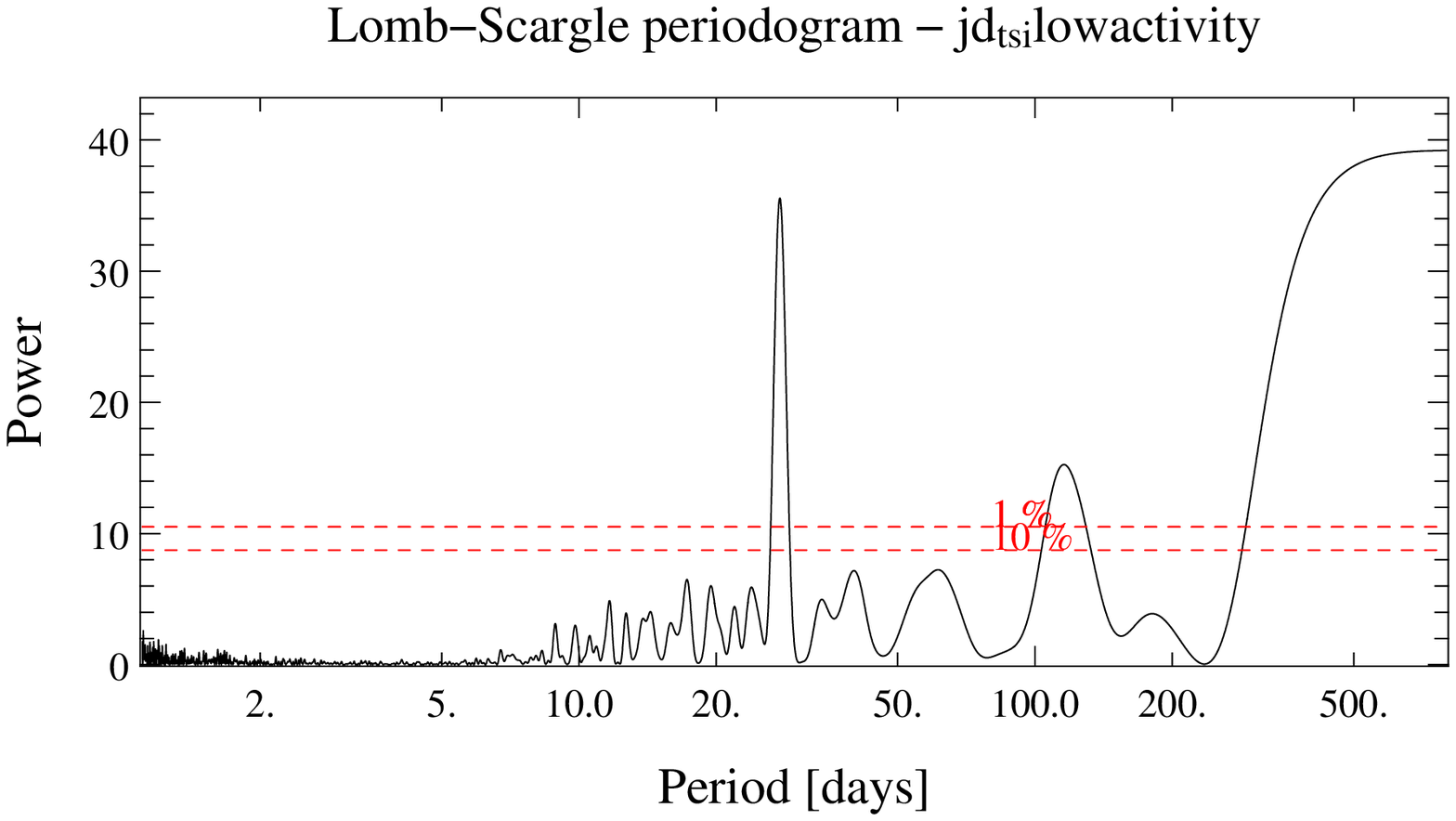} \\
\includegraphics[angle=90,width=.3\hsize]{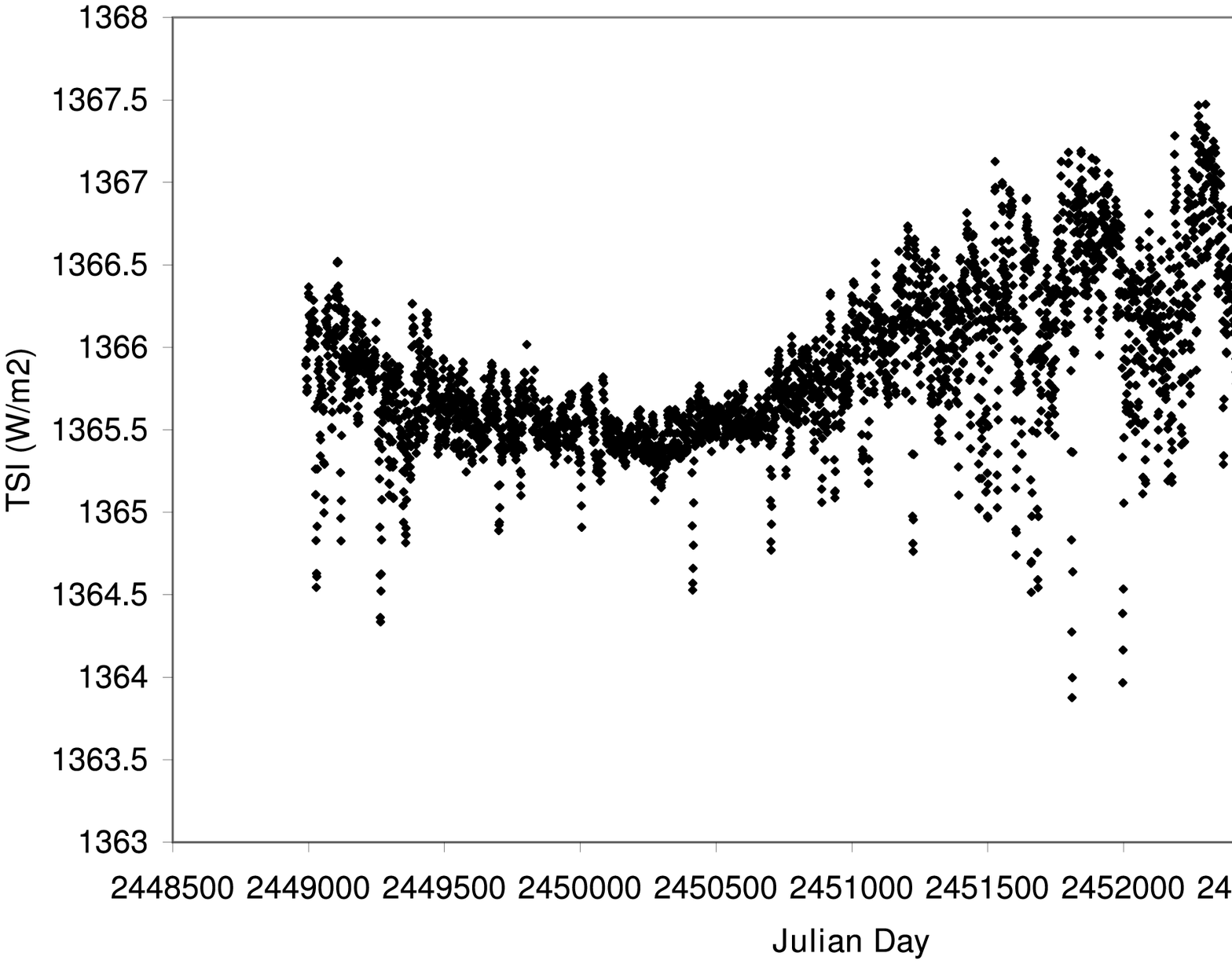}& \includegraphics[angle=90,width=0.3\hsize]{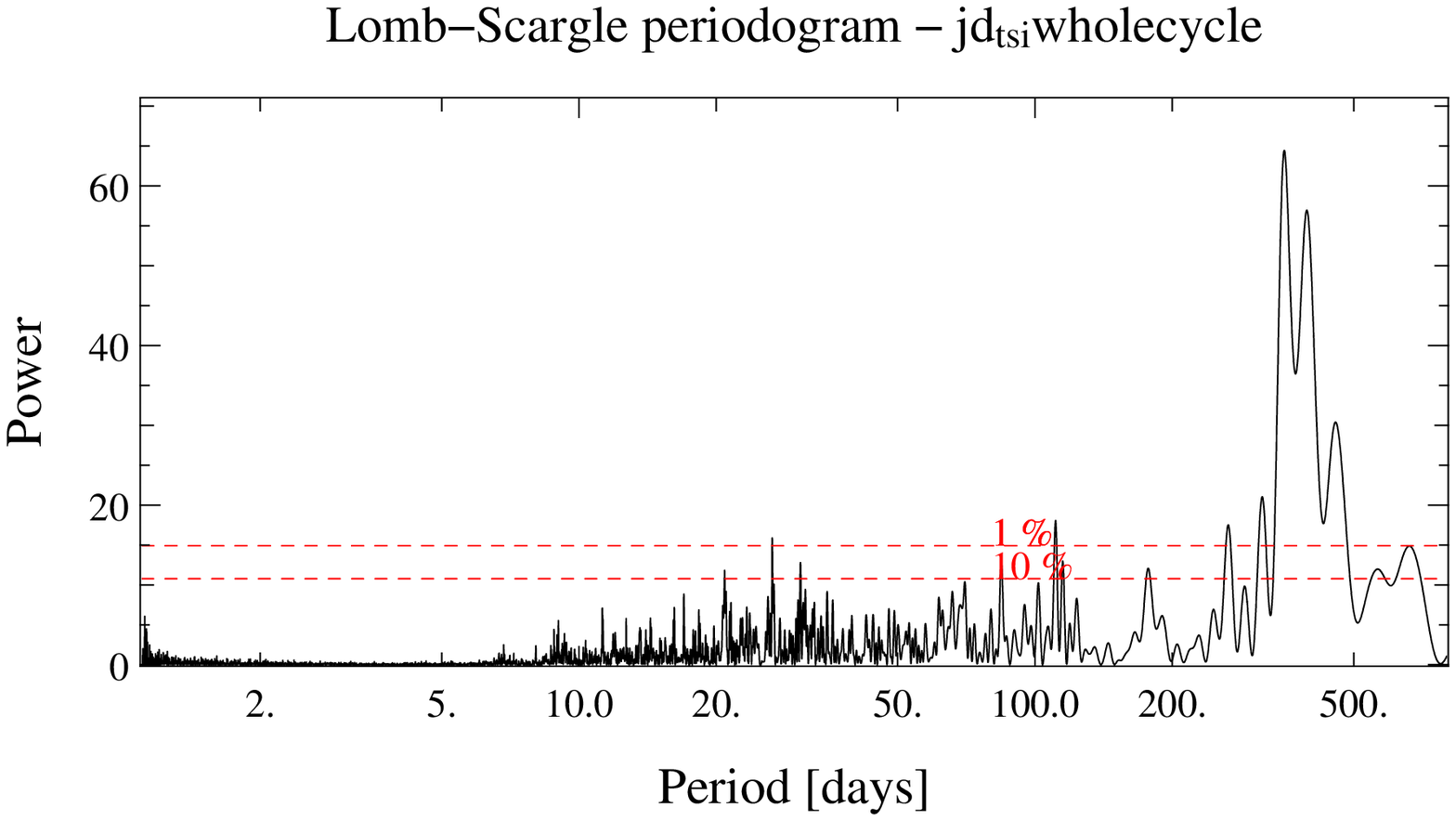}
 \end{tabular} 
 \caption{TSI temporal variations, and corresponding periodograms when considering the whole period (Left), the low activity period selected (Middle), and the high activity period selected (Right).} 
 \label{jd_tsi}
\end{figure}

\begin{figure}[ht!]
  \centering
 \includegraphics[angle=0,width=\hsize]{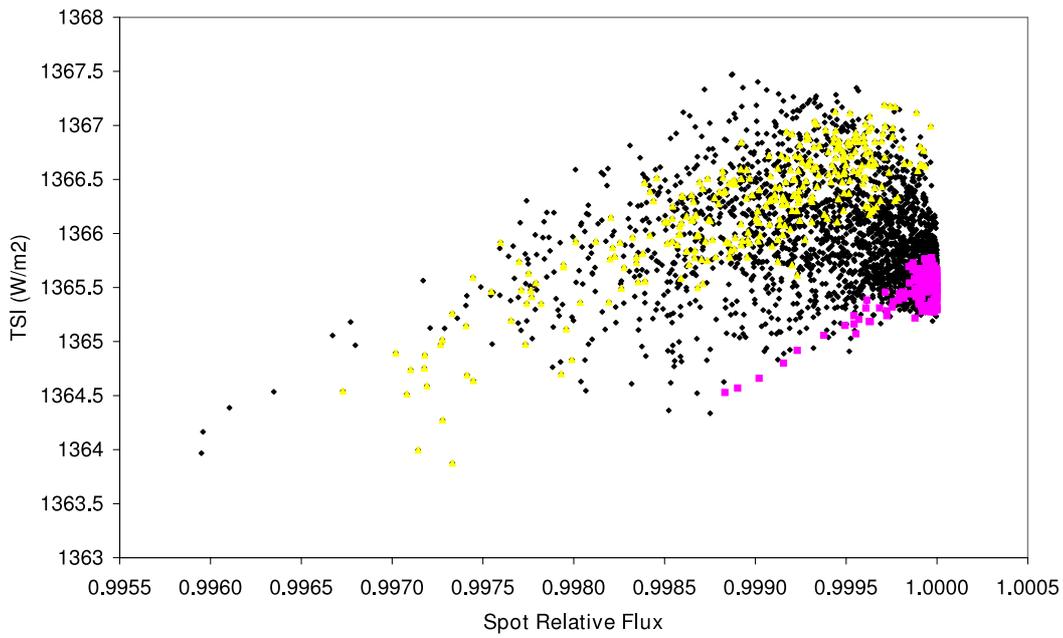}
 \caption{ TSI and spotted Sun photometry over the whole period. The values corresponding to the high and low activity periods selected for reference  (see text) are indicated resp. by triangles and squares.} 
  \label{tsi_rv_phot}
\end{figure}

\begin{figure}[ht!]
  \centering
  \begin{tabular}{cc}
 \includegraphics[angle=90,width=.3\hsize]{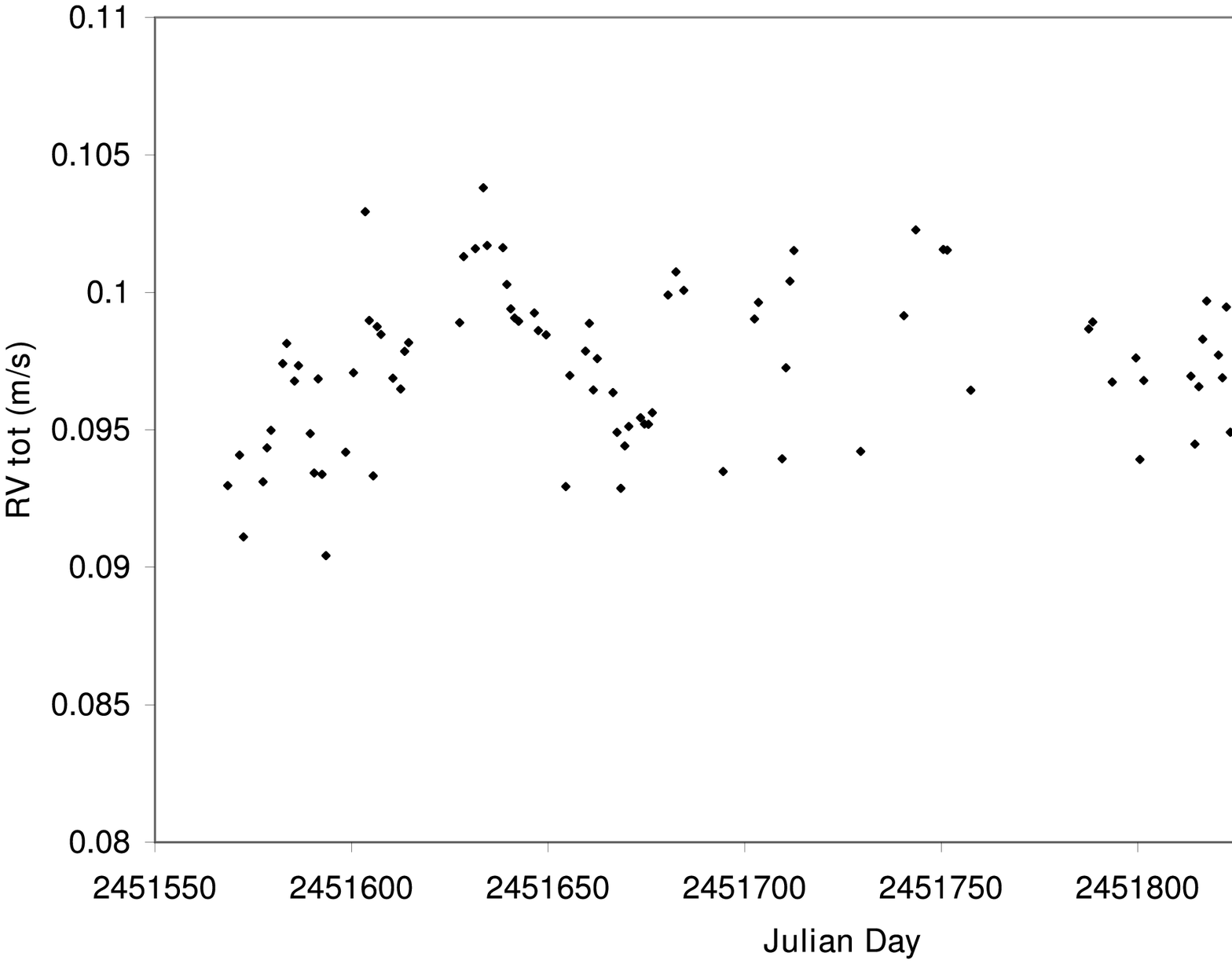}&  \includegraphics[angle=90,width=0.3\hsize]{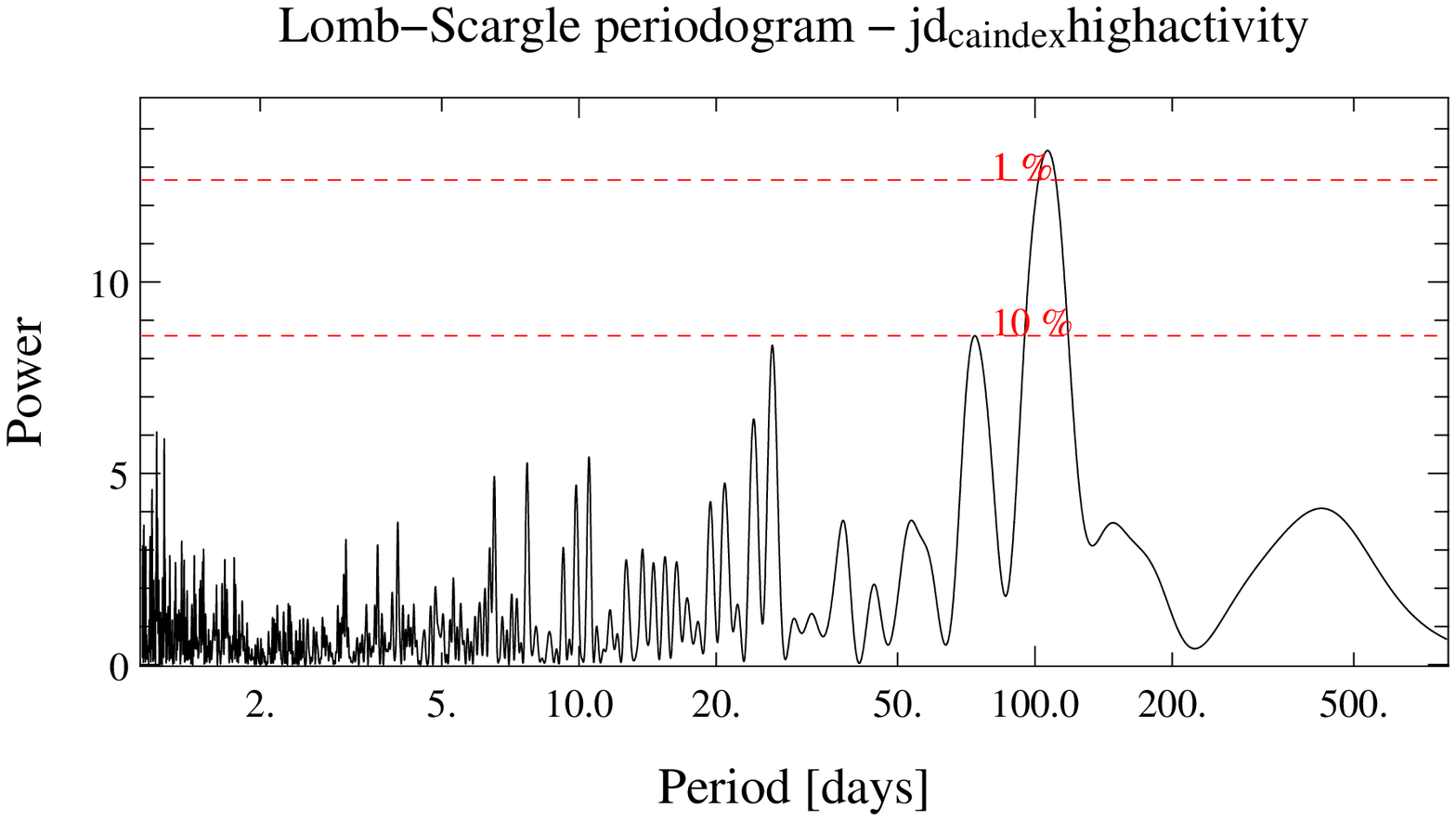}\\
\includegraphics[angle=90,width=.3\hsize]{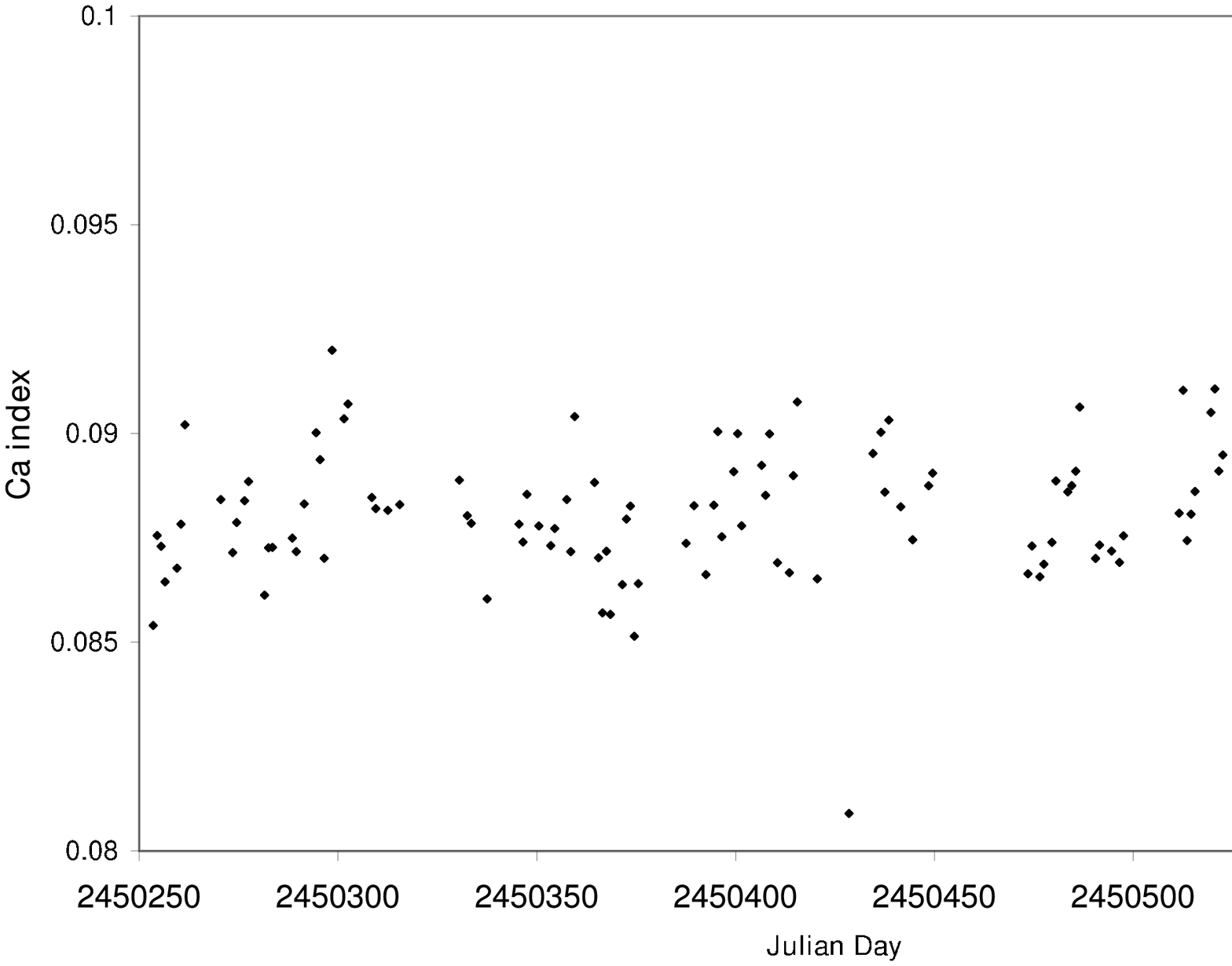}& \includegraphics[angle=90,width=0.3\hsize]{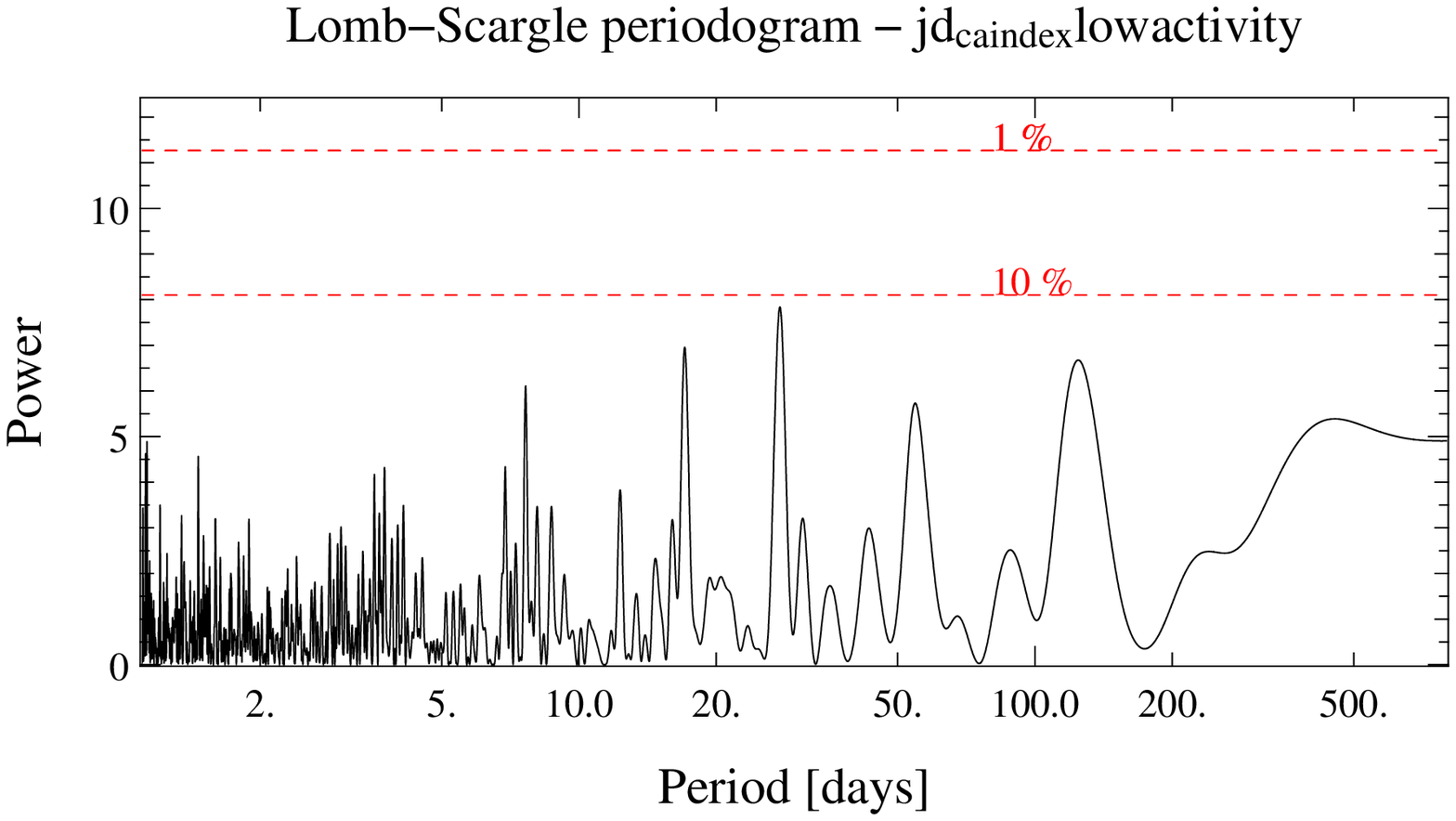} \\
\includegraphics[angle=90,width=.3\hsize]{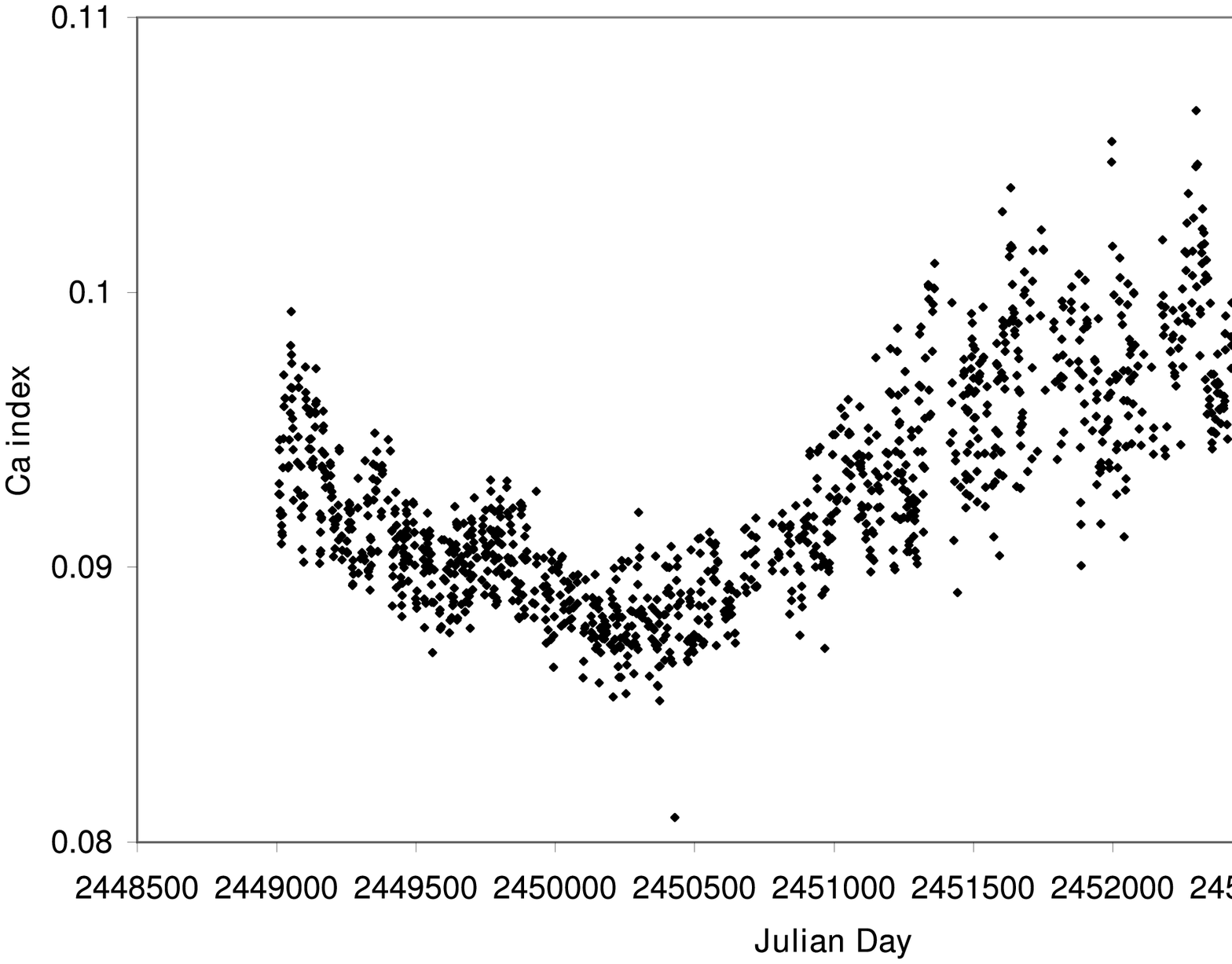}& \includegraphics[angle=90,width=0.3\hsize]{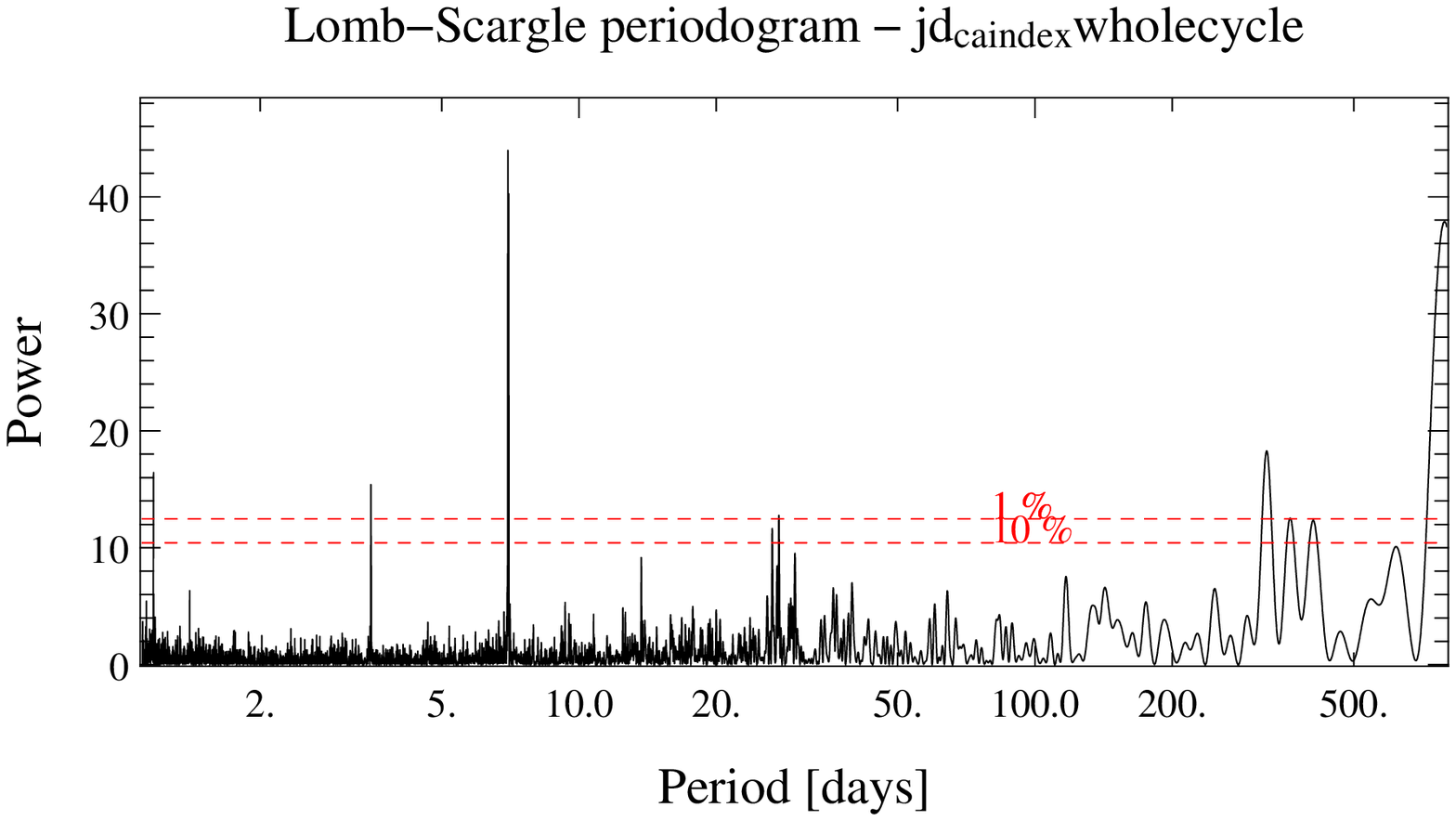}
 \end{tabular}   
\caption{ Temporal variations of the Ca index, and corresponding periodograms when considering  the whole period (Left), the low activity period selected (Middle), and the high activity period selected (Right).} 
  \label{jd_caindex}
\end{figure}

\begin{figure}[ht!]
  \centering
 \includegraphics[angle=0,width=\hsize]{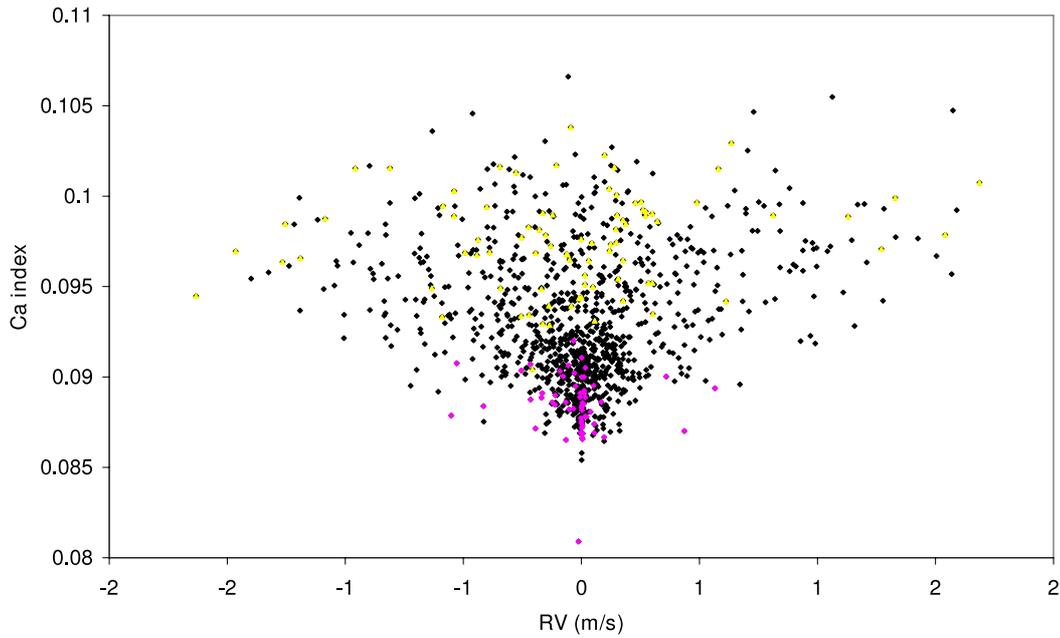}\\
 \caption{Simulated RV and measured Ca index over the whole period. The values corresponding to the high and low activity periods selected for 
reference (see text) are indicated resp. by triangles and squares.} 
  \label{rv_caindex}
\end{figure}

\begin{figure}[ht!]
  \centering
  \includegraphics[angle=0,width=\hsize]{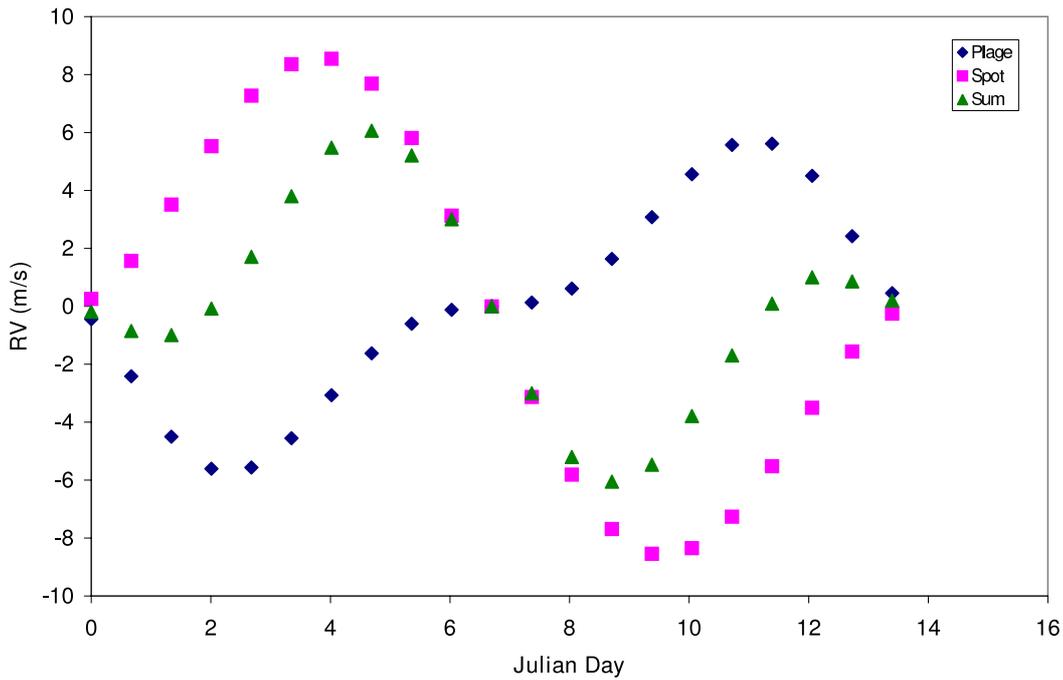}
\caption{Contributions to the RV variations of a spot (squares) and a plage  (diamonds) with 
respective fp = 2 percent and fp = 20 percent and temperatures respectively 550 K lower and 10 K higher 
than the Sun surface. The total RV variations are represented by triangles.} 
  \label{contrib_spot_plage_tot}
\end{figure}

\vfill\eject

\begin{figure}[ht!]
  \centering
 \begin{tabular}{cccc}
\includegraphics[angle=90,width=.3\hsize]{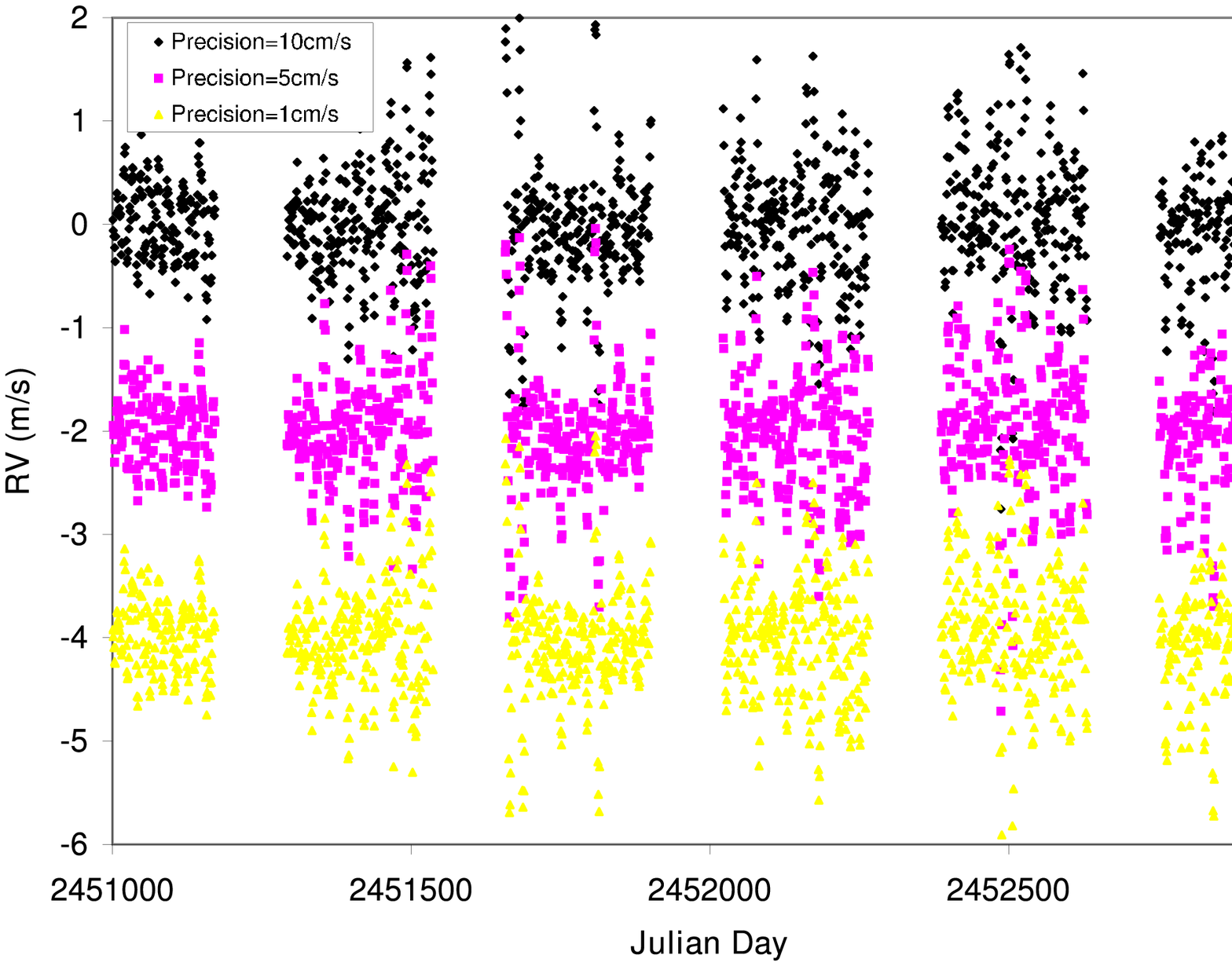} & \includegraphics[angle=90,width=0.3\hsize]{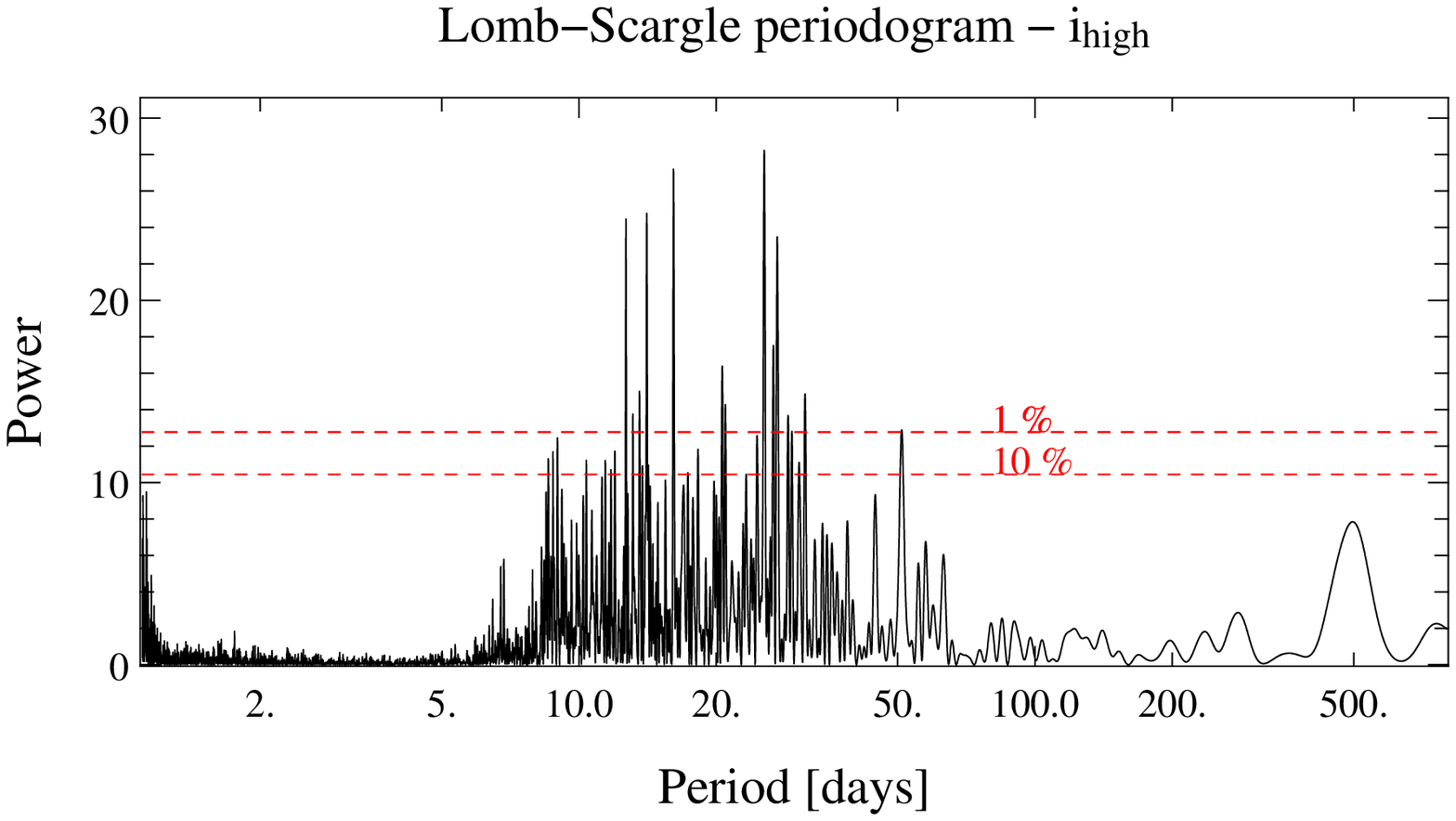}& \includegraphics[angle=90,width=0.3\hsize]{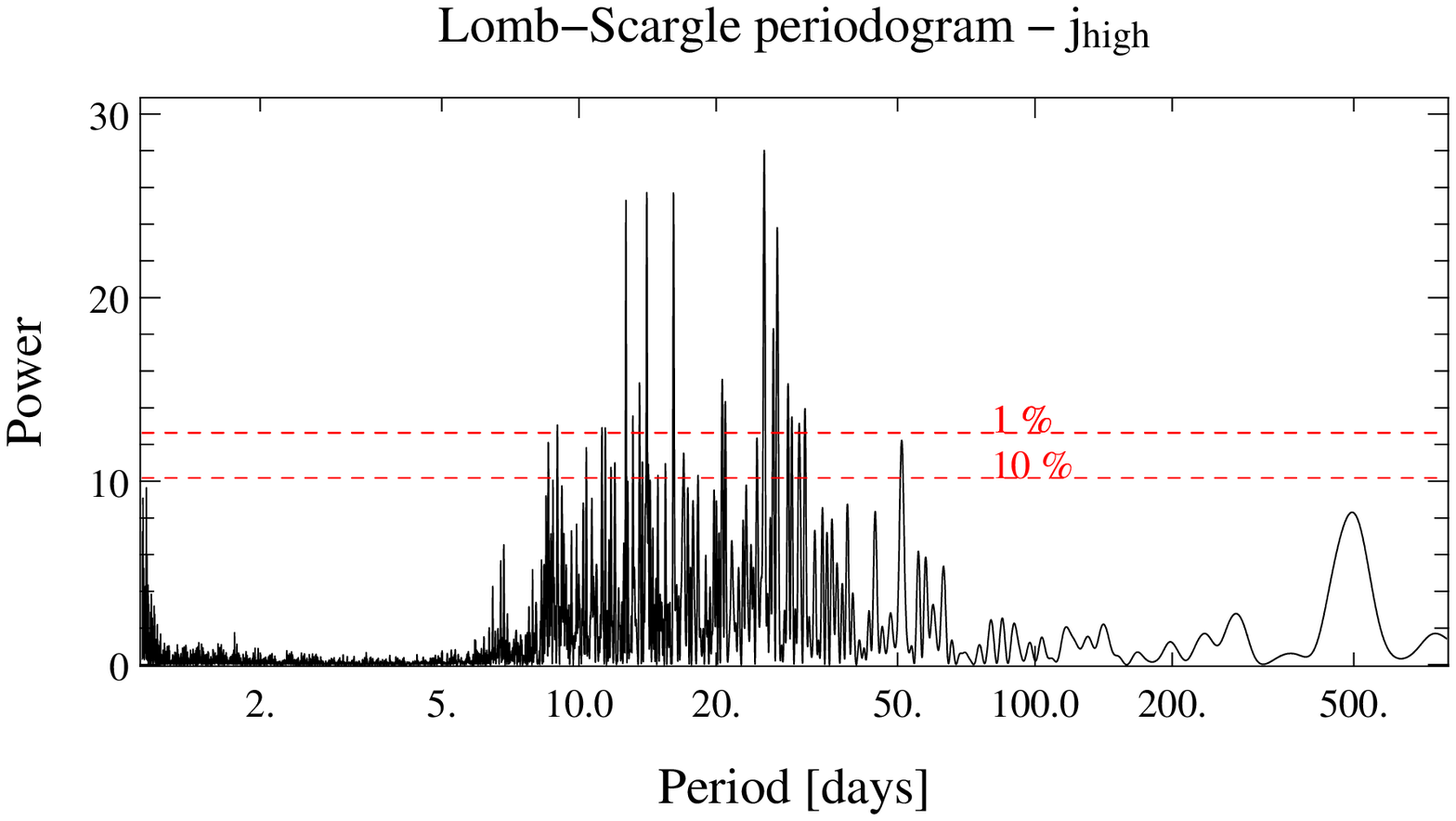}& \includegraphics[angle=90,width=0.3\hsize]{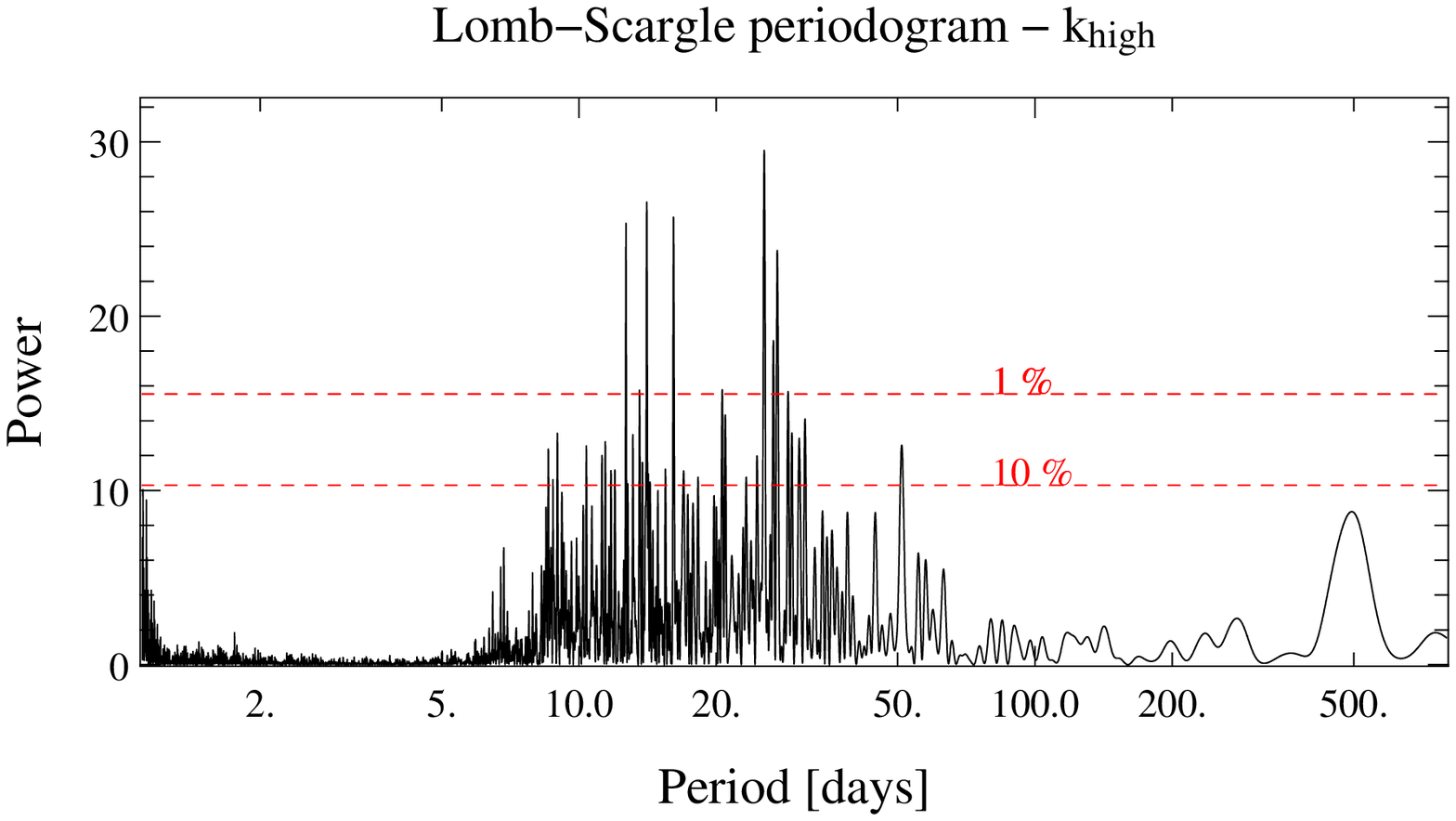}\\
 \includegraphics[angle=90,width=.3\hsize]{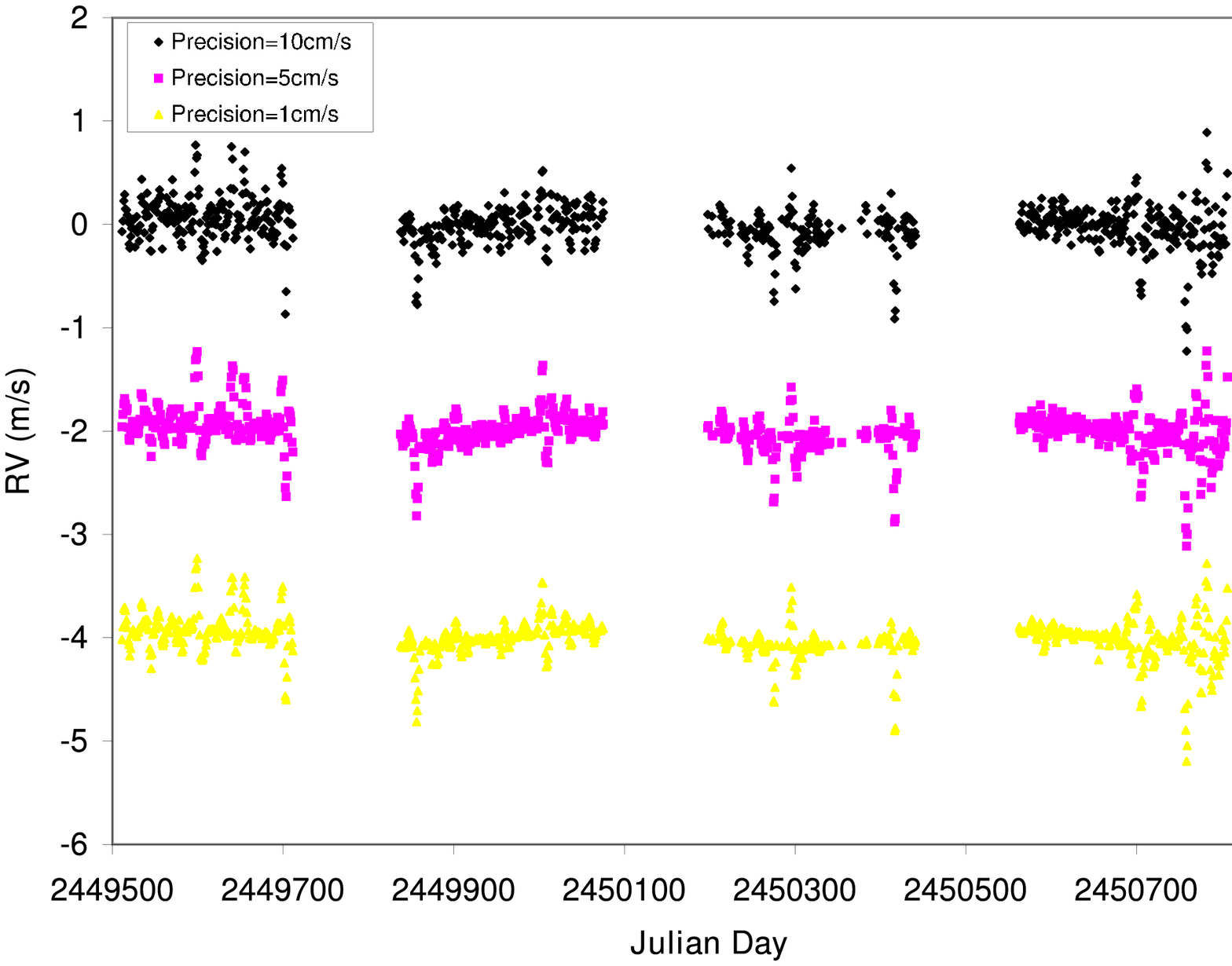}  & \includegraphics[angle=90,width=0.3\hsize]{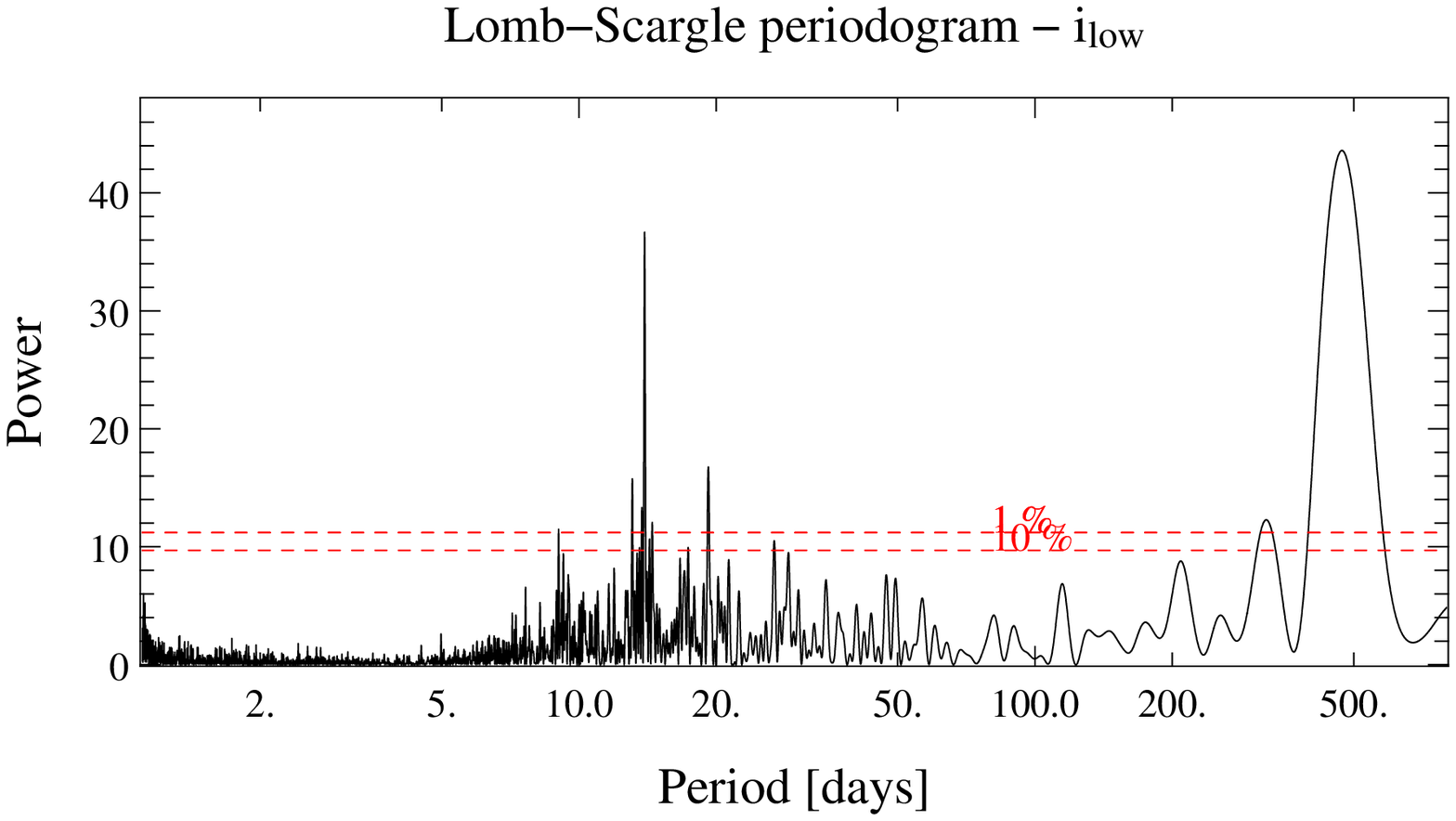}& \includegraphics[angle=90,width=0.3\hsize]{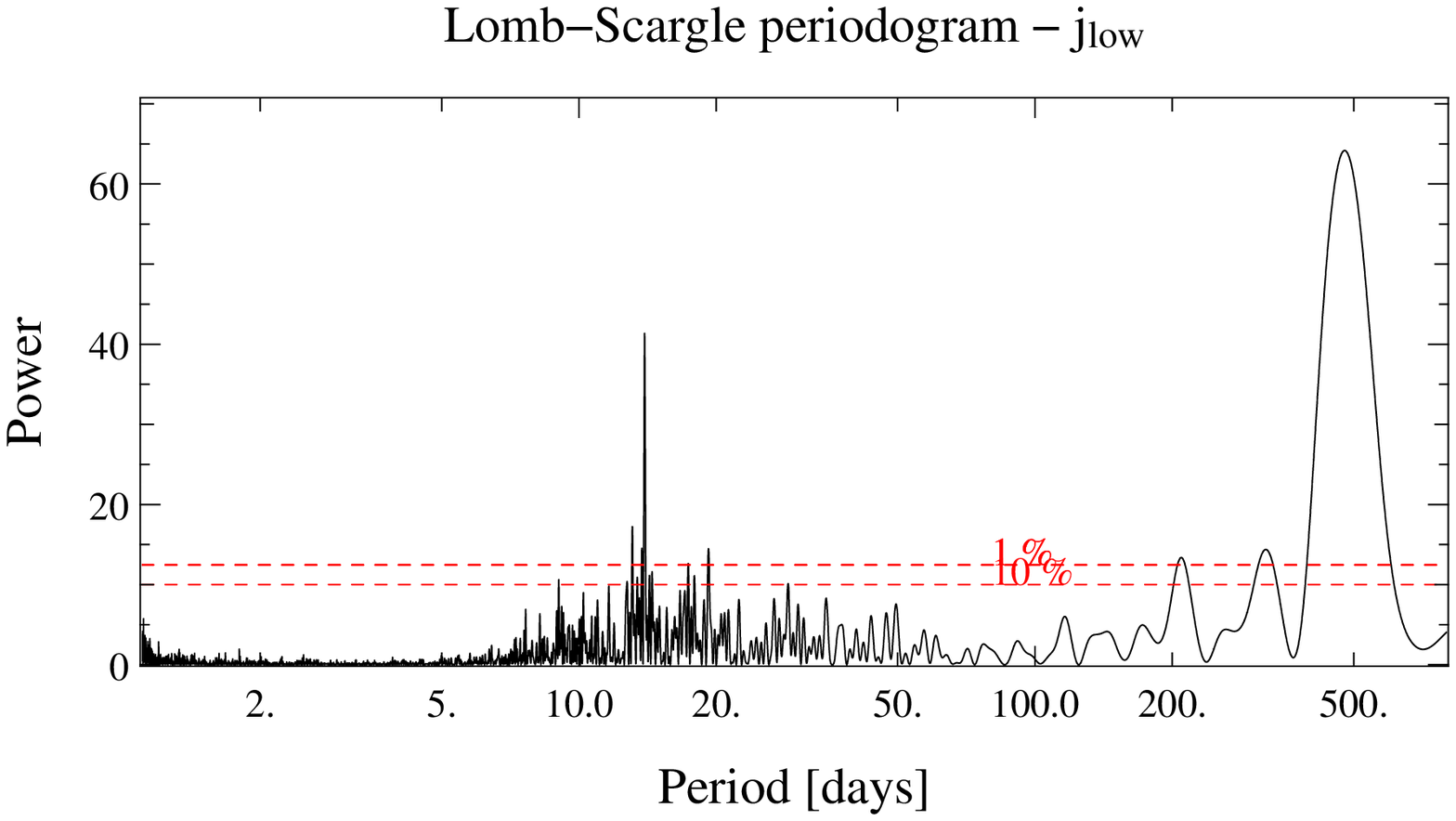}& \includegraphics[angle=90,width=0.3\hsize]{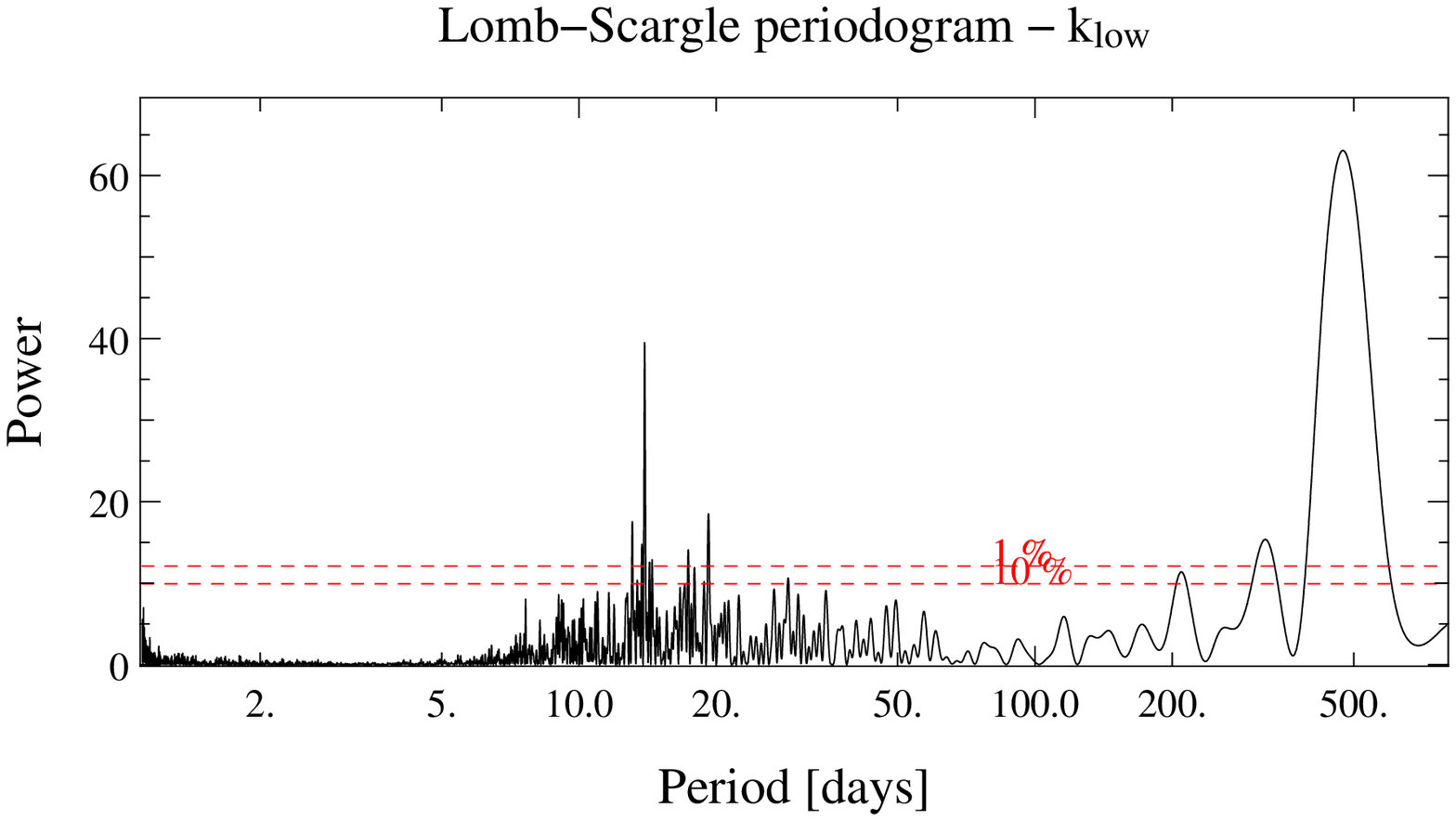}\\
 \includegraphics[angle=90,width=0.3\hsize]{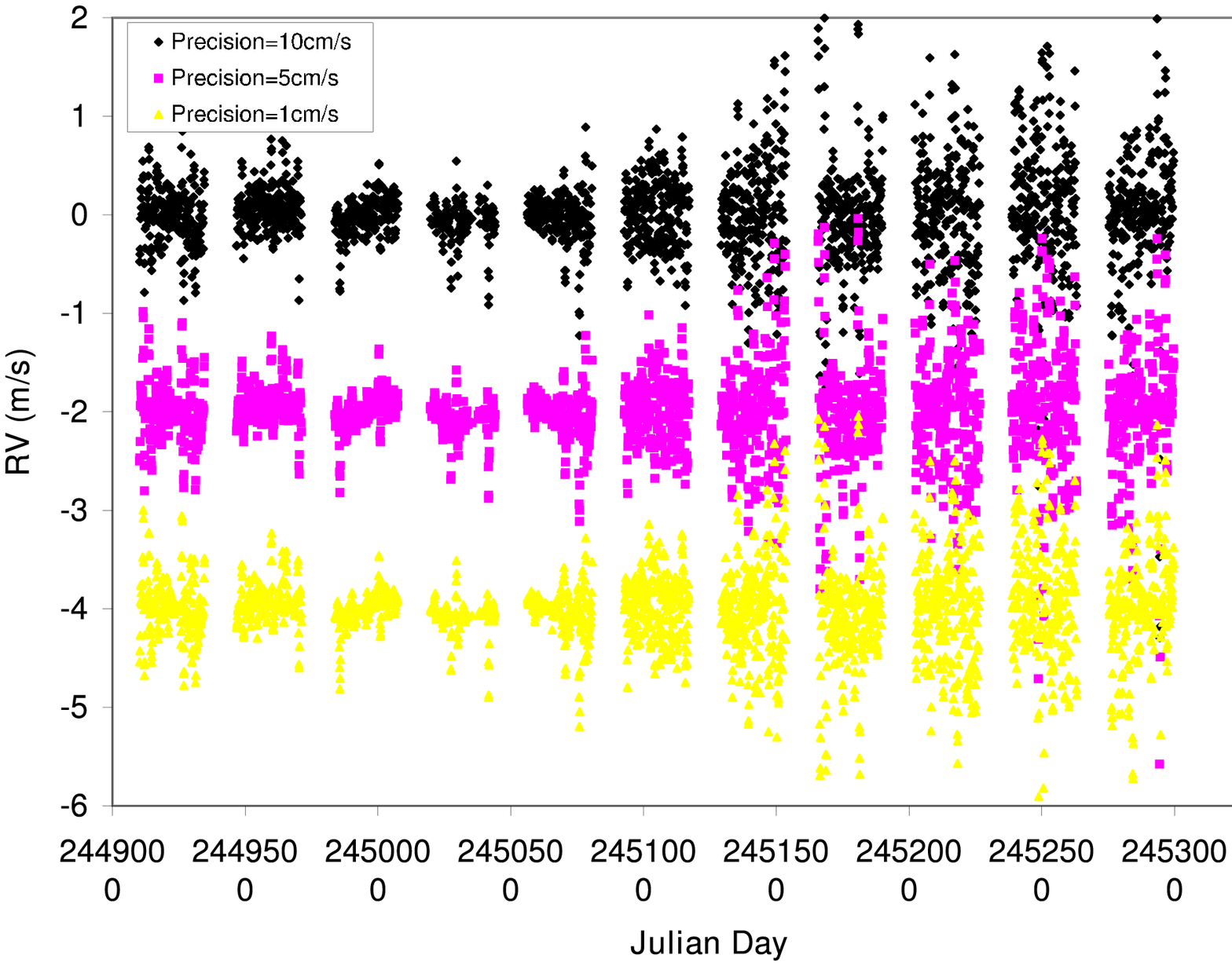} & \includegraphics[angle=90,width=0.3\hsize]{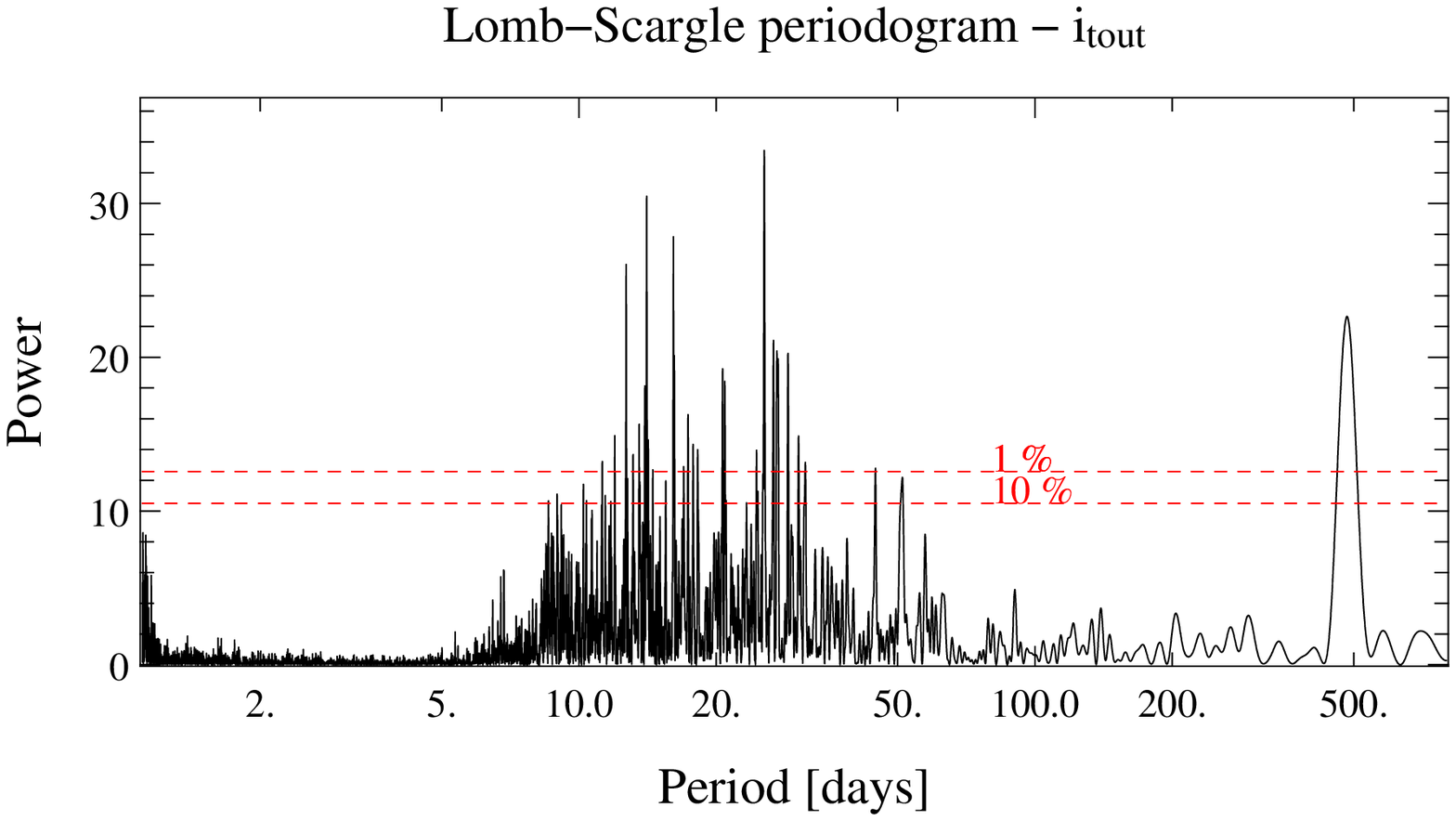}& \includegraphics[angle=90,width=0.3\hsize]{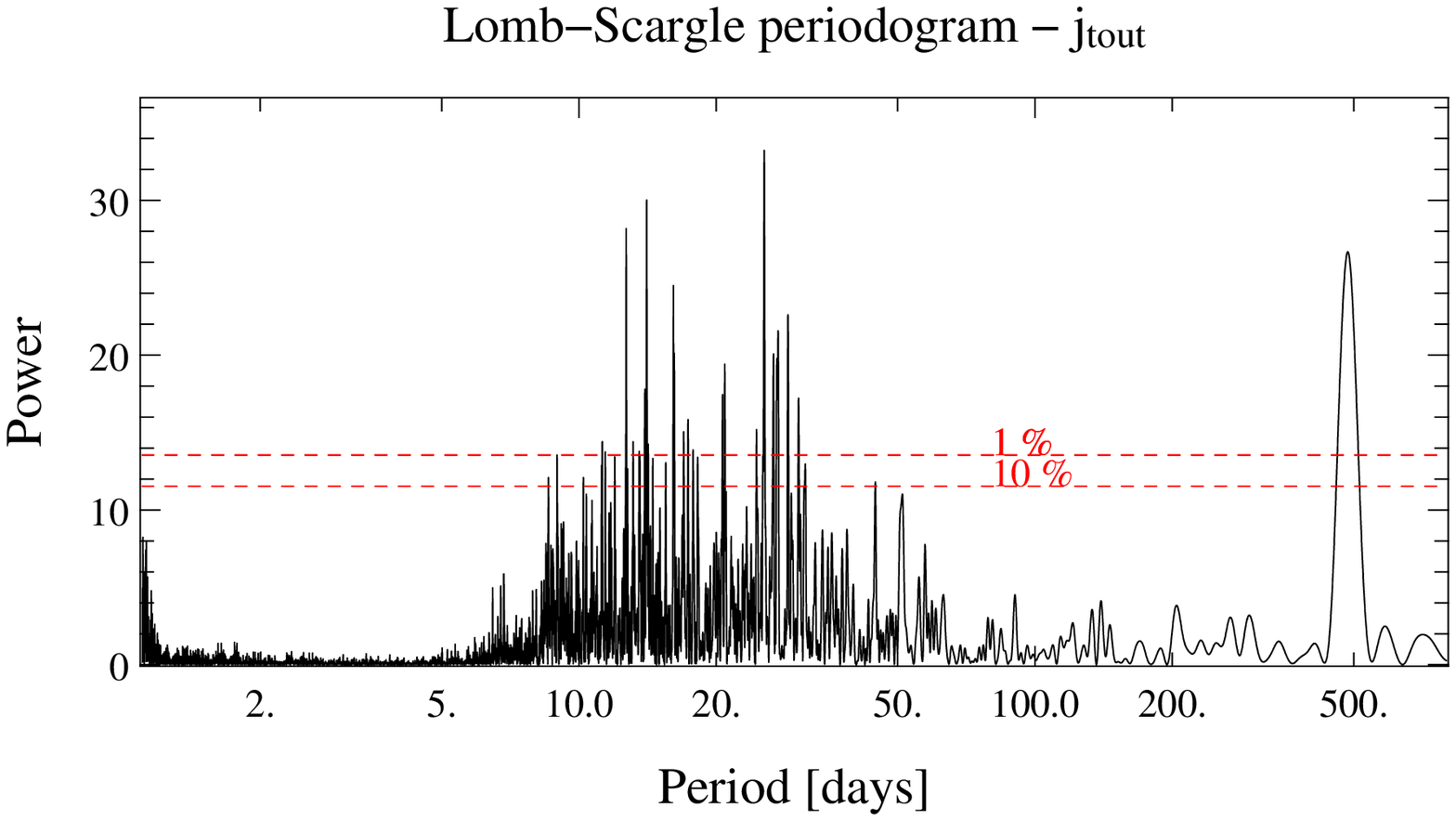}& \includegraphics[angle=90,width=0.3\hsize]{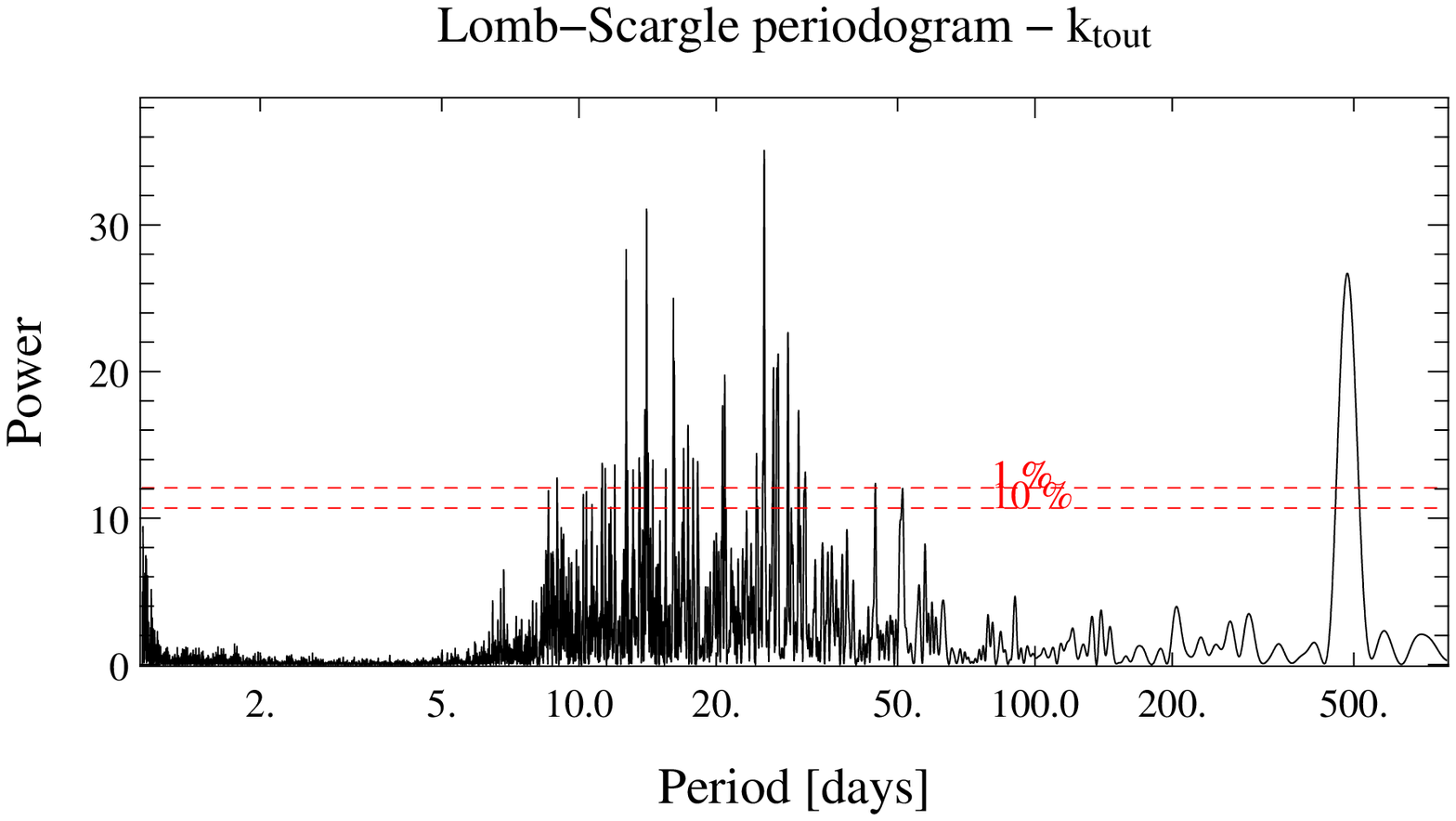}\\
 \end{tabular} 
 \caption{ Top: RV of a spotted solar-type star surrounded by a 1 \me planet orbiting at 1.2 AU. The star is assumed to be observable 8 months per year. The RV curves correspond, from top to bottom, to precisions of 10 cm/s, 5cm/s and 1 cm/s (note that the curves for the precisions of 5 and 1 cm/s have been vertically shifted for clarity purposes). The full cycle (Left) as well as a low activity (Middle) and high activity (Right) periods are considered. The corresponding periodograms are provided: first line: 10 cm/s; second line: 5cm/s and third line 1cm/s. }
\label{1p2_4m}
\end{figure}

\clearpage
\begin{figure}[ht!]
  \centering
 \begin{tabular}{cccc}
\includegraphics[angle=90,width=0.3\hsize]{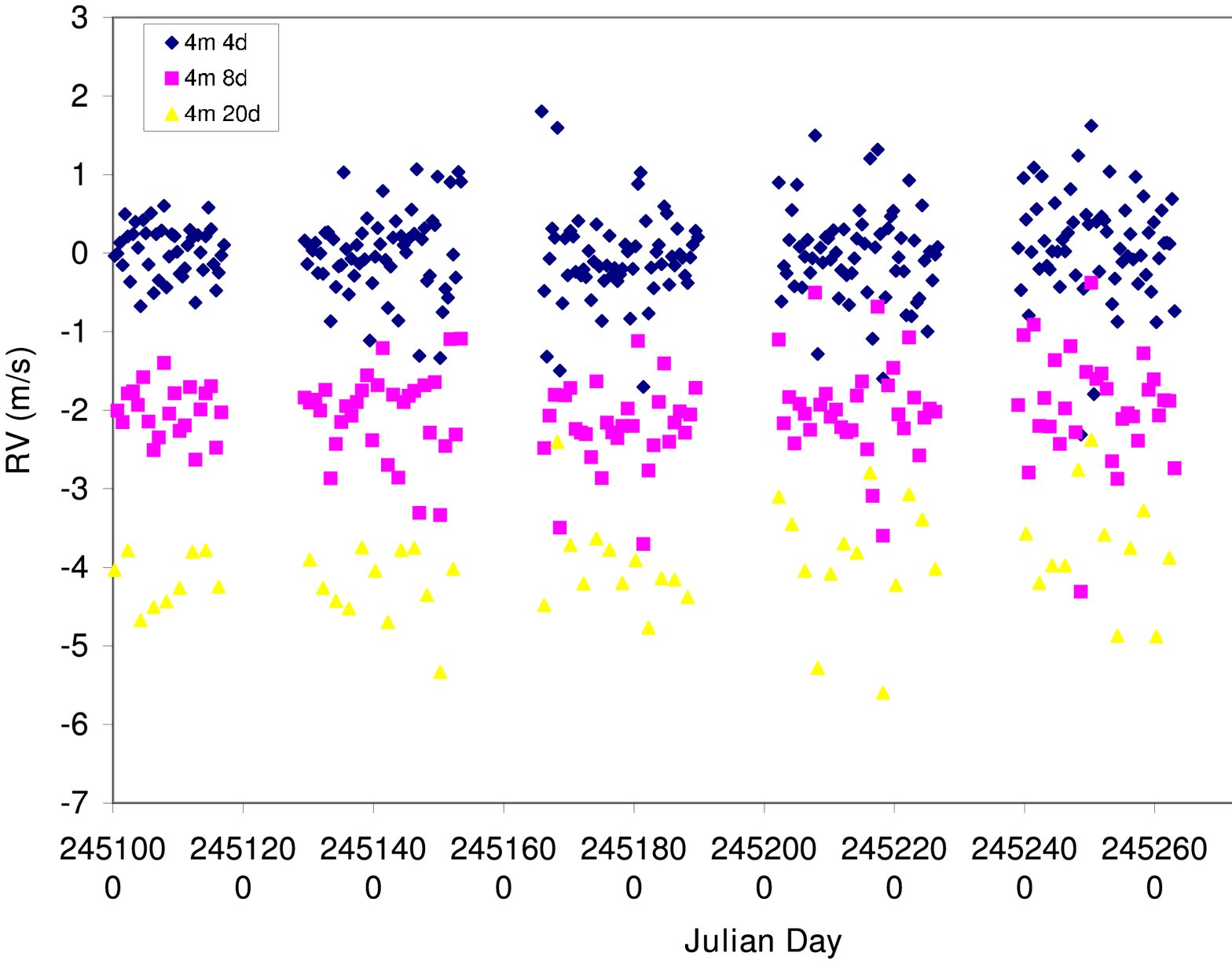}&
\includegraphics[angle=90,width=0.3\hsize]{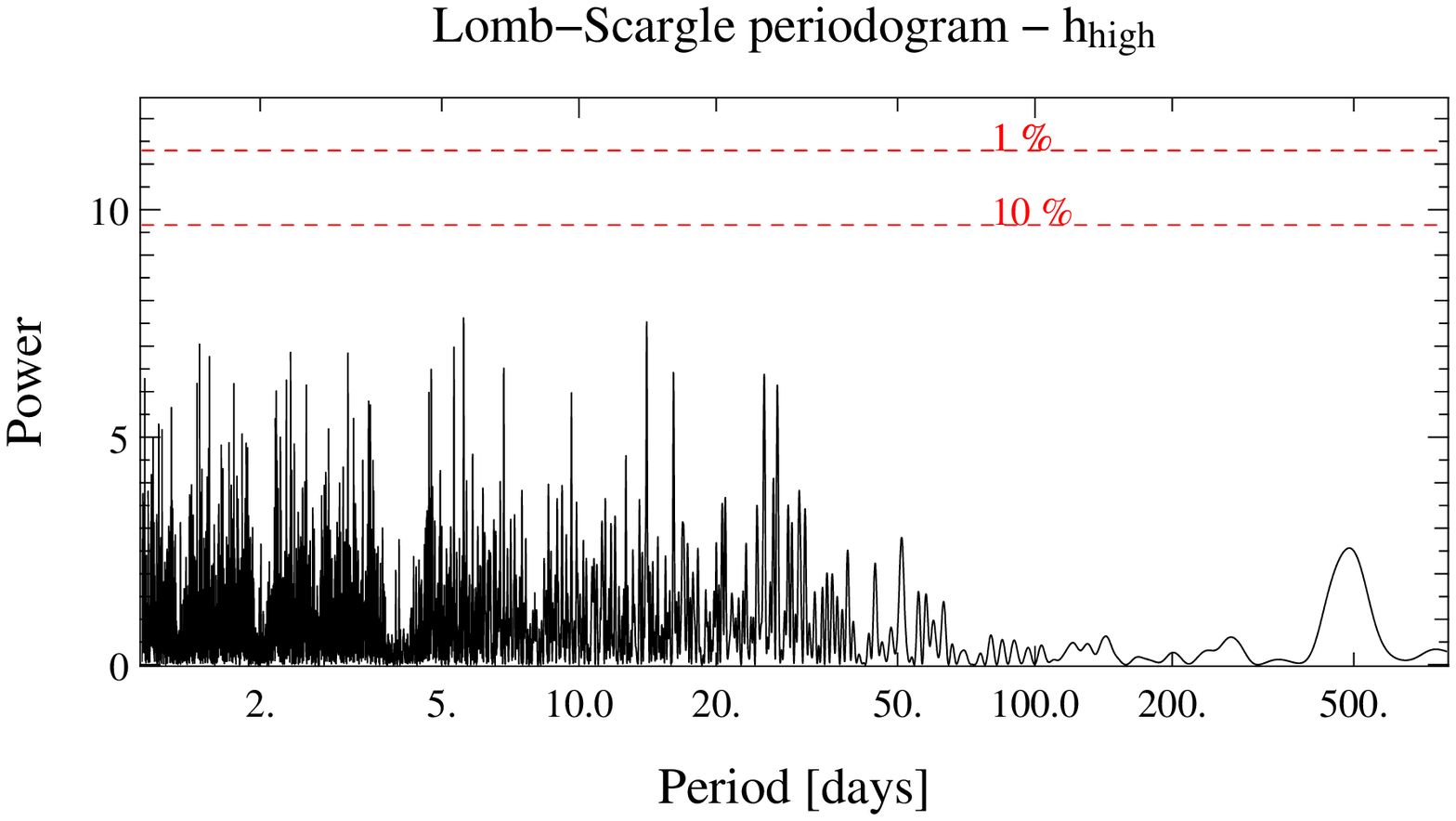}&
\includegraphics[angle=90,width=0.3\hsize]{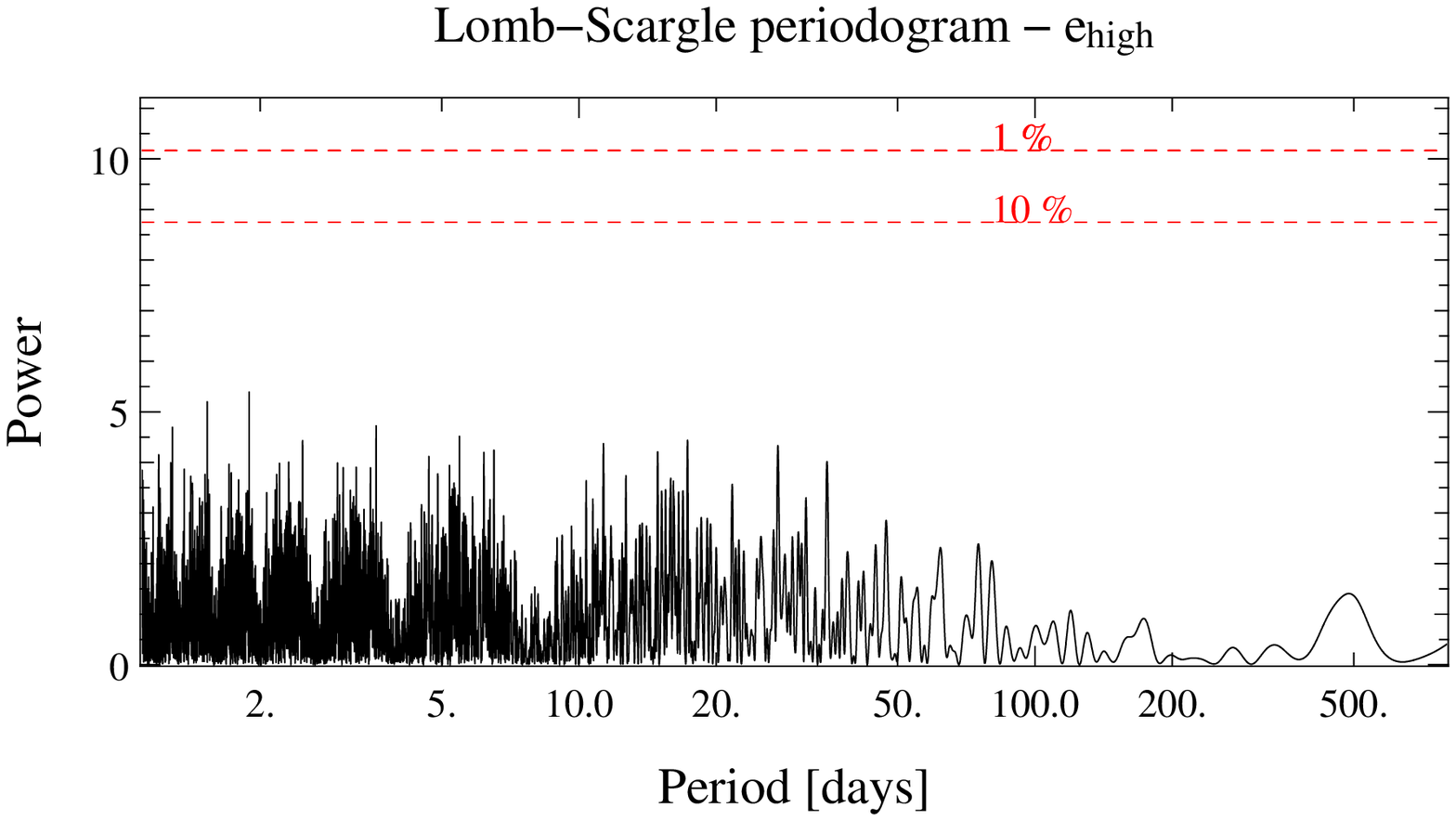}&
\includegraphics[angle=90,width=0.3\hsize]{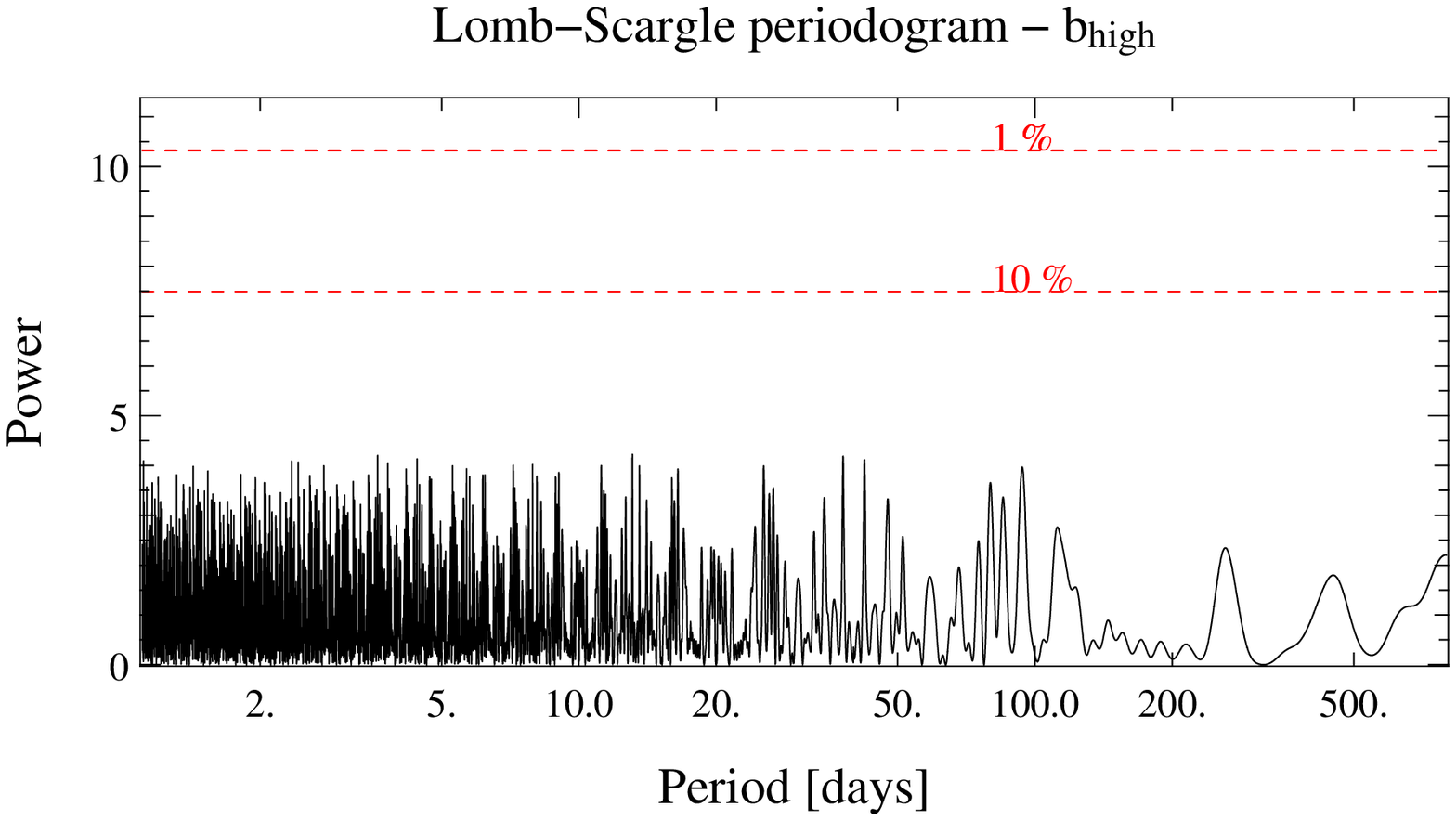}\\
\includegraphics[angle=90,width=0.3\hsize]{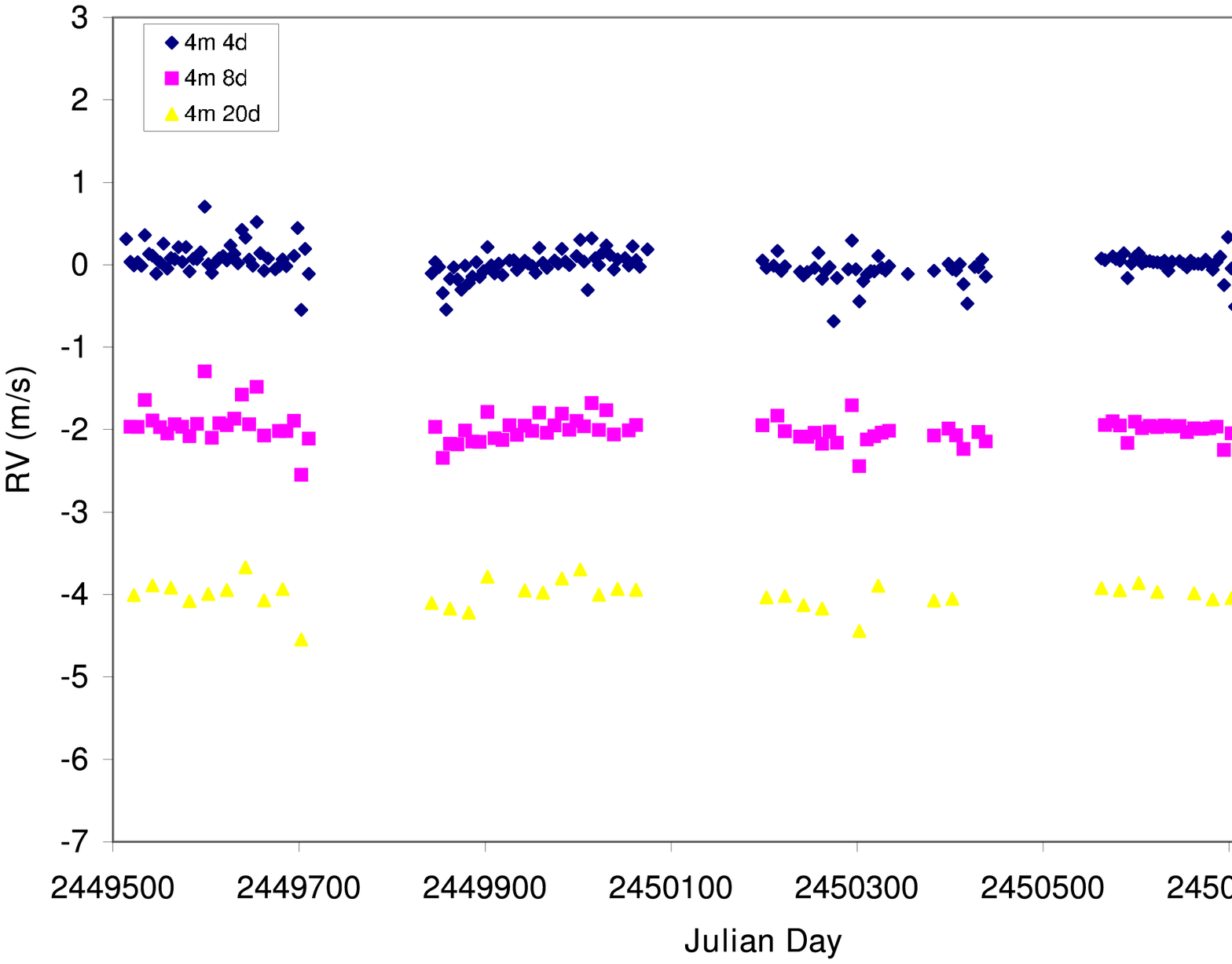}&
\includegraphics[angle=90,width=0.3\hsize]{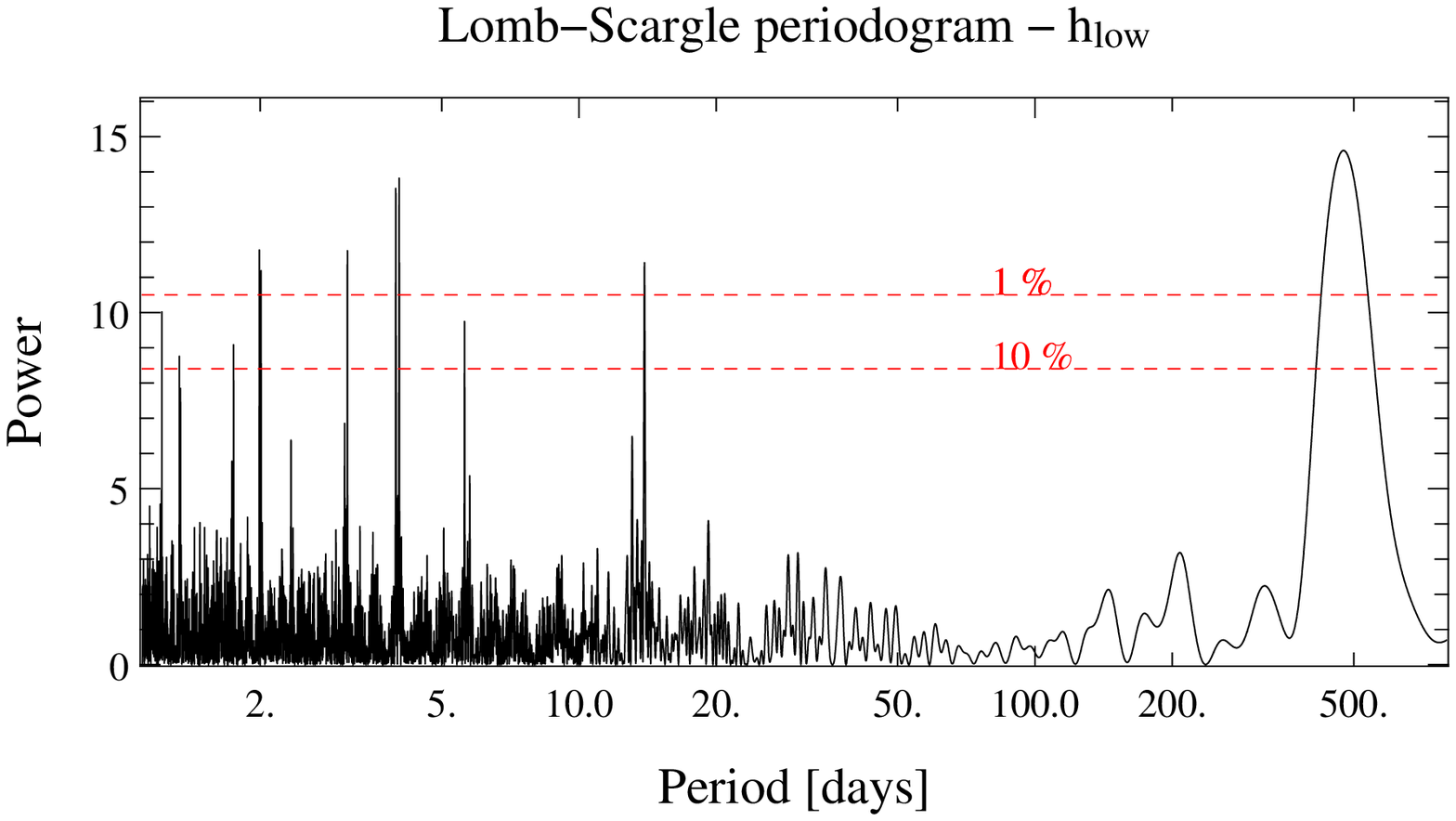}&
\includegraphics[angle=90,width=0.3\hsize]{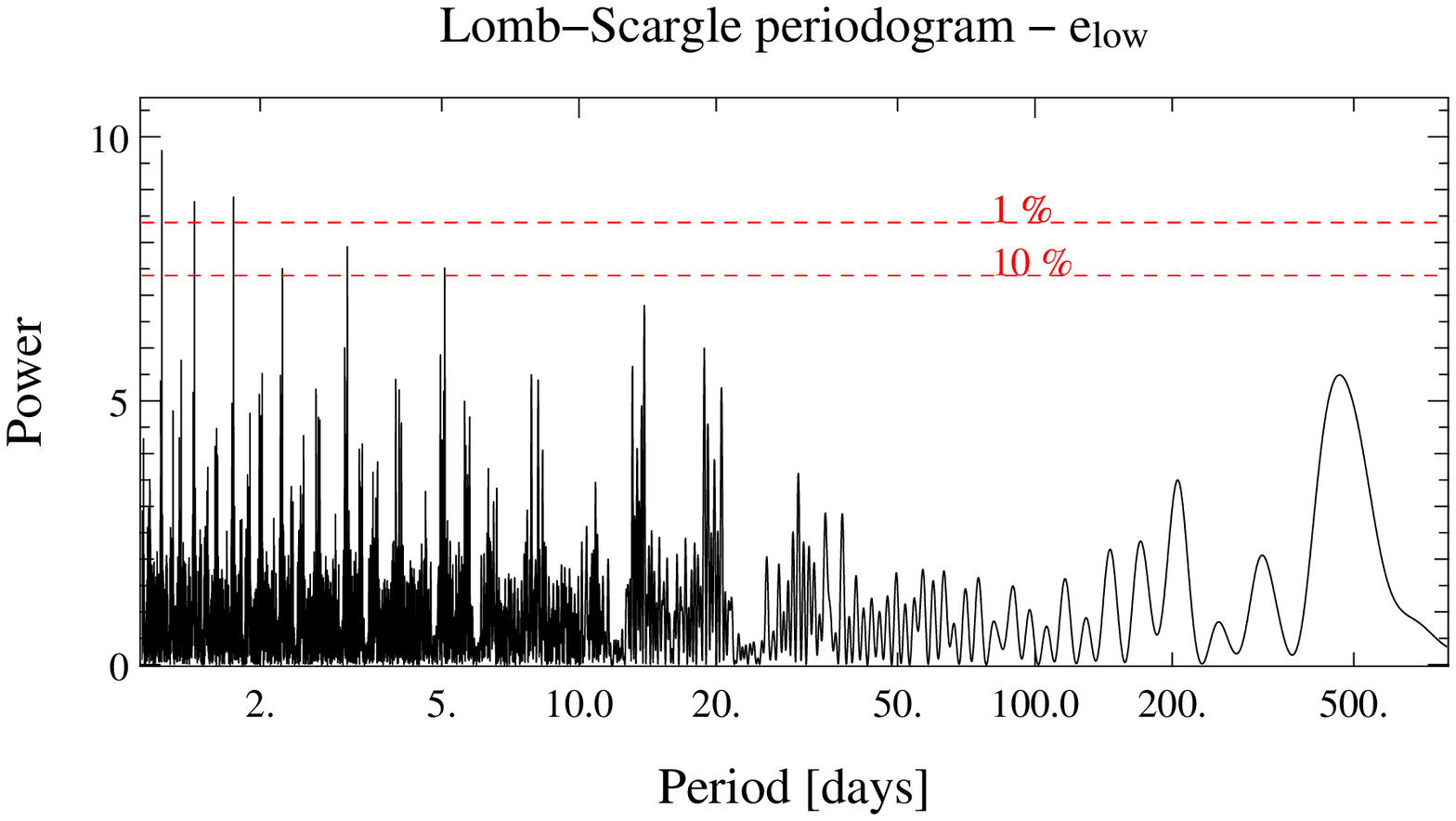}&
\includegraphics[angle=90,width=0.3\hsize]{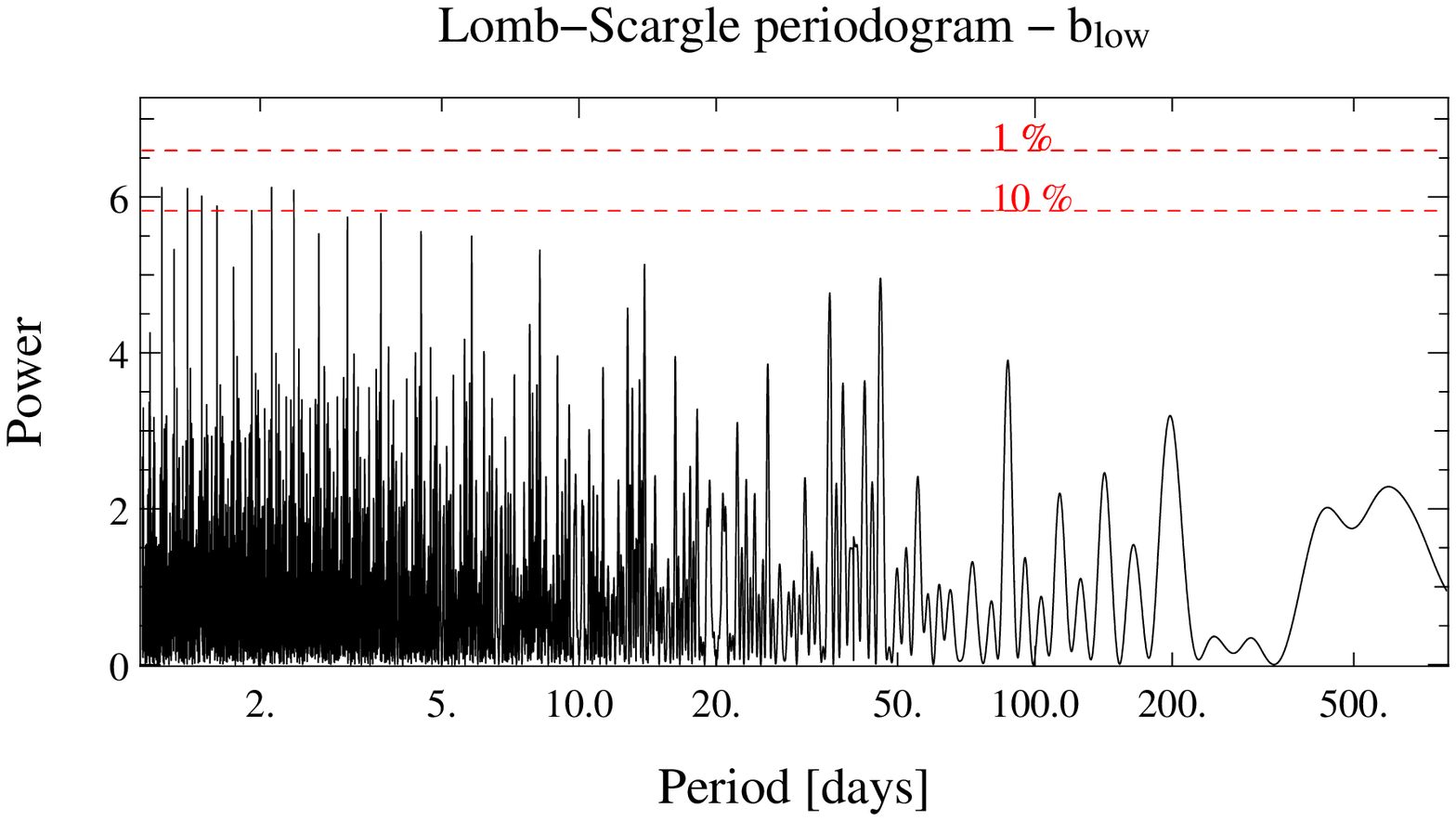}\\
\includegraphics[angle=90,width=0.3\hsize]{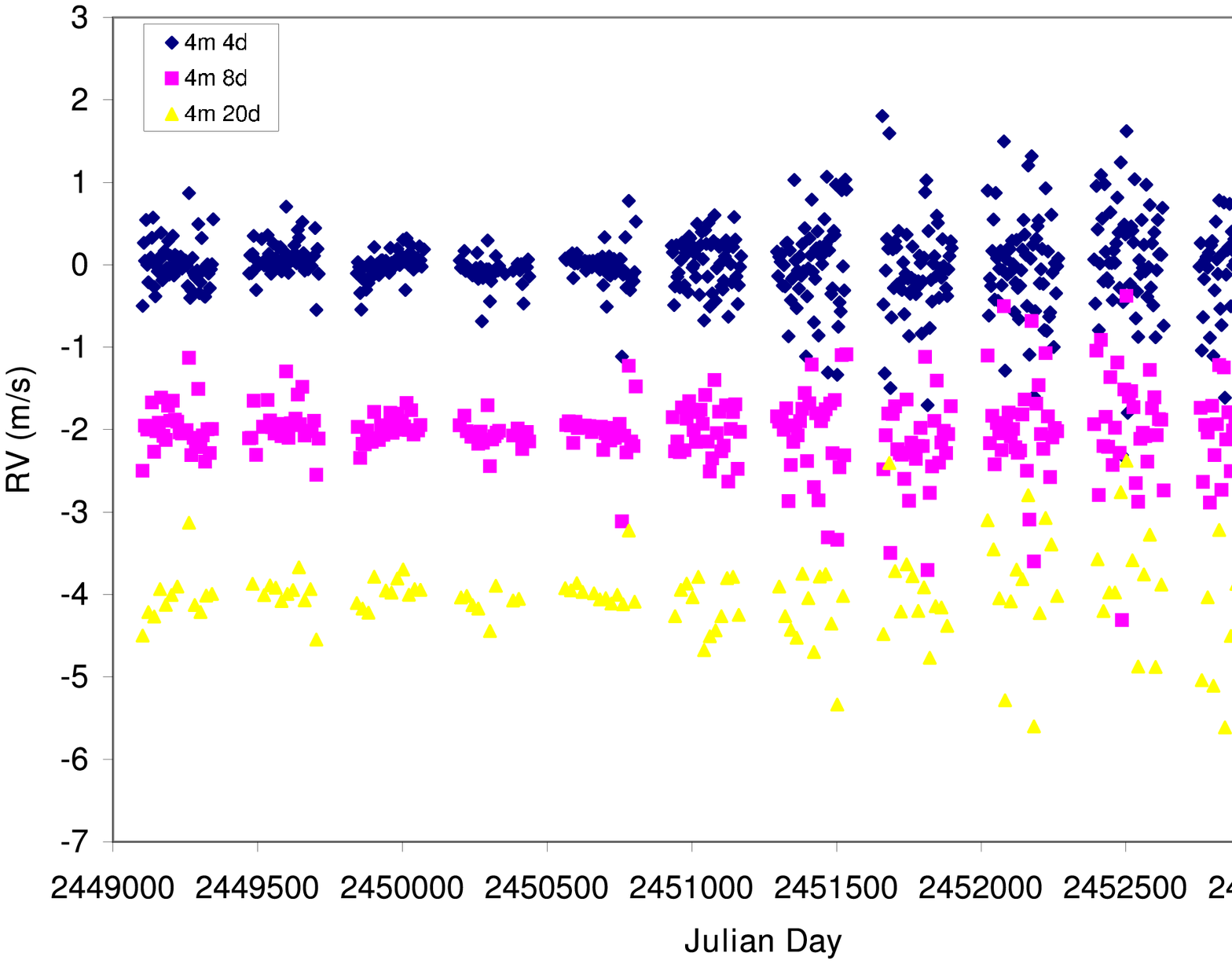}&
\includegraphics[angle=90,width=0.3\hsize]{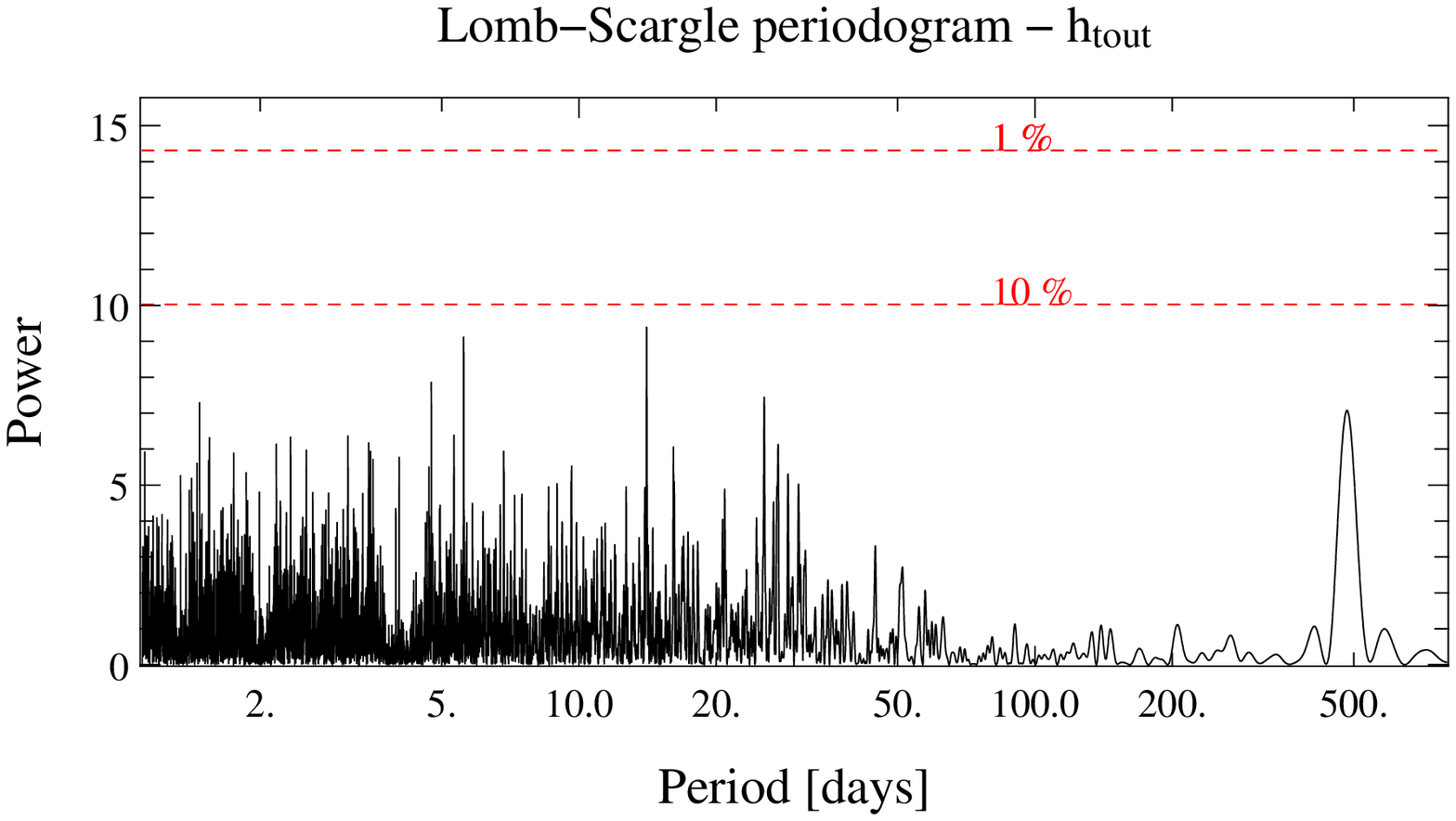}&
\includegraphics[angle=90,width=0.3\hsize]{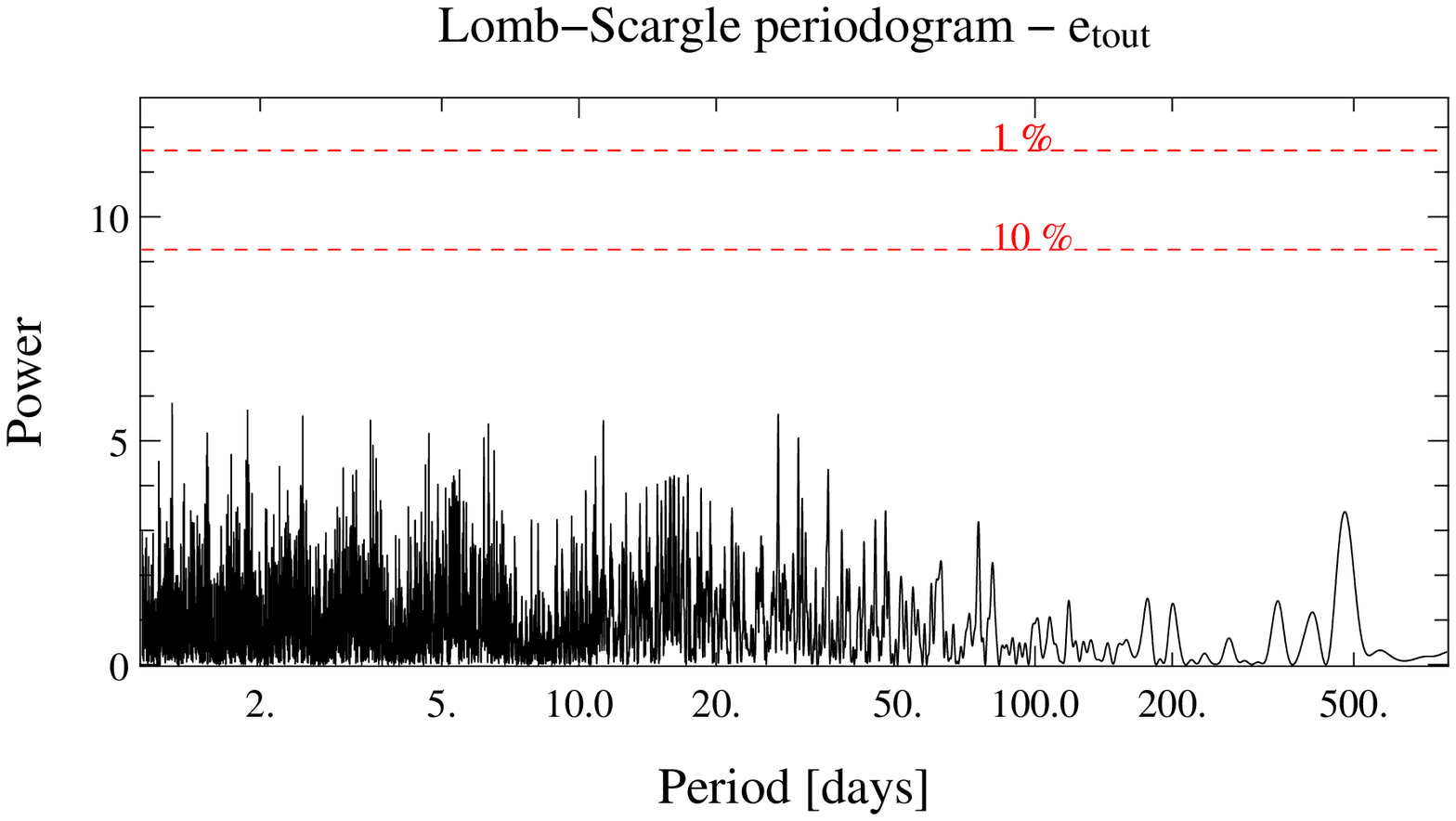}&
\includegraphics[angle=90,width=0.3\hsize]{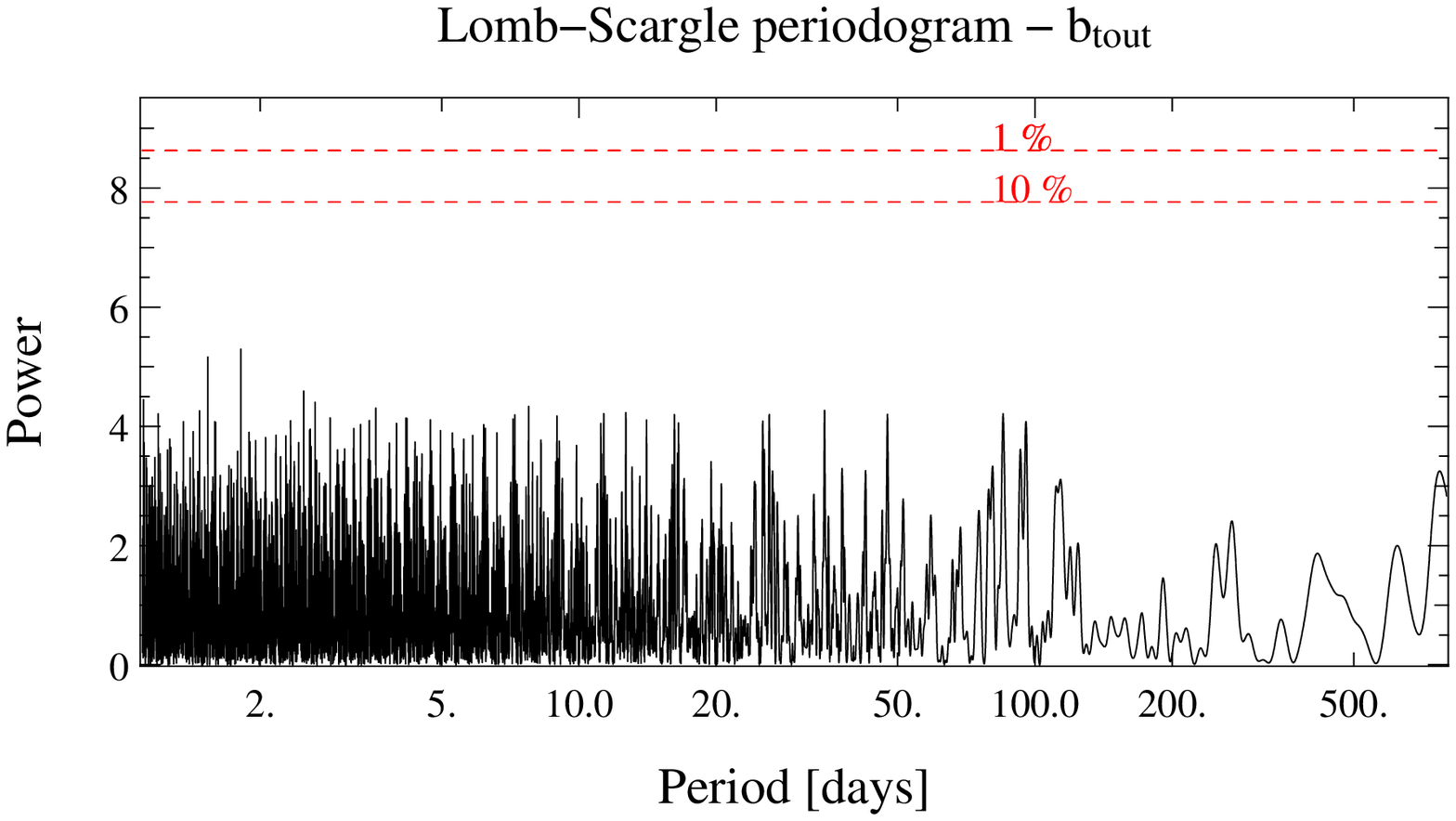}\\
\end{tabular} 
 \caption{Impact of the data sampling: Top: RV of a spotted solar-type star surrounded by a 1 \me planet orbiting at 1.2 AU. The star is assumed to be observable 8 months per year and observed every 4 days (top data); 8 days (middle data) or 20 days (lower data). In all casses, we assume a precision of 5 cm/s. The RV curves have been vertically shifted for clarity purposes). The full cycle (Left) as well as a low activity (Middle) and high activity (Right) periods are considered. The corresponding periodograms are provided: first line: temporal sampling is 4 days; second line: 8 days and third line 20 days. }
\label{1p2_tempsampling}
\end{figure}

\vfill\eject

\begin{figure}[ht!]
  \centering
\begin{tabular}{cc}
\includegraphics[angle=-90,width=0.4\hsize]{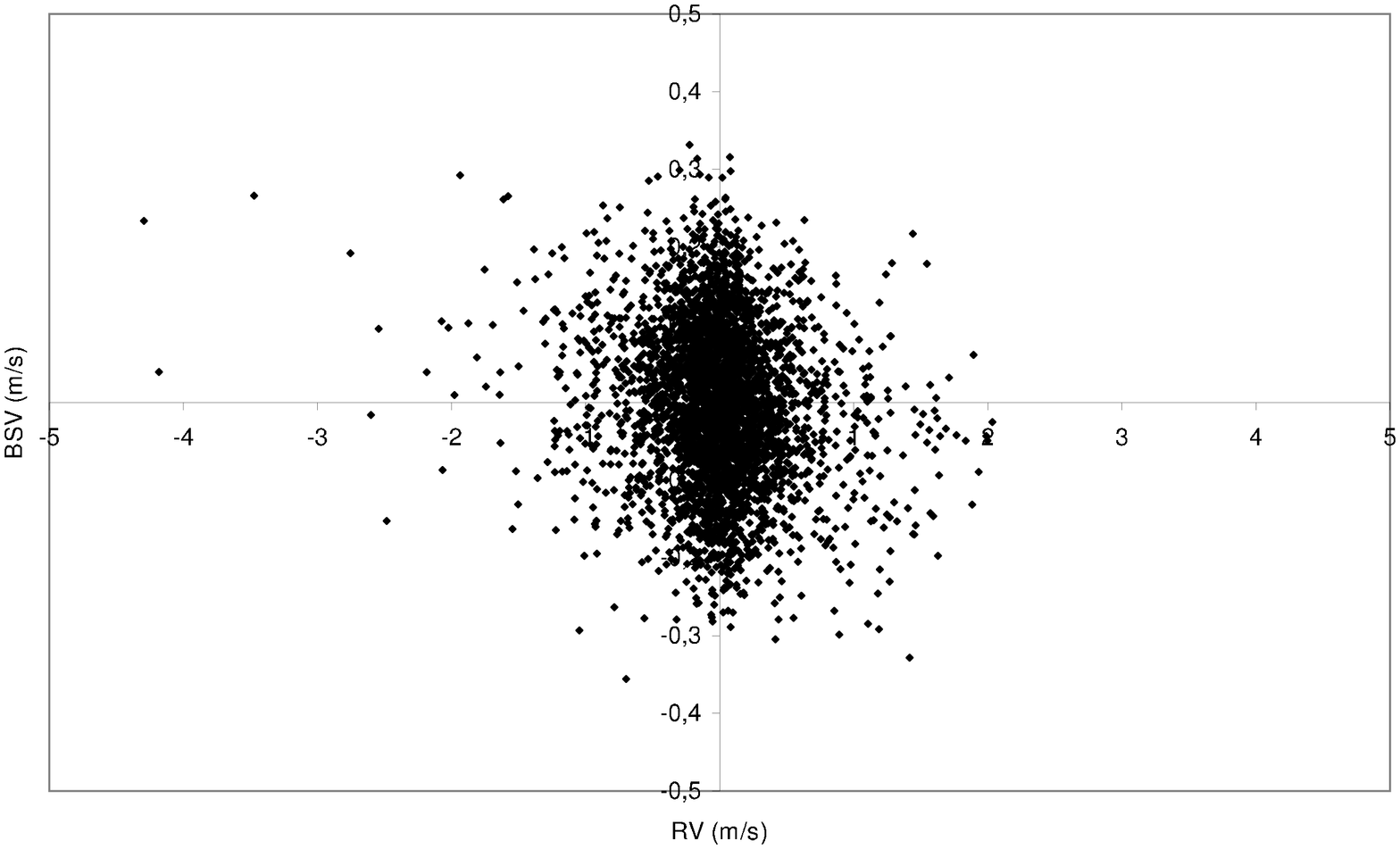}&\includegraphics[angle=-90,width=0.4\hsize]{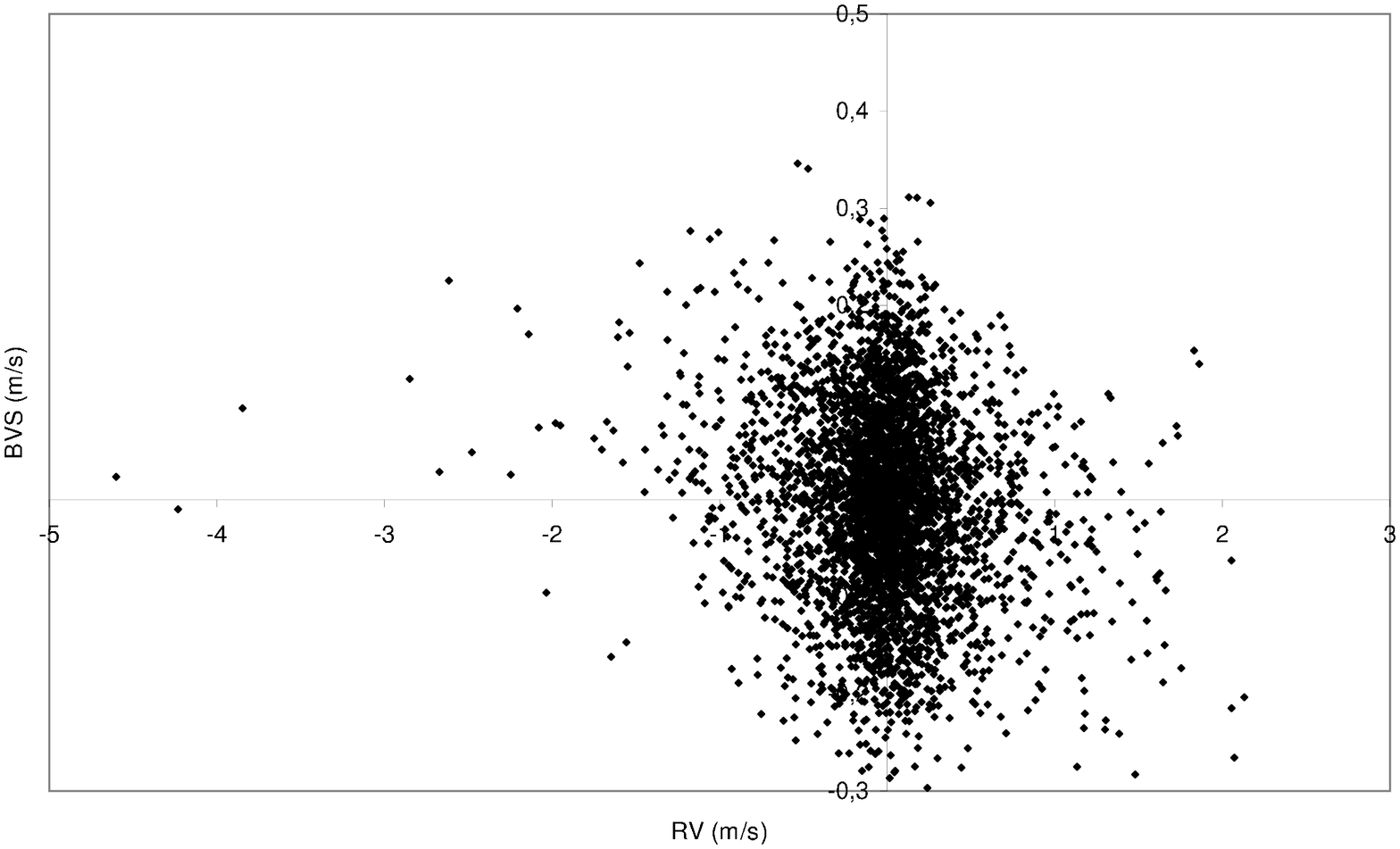}\\
\includegraphics[angle=-90,width=0.4\hsize]{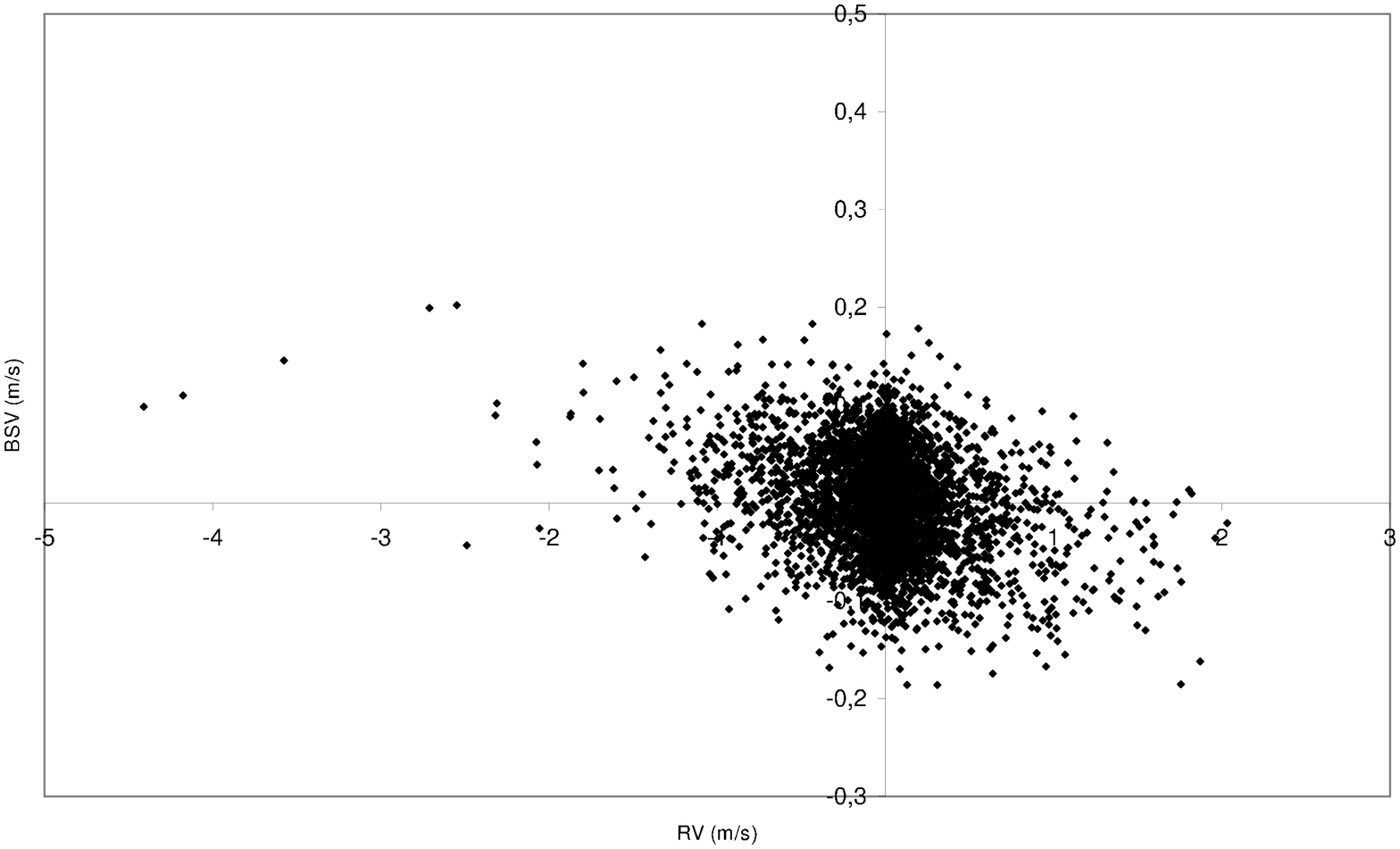}&\includegraphics[angle=-90,width=0.4\hsize]{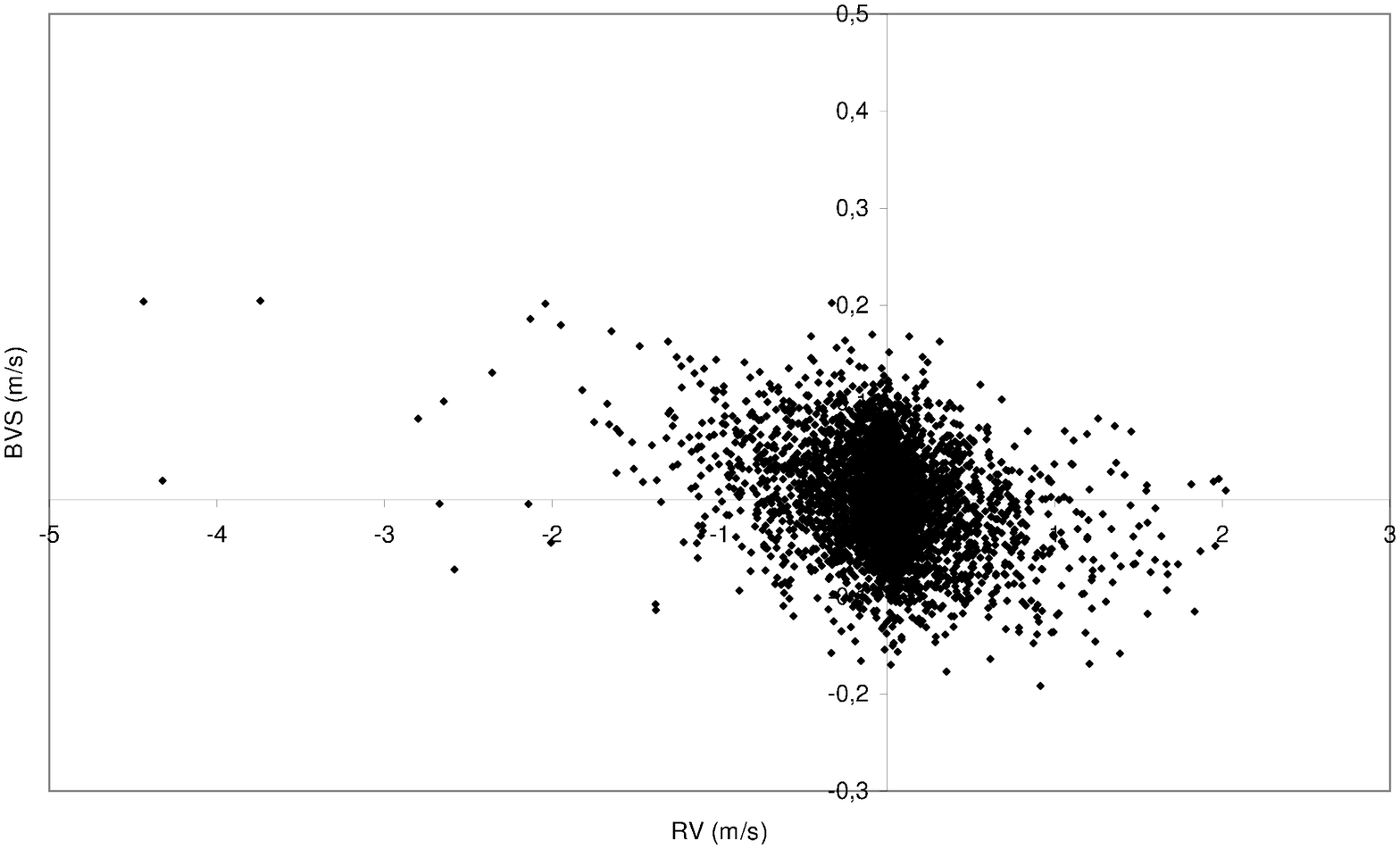}\\
\includegraphics[angle=-90,width=0.4\hsize]{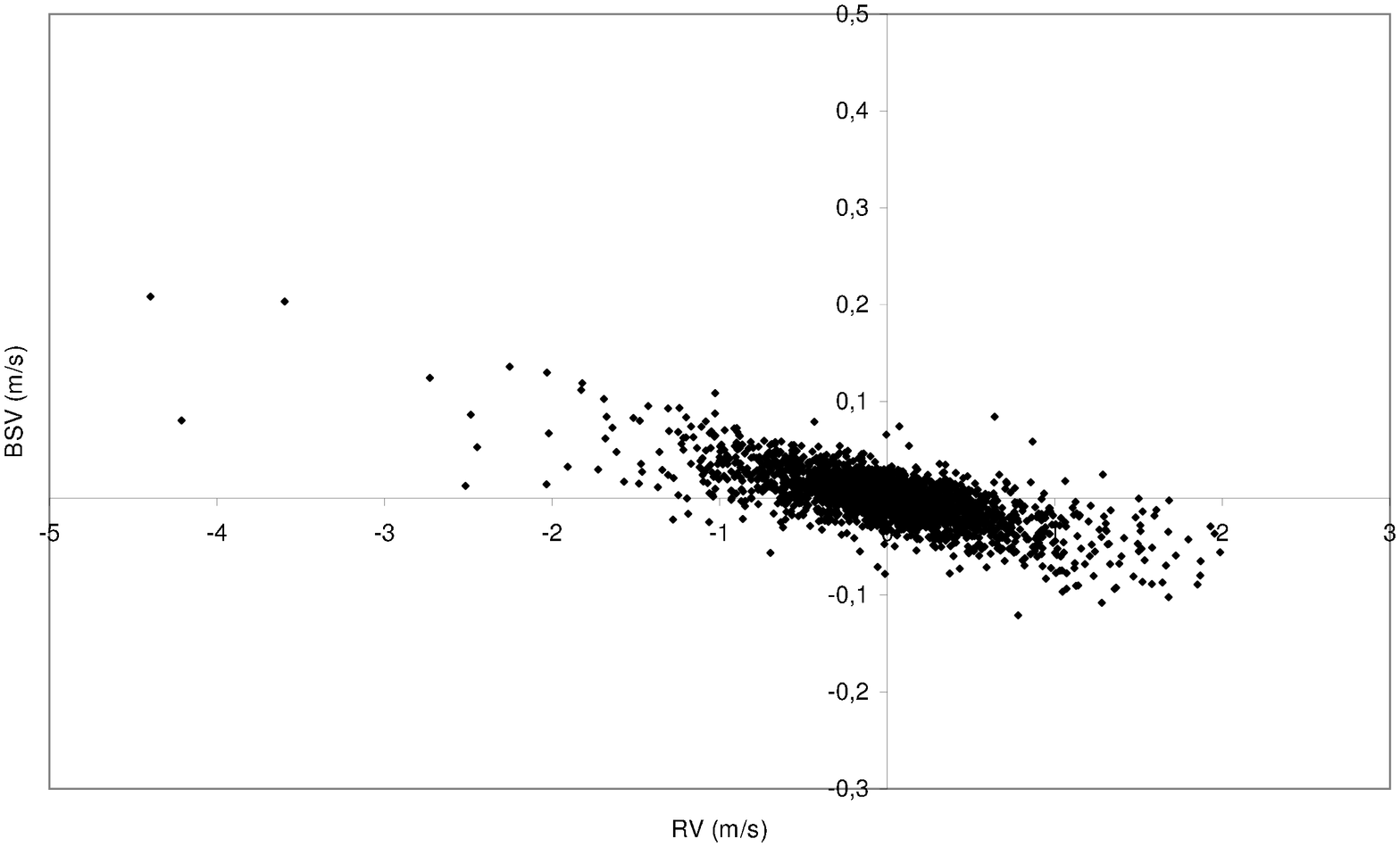}&\includegraphics[angle=-90,width=0.4\hsize]{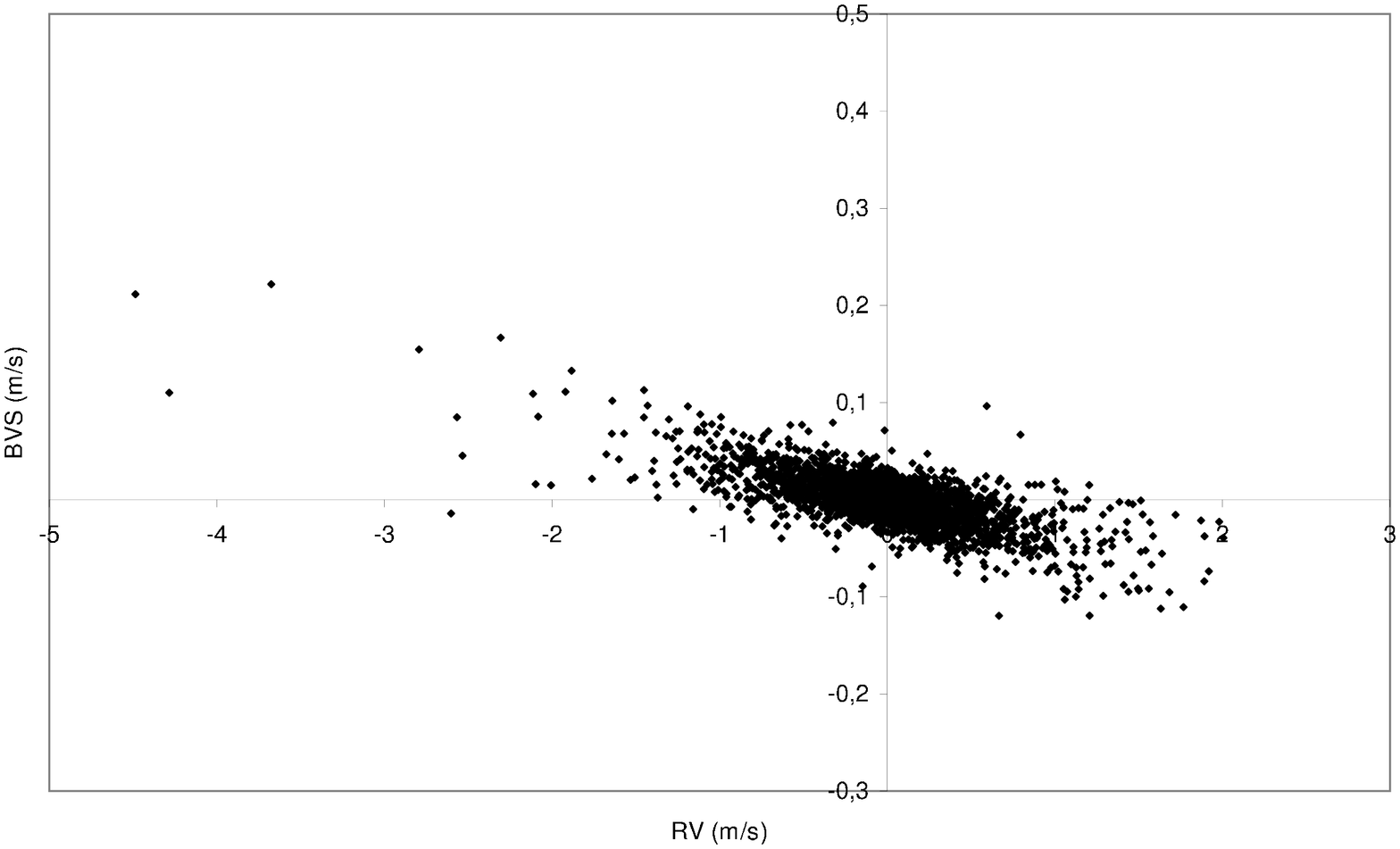}\\
\includegraphics[angle=-90,width=0.4\hsize]{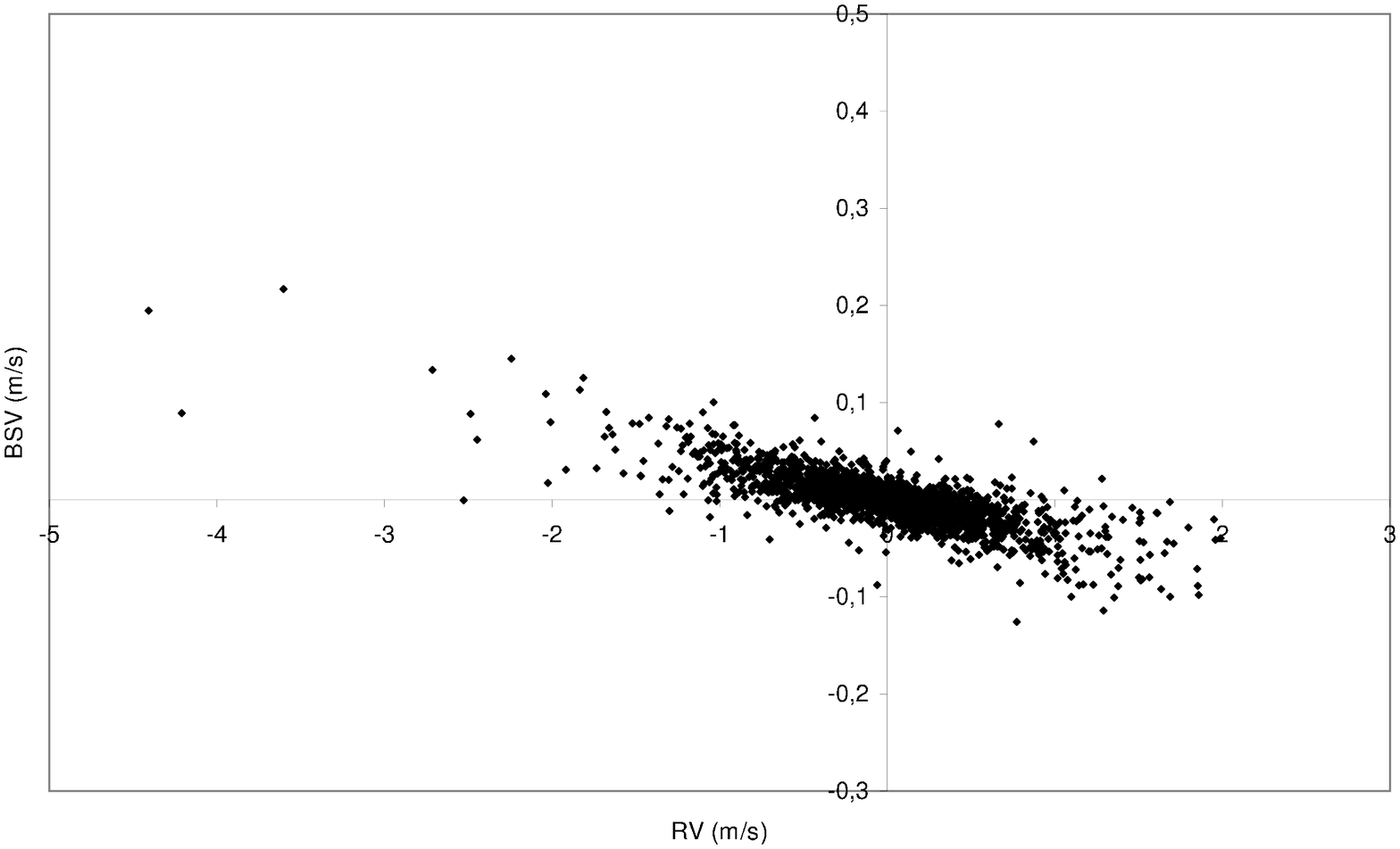}&\includegraphics[angle=-90,width=0.4\hsize]{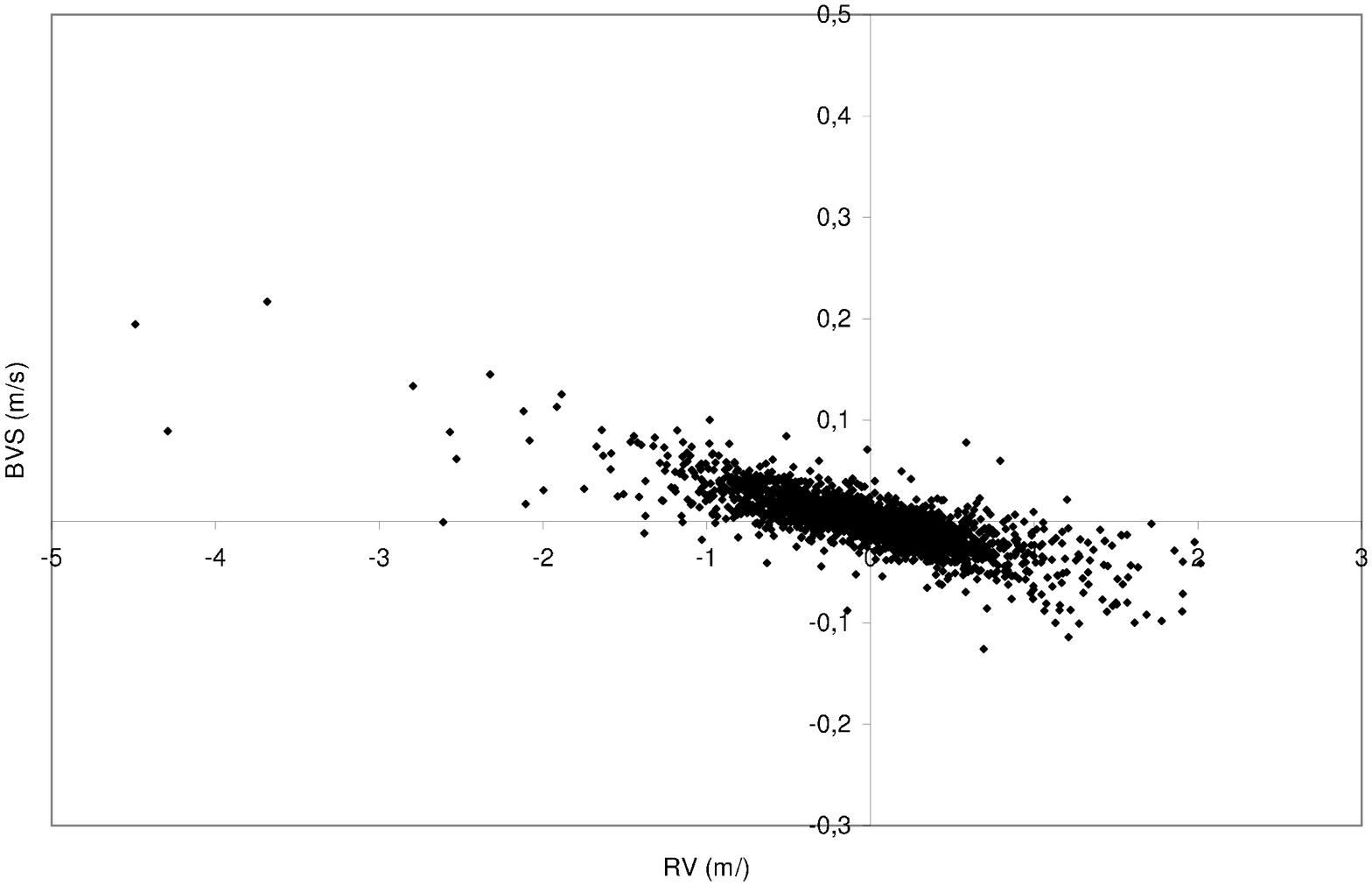}\\
\end{tabular} 
 \caption{Left: (RV; BVS) diagrams for a spotted solar-type star surrounded by a 1 \me planet on a circular orbit at 1.2 AU. The star is assumed to be observable 8 months per year. From top to bottom, we consider precisions of 1 cm/s, 5cm/s and 10 cm/s, as well as an infinite precision. The spot induced RV-BVS correlation is quite clear in the data without noise, and with a 1cm/s precision; it is marginally seen when considering a precision of 5 cm/s and is lost when considering a 10 cm/s precision. Right: same, assuming that the spotted star is not surrounded by a planet. The correlation is quite clear in the data without noise, or with a low noise (1 cm/s), and is not so clear when considering a noise of 5 cm/s and lost when considering a 10 cm/s noise. The comparison of the left and right figures shows that in this case the planet does not impact the RV-BSV correlation.}
\label{rvtot_bsv_noise_550}
\end{figure}


\vfill\eject

\begin{figure}[ht!]
  \centering
 \begin{tabular}{ccc}
\includegraphics[angle=90,width=0.3\hsize]{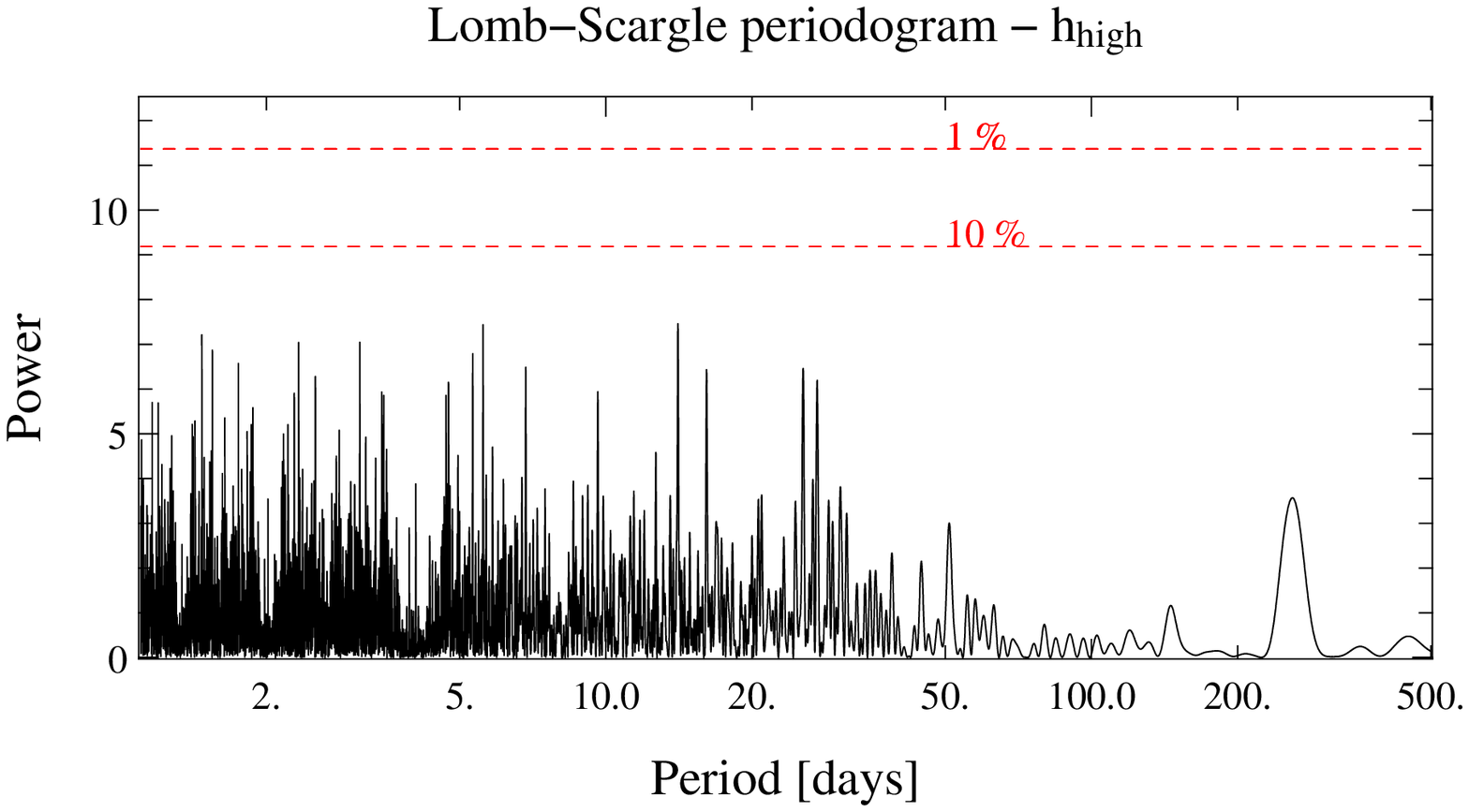}&\includegraphics[angle=90,width=0.3\hsize]{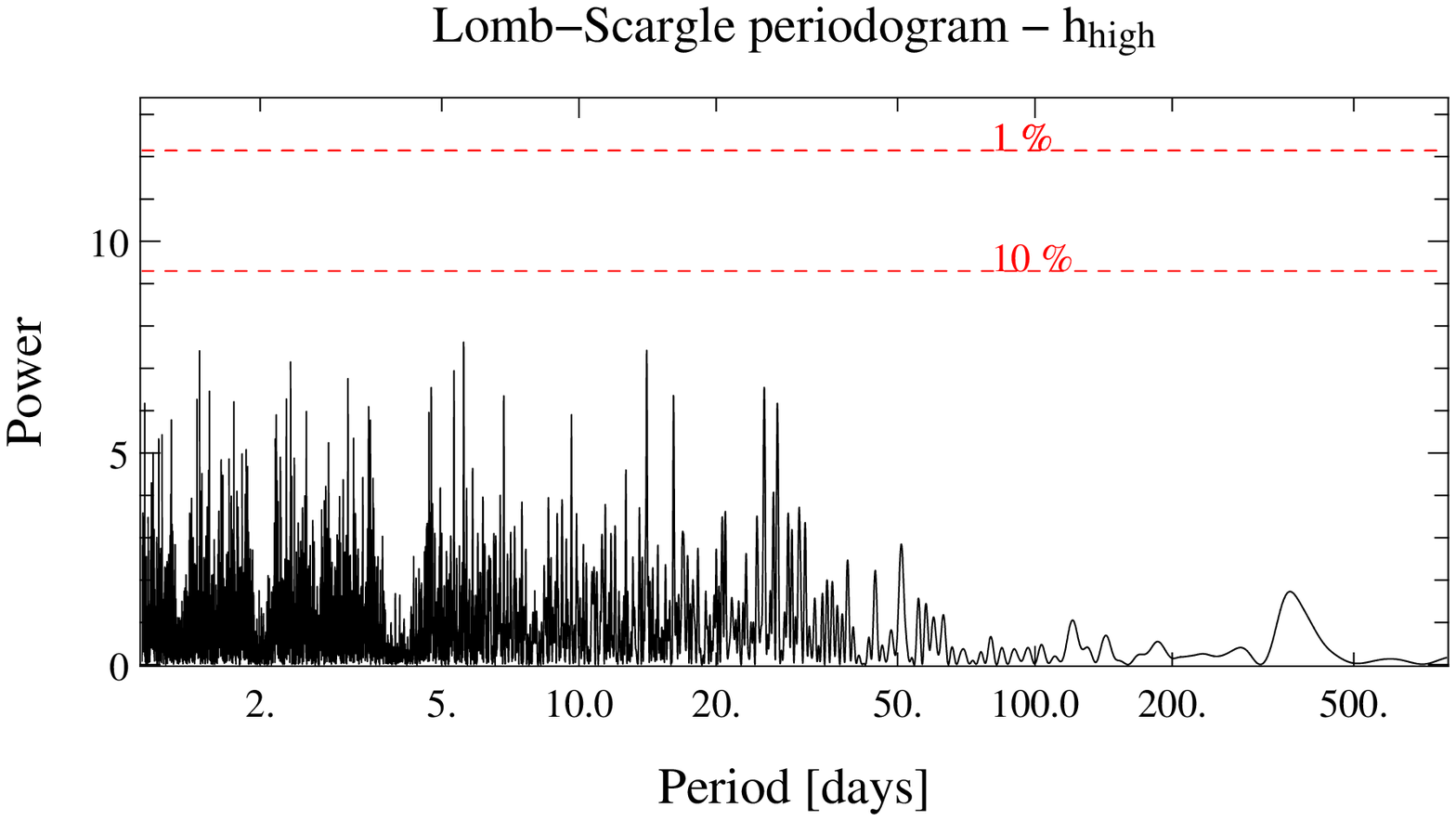}&\includegraphics[angle=90,width=0.3\hsize]{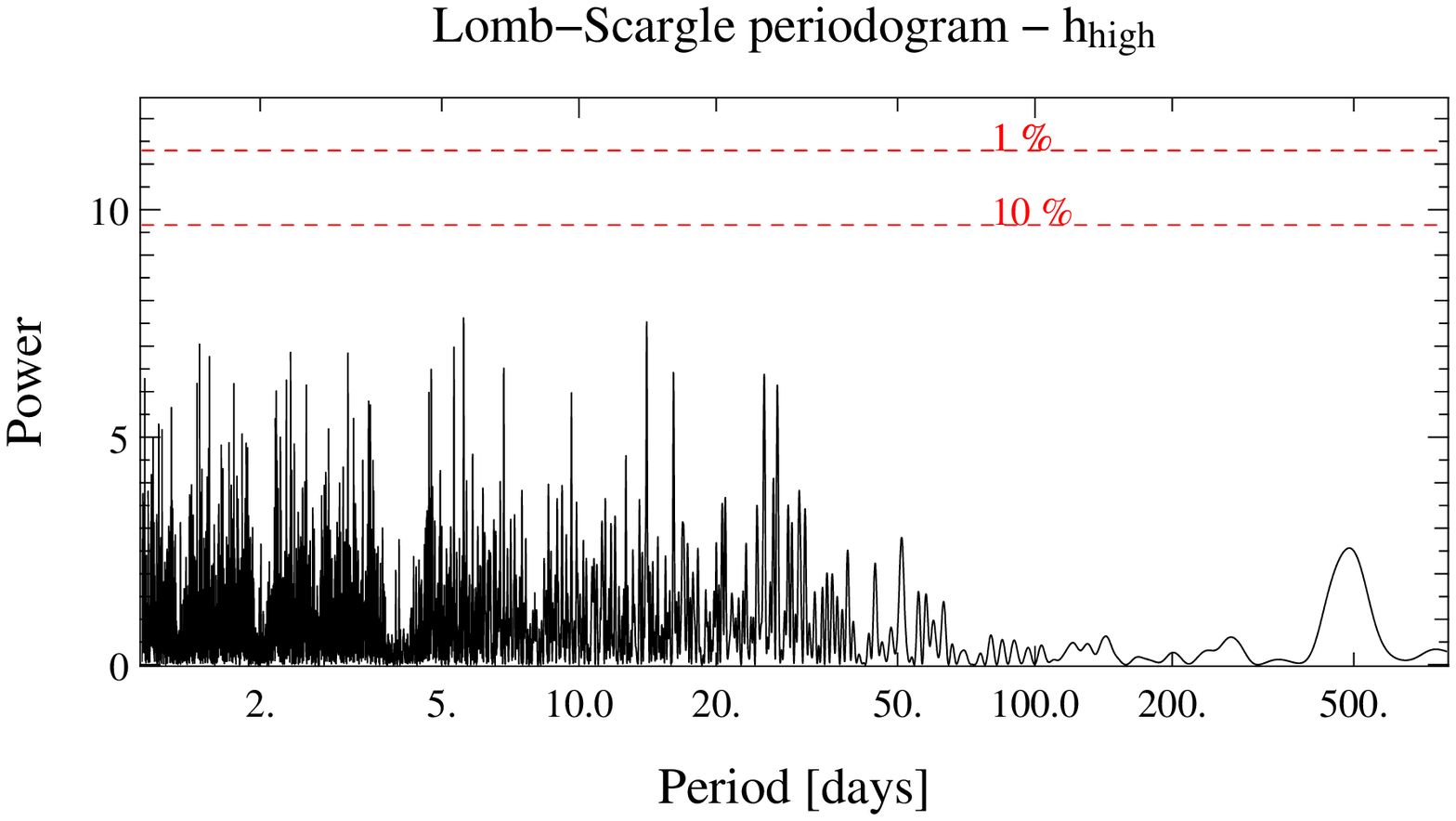}\\
\includegraphics[angle=90,width=0.3\hsize]{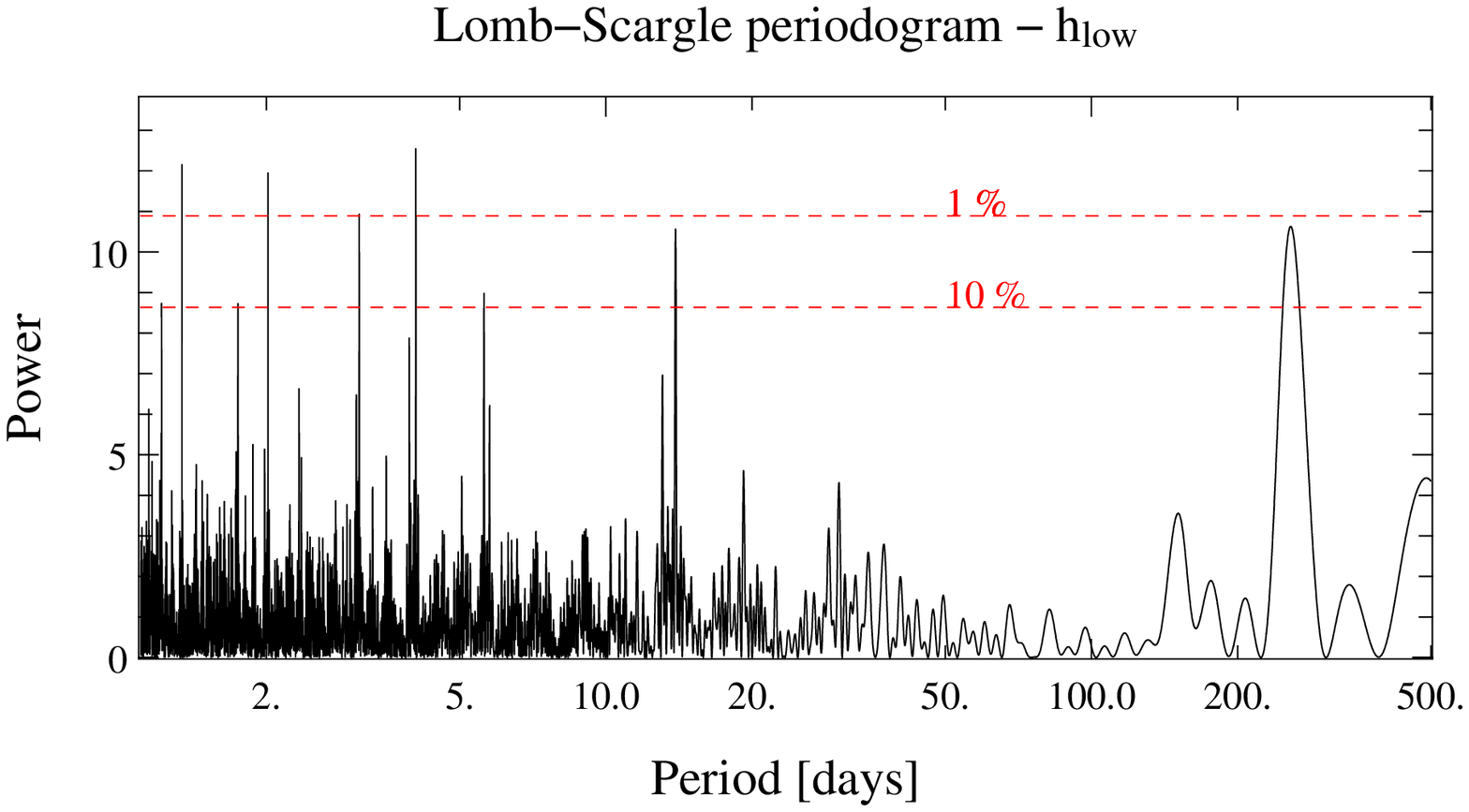}&\includegraphics[angle=90,width=0.3\hsize]{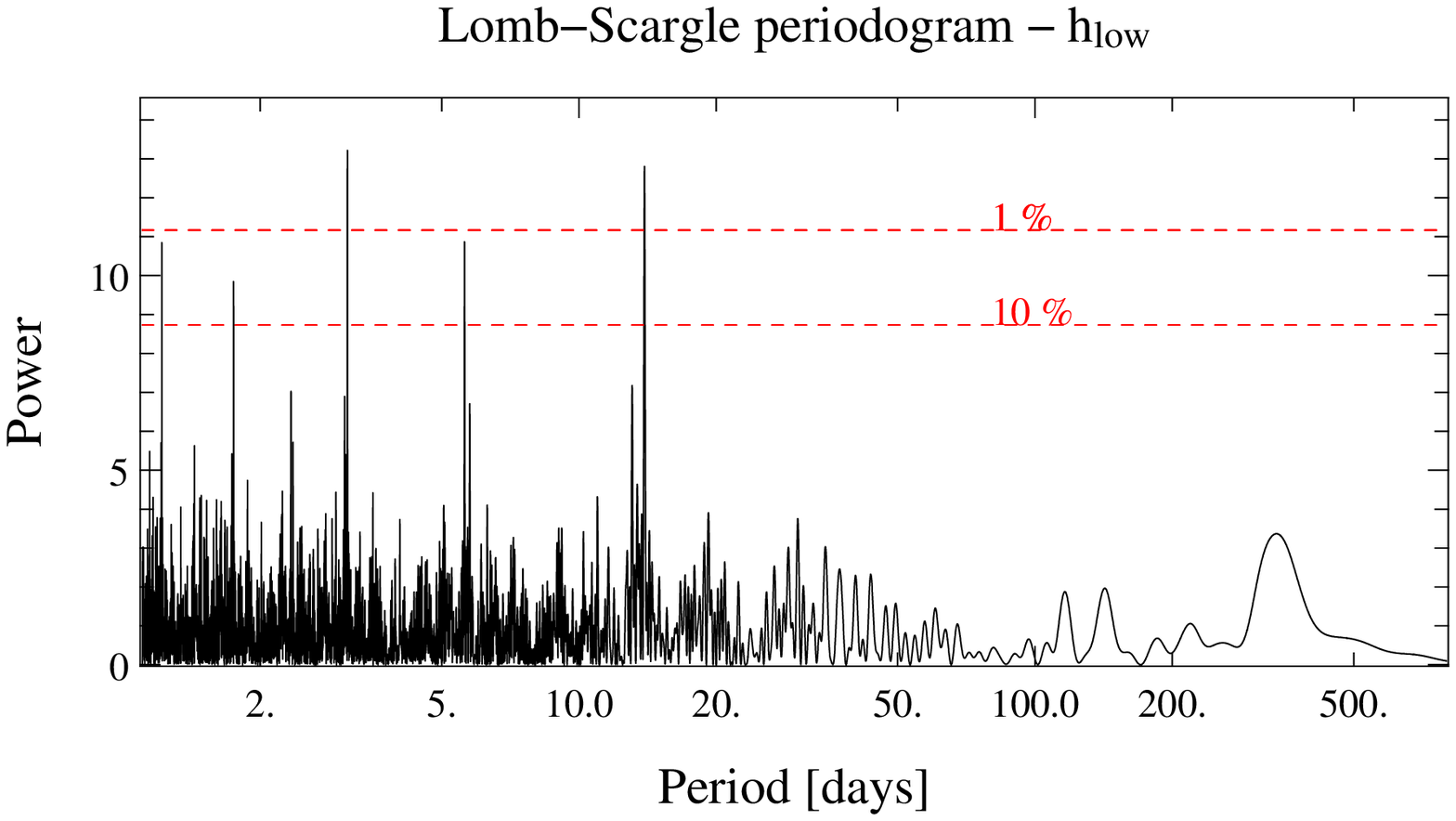}&\includegraphics[angle=90,width=0.3\hsize]{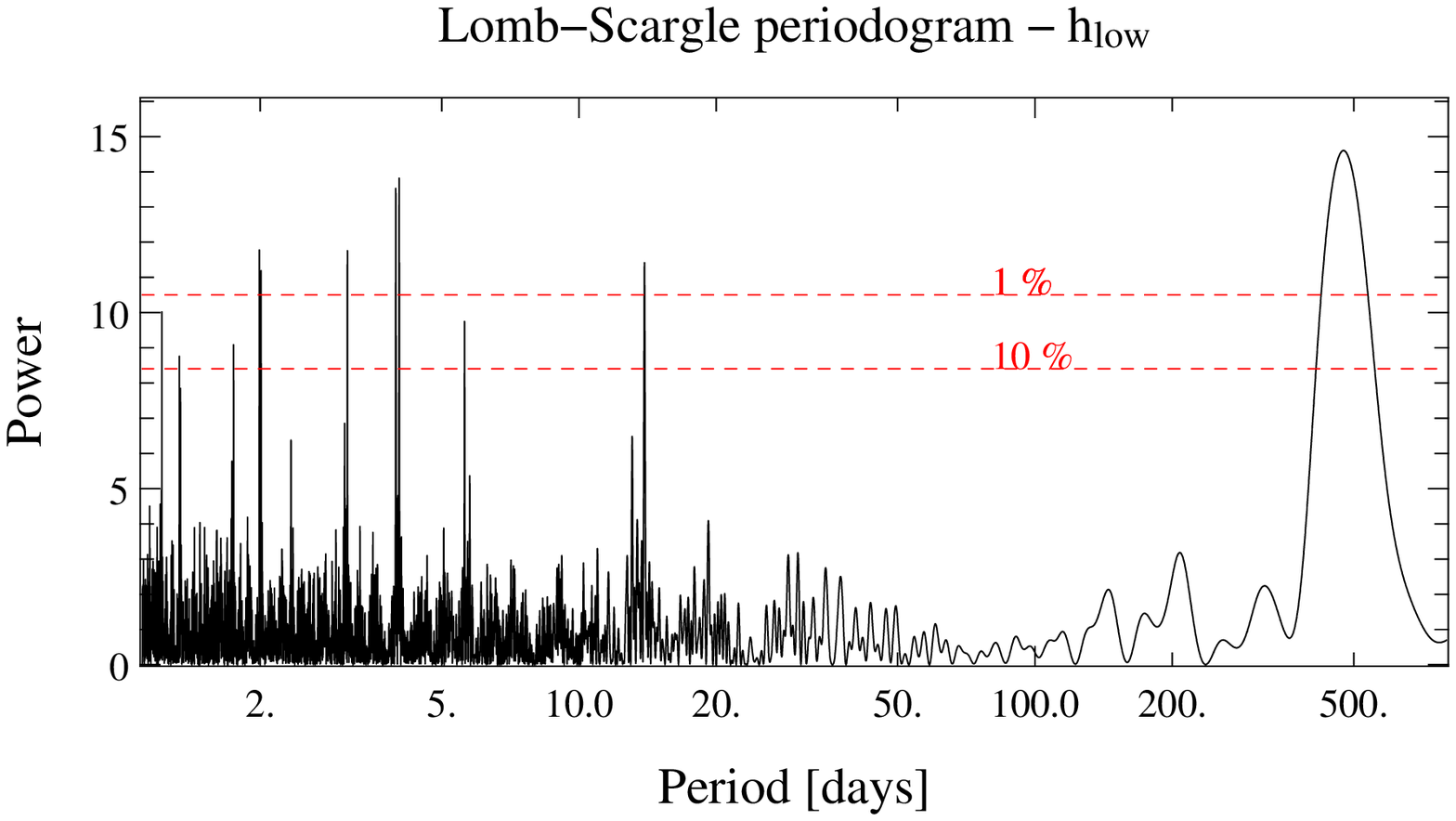}\\
\includegraphics[angle=90,width=0.3\hsize]{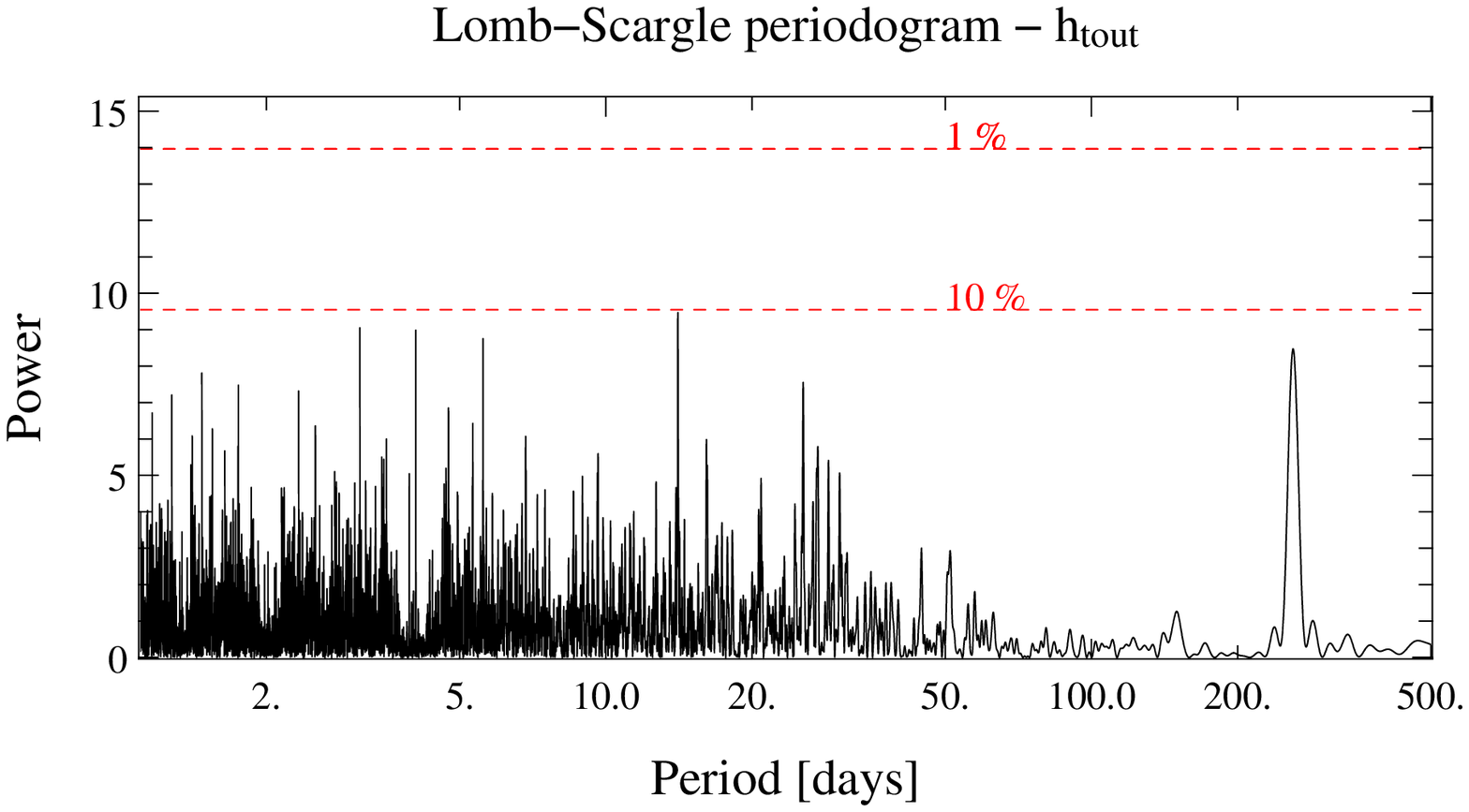}&\includegraphics[angle=90,width=0.3\hsize]{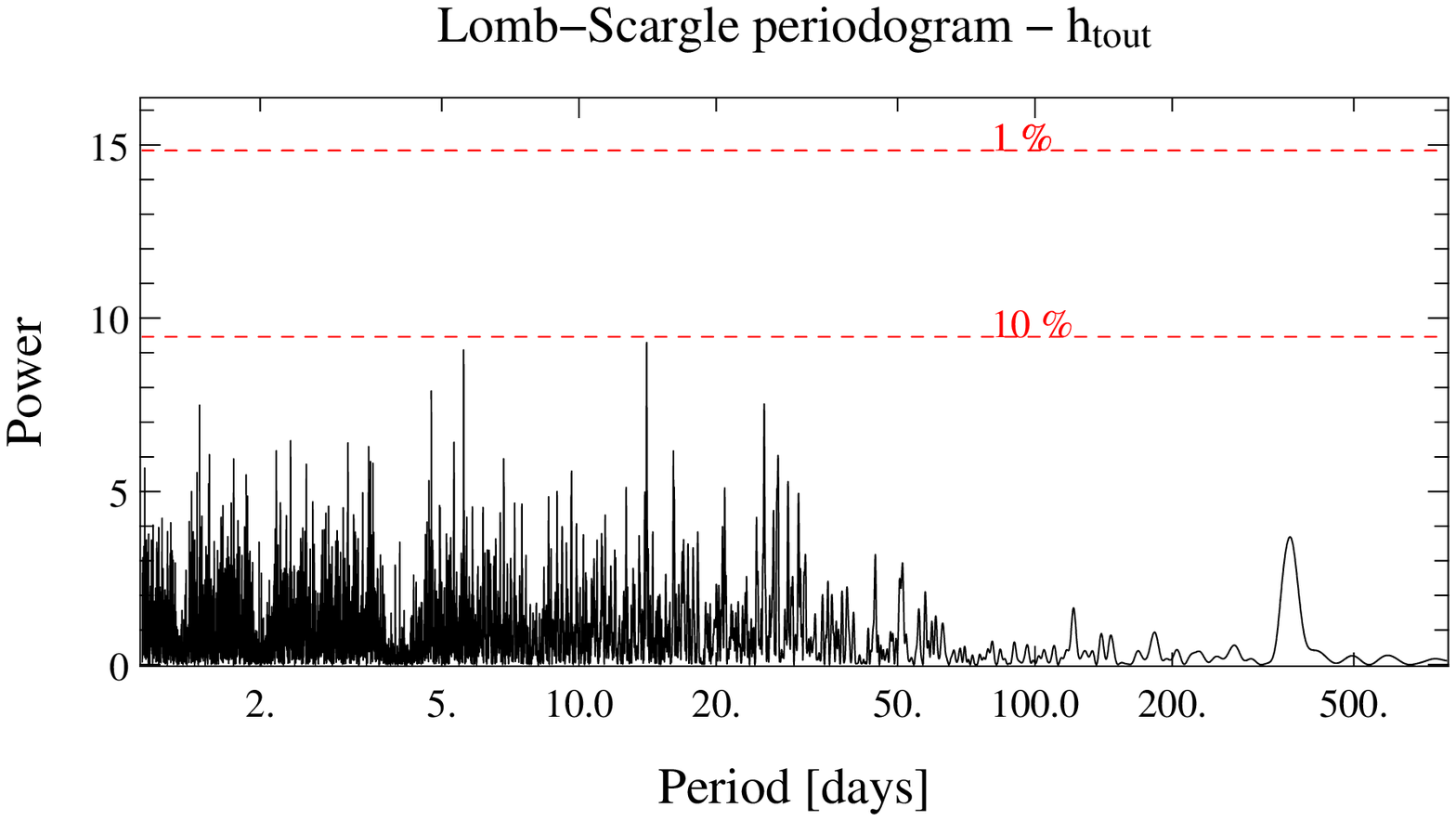}&\includegraphics[angle=90,width=0.3\hsize]{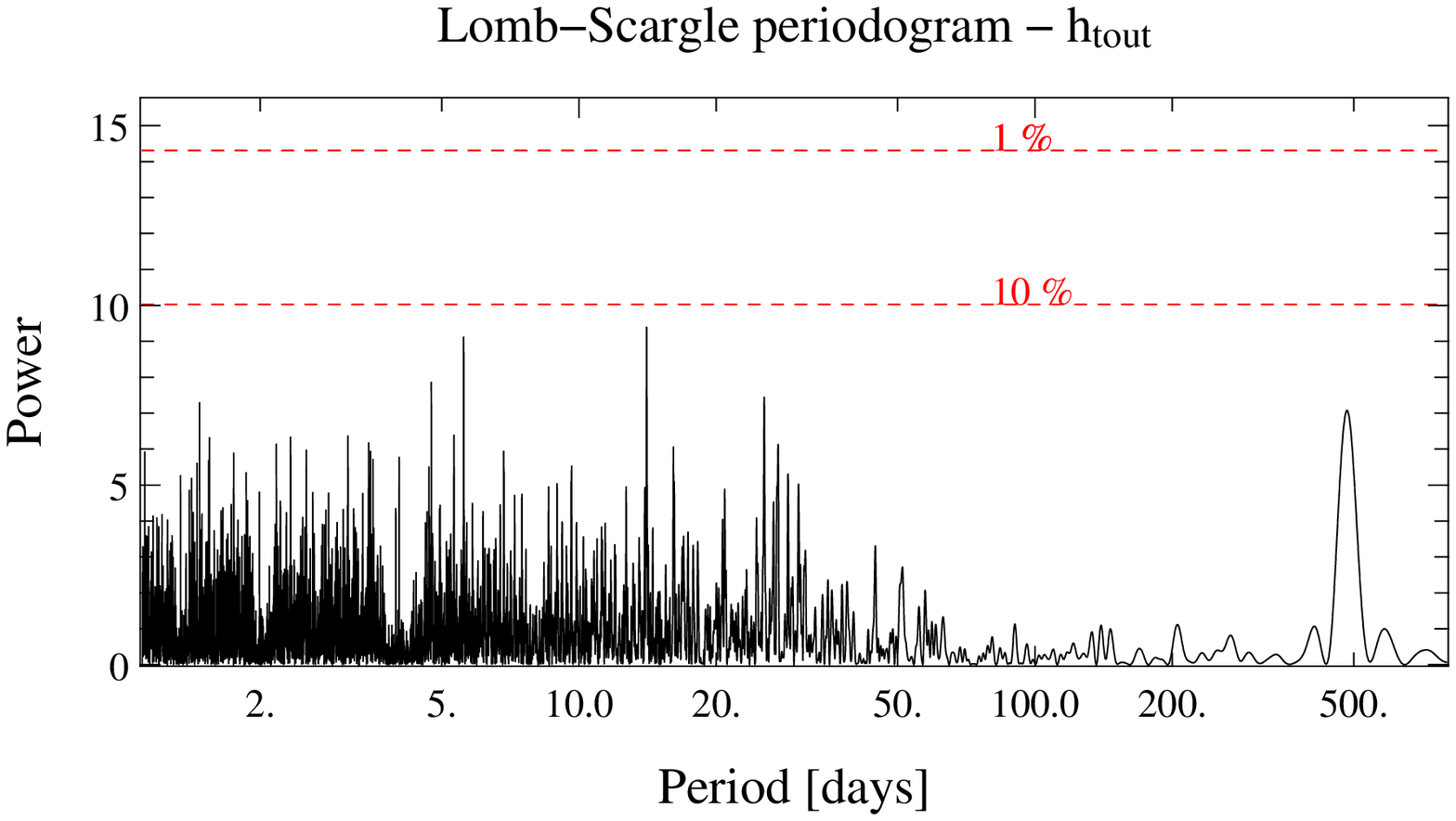}\\
 \end{tabular} 
 \caption{Impact of the orbital radius. Periodograms corresponding to the RV of a spotted solar type star surrounded by a 1\me planet orbiting at resp. 0.8 (Top), 1 (Middle) and 1.2 (Bottom) AU. The star is again assumed to be observable 8 months per year, and to be actually observed every 4 days, during the whole cycle (Left), the low (Middle) and high (Right) activity periods, with a precision of 5 cm/s.}
\label{4m_4d_5cms}
\end{figure}

\begin{figure}[ht!]
  \centering
\begin{tabular}{cc}
\includegraphics[angle=90,width=0.3\hsize]{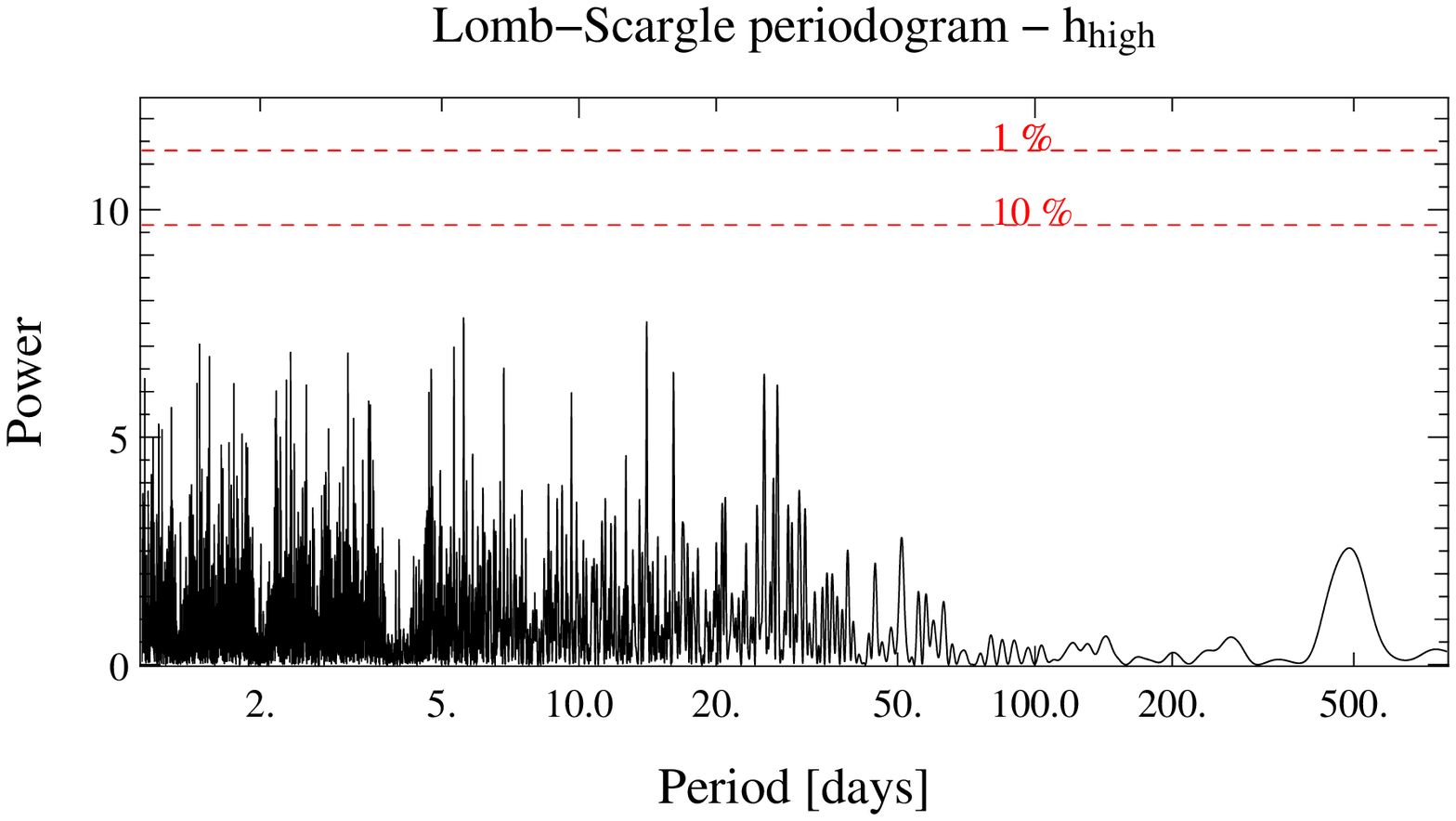}& \includegraphics[angle=90,width=0.3\hsize]{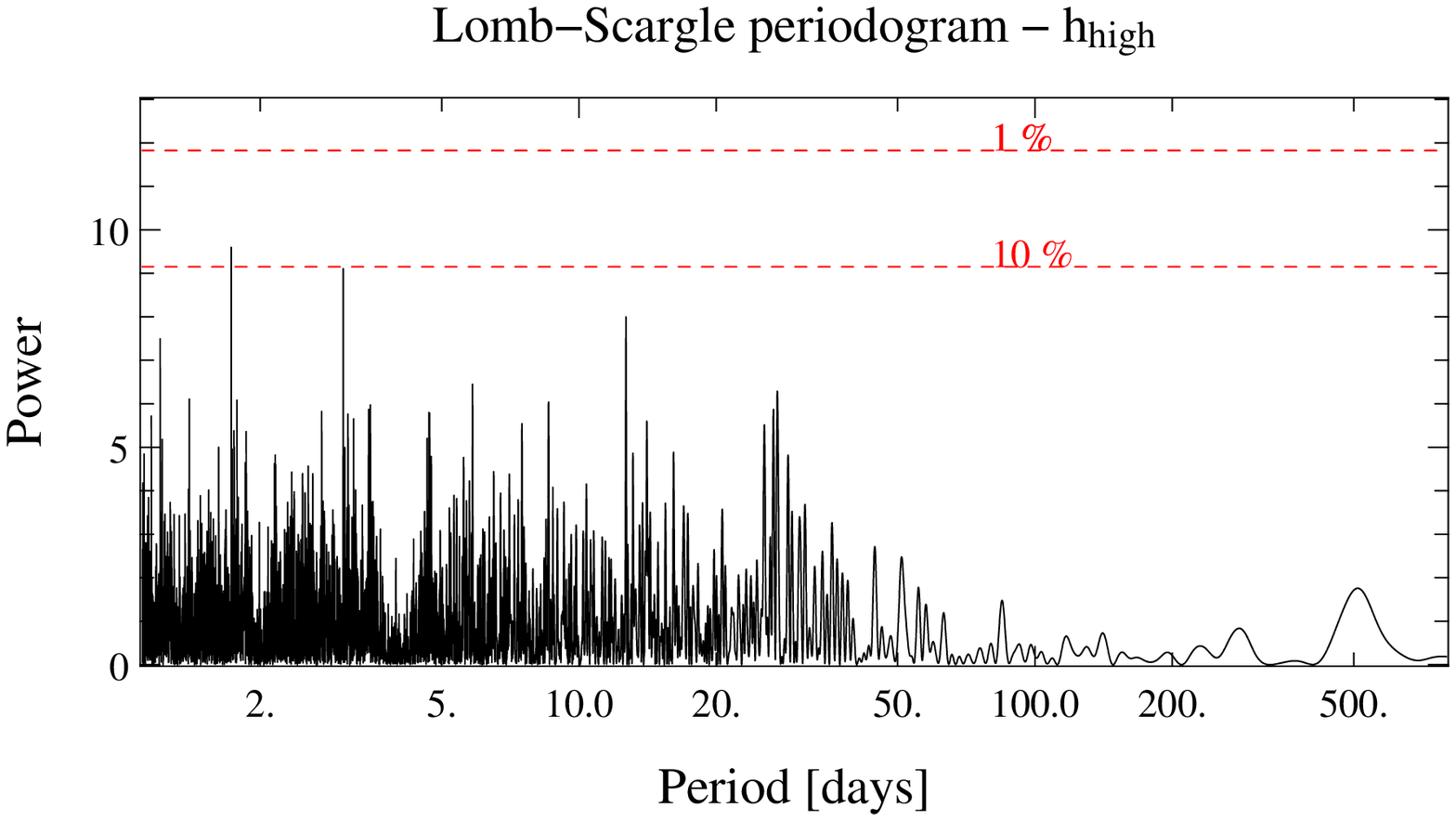} \\
\includegraphics[angle=90,width=0.3\hsize]{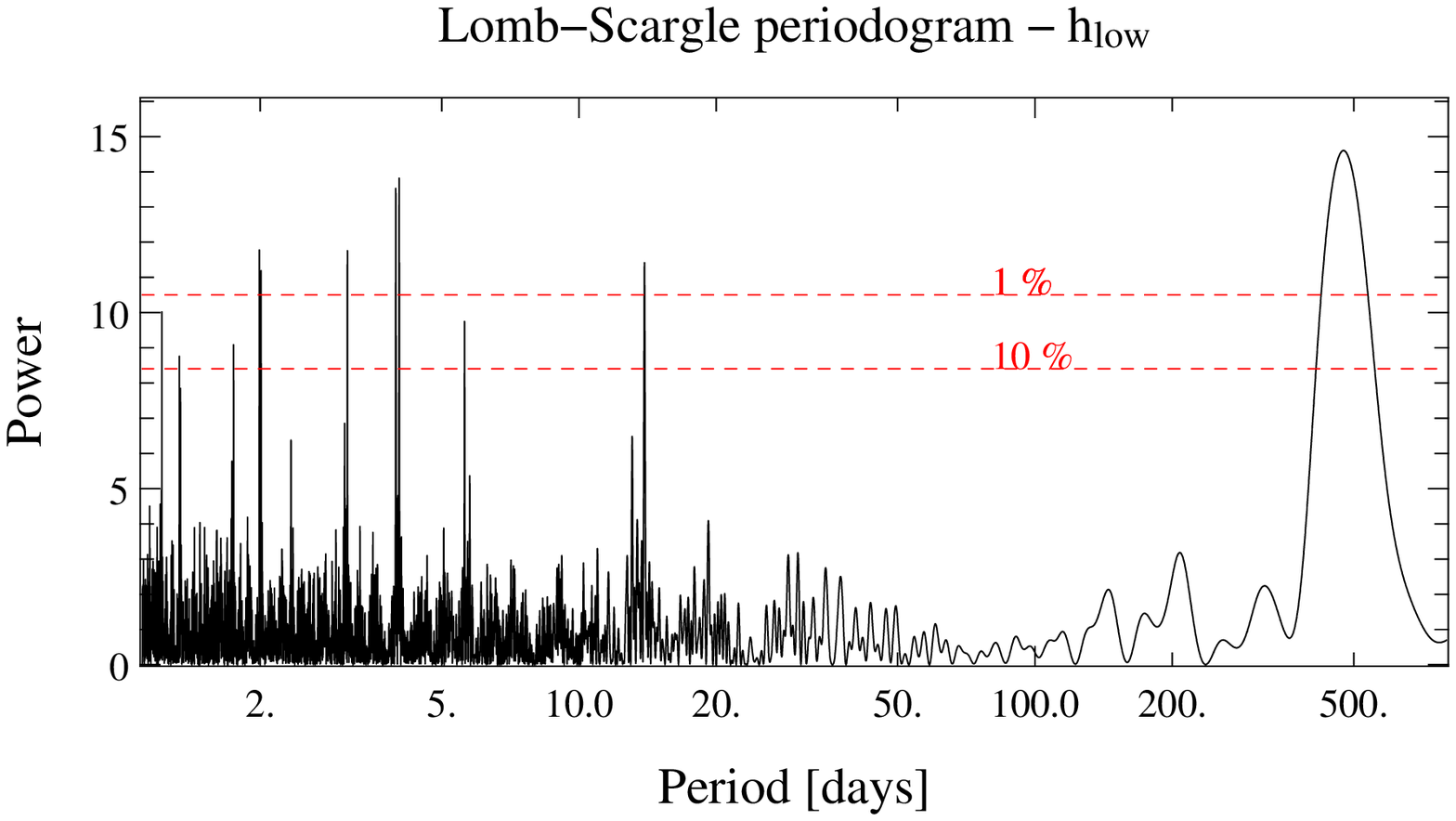} & \includegraphics[angle=90,width=0.3\hsize]{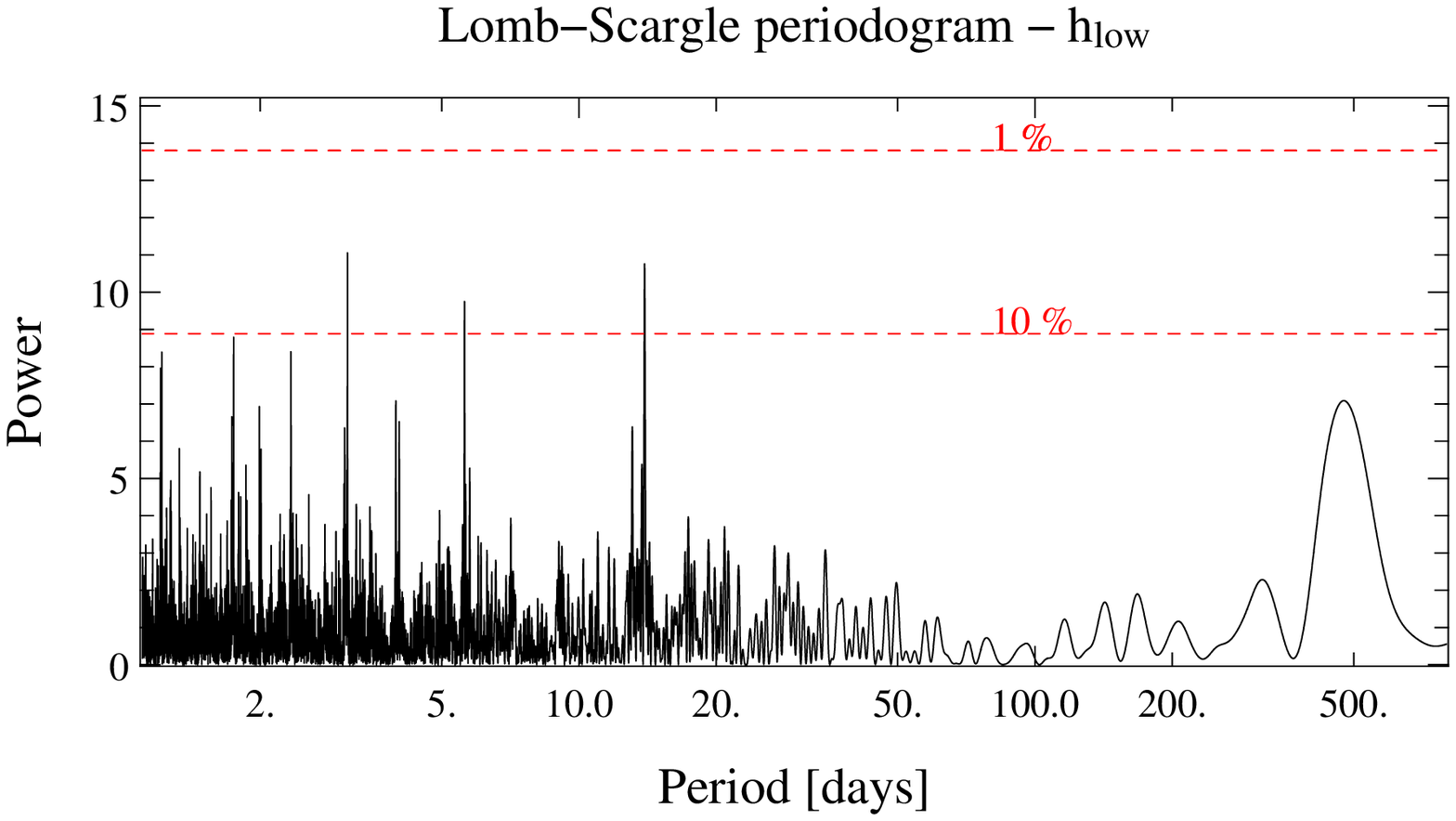} \\
\includegraphics[angle=90,width=0.3\hsize]{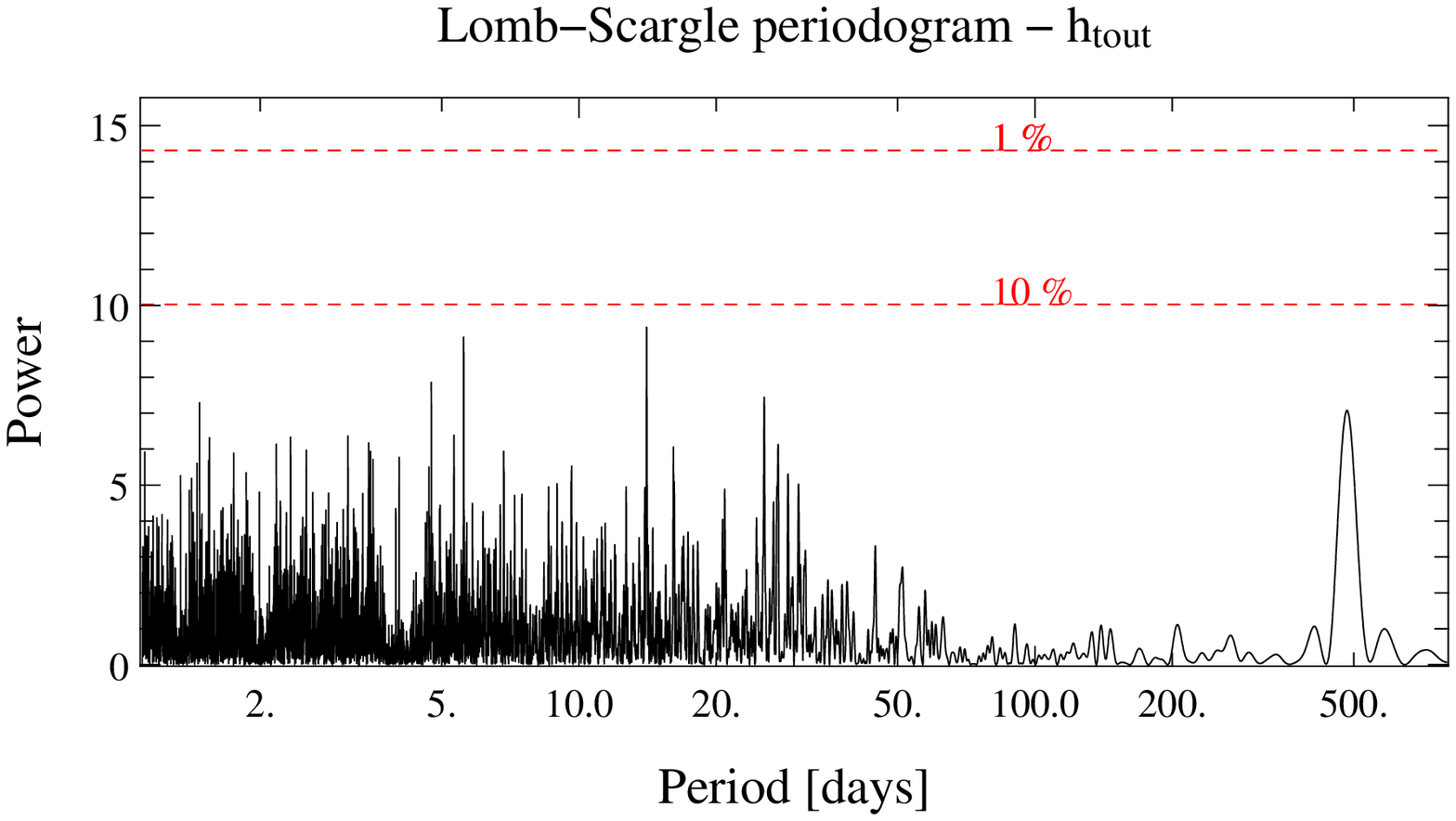}& \includegraphics[angle=90,width=0.3\hsize]{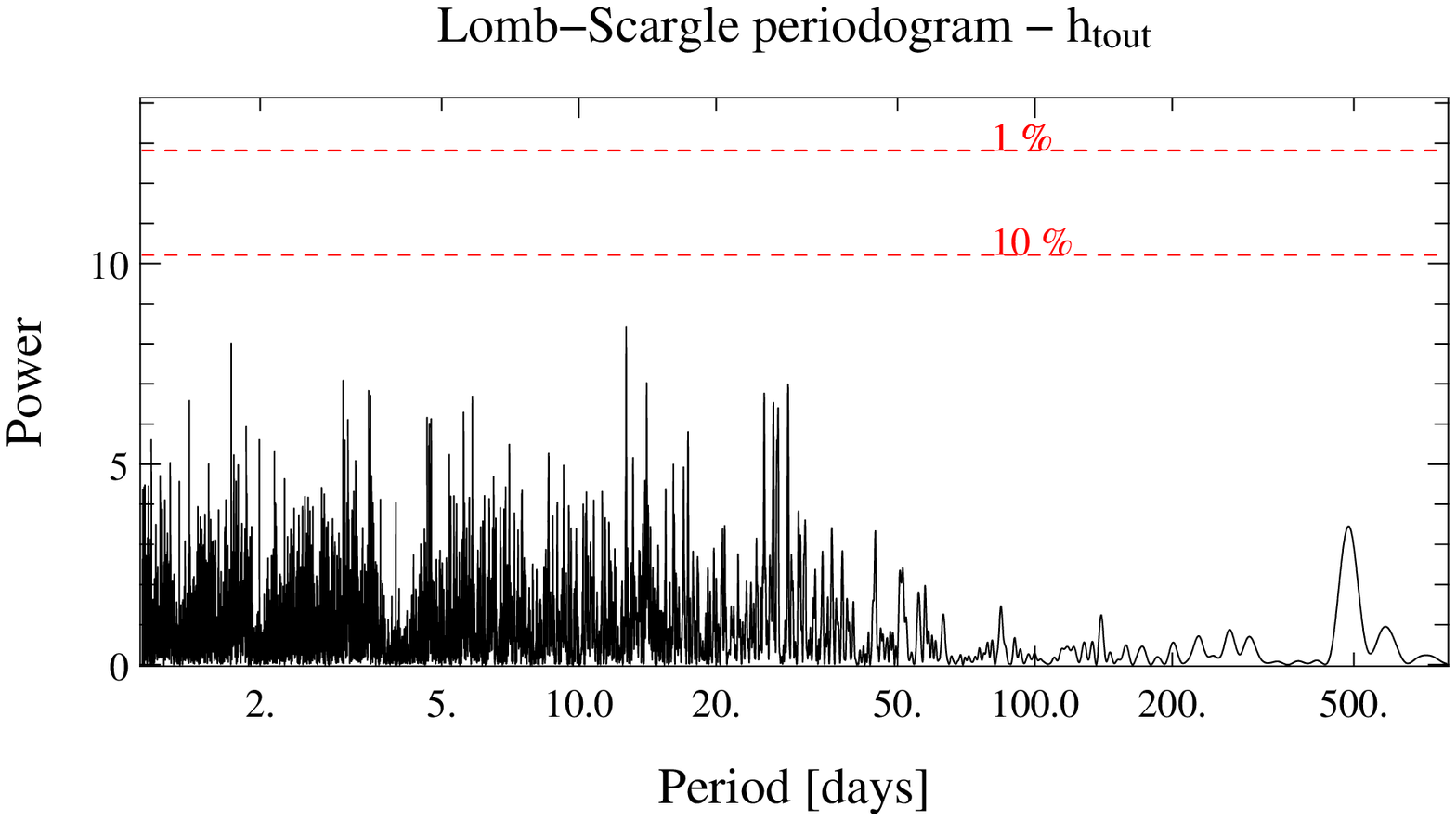}\\
 \end{tabular} 
 \caption{ Impact of the spots temperature. periodograms corresponding to the RV of a spotted solar type star surrounded by a 1\me\  planet orbiting at 1.2  AU. The spot temperature is either  (Top) or 1200 K (Bottom) below the star effective temperature. The star is again assumed to be observable 8 months per year, and to be actually observed every 4 days, with a precision of 5 cm/s. From Left to Right, the whole cycle, the low activity period and the high activity period are considered.}
\label{4m_4d_impacttemp}
\end{figure}

\vfill\eject

\begin{figure}[ht!]
  \centering
\includegraphics[angle=0,width=\hsize]{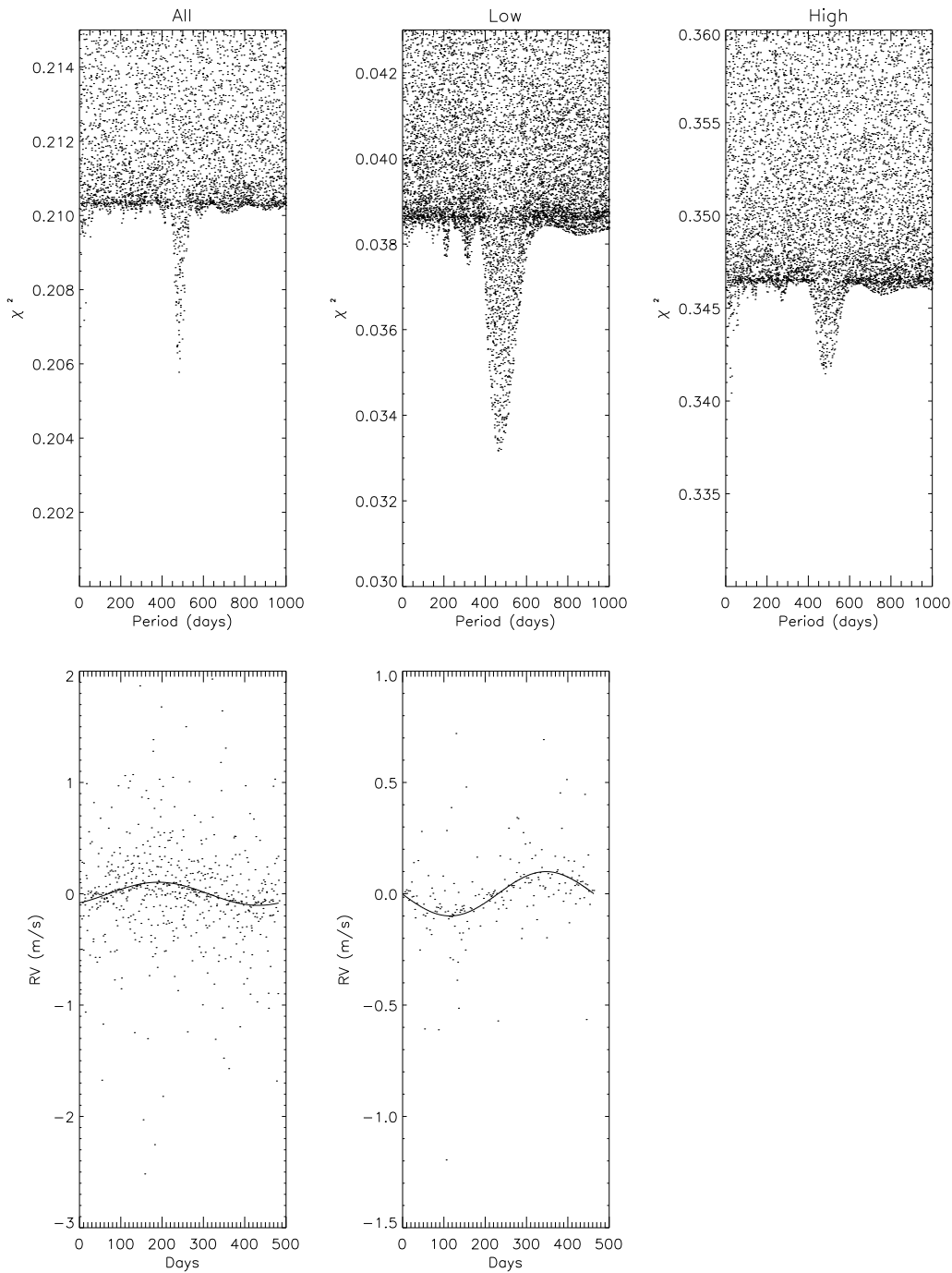}
\caption{ Fit of the RV data and parameters found in the case of a 1\me\  planet orbiting at 1.2 AU. The spotted star is assumed to be observed 8 months per year, with a 4 days temporal sampling (see the corresponding RV curves  in Figure~\ref{rv_all}). A 1 cm/s noise level is considered. Left) 
the whole cycle is considered; Middle) the low activity period is considered, and Right) the high activity periods is considered. Top : $\chi2$ versus period for a large number of realizations. The minimum  $\chi2$  corresponds to the planet period. Bottom : observed (dots) and fitted (solid line) RV versus time after folding of the time scale according to the fitted period.}
\label{fits_1p2_4m_4d_1cms}
\end{figure}

\end{document}